\documentclass[10pt,letterpaper]{article}
\UseRawInputEncoding
\usepackage[top=0.85in,left=2.75in,footskip=0.75in]{geometry}

\usepackage{amsmath,amssymb}

\usepackage{changepage}

\usepackage{textcomp,marvosym}

\usepackage{cite}

\usepackage{amsthm}
\newtheorem{theorem}{Theorem}

\newtheorem{corollary}{Corollary}
\newtheorem{definition}{Definition}

\newtheorem{proposition}{Proposition}  

\usepackage{tabularx}

\usepackage{nameref}

\usepackage[right]{lineno}

\usepackage[nopatch=eqnum]{microtype}
\DisableLigatures[f]{encoding = *, family = * }

\usepackage[table]{xcolor}

\usepackage{array}

\newcolumntype{+}{!{\vrule width 2pt}}

\newlength\savedwidth

\raggedright
\usepackage[aboveskip=1pt,labelfont=bf,labelsep=period,justification=raggedright,singlelinecheck=off]{caption}

\makeatletter
\renewcommand{\@biblabel}[1]{\quad#1.}
\makeatother

\usepackage{lastpage,fancyhdr,graphicx}
\usepackage{epstopdf}
\fancyheadoffset[L]{2.25in}
\fancyfootoffset[L]{2.25in}

\usepackage{siunitx}
\usepackage{csquotes}
\usepackage{algorithmicx}
\usepackage{algcompatible}
\usepackage{booktabs}
\usepackage{longtable}
\usepackage{adjustbox}
\usepackage{array}
\usepackage{url}
\usepackage{titlesec}
\usepackage{authblk}
\usepackage{graphicx}
\usepackage{algorithm}
\usepackage{algpseudocode}
\usepackage{tabularx}
\usepackage{float}
\usepackage{placeins}
\usepackage{comment}
\usepackage{multirow}
\usepackage{tikz}
\usetikzlibrary{decorations.pathreplacing, arrows.meta}
\usepackage{rotating}

\newcommand{\orcid}[1]{\href{#1}{ORCID}}

\usepackage[hidelinks]{hyperref}
\usepackage{cleveref}
\crefformat{figure}{#2Fig~#1#3}
\Crefformat{figure}{#2Fig~#1#3}
\crefformat{table}{#2Table~#1#3}
\Crefformat{table}{#2Table~#1#3}
\crefformat{section}{#2Section~#1#3}
\Crefformat{section}{#2Section~#1#3}
\crefformat{algorithm}{#2Algorithm~#1#3}
\Crefformat{algorithm}{#2Algorithm~#1#3}
\crefformat{appendix}{#2Appendix~#1#3}
\Crefformat{appendix}{#2Appendix~#1#3}

\makeatletter
\renewcommand{\theequation}{\arabic{section}.\arabic{equation}}
\@addtoreset{equation}{section}
\makeatother

\hypersetup{unicode=true}
\titleformat{\subsection}
  {\mdseries\itshape\large}
  {\thesubsection}{1em}{}

\usepackage{chngcntr}

\begin{document}
\vspace*{0.2in}

\begin{flushleft}
{\Large
\textbf{Calibration-guided data fusion: A framework for understanding data contribution in multi-source inference through efficient Bayesian optimization with application to virus dynamics parameter estimation}
}
\newline
\\
Yingying Xu\textsuperscript{1,2,3,4*},
\\
\bigskip
\textbf{1} Independent Researcher
\\
\textbf{2} Department of Mathematics and Statistics, University of Helsinki, Helsinki, Finland
\\
\textbf{3} Finnish Center for Artificial Intelligence (FCAI), Espoo, Finland
\\
\textbf{4} RIKEN Center for Interdisciplinary Theoretical and Mathematical Sciences (iTHEMS), Wako, Japan
\\
\bigskip

* yingying.xu.kyo@gmail.com

\end{flushleft}

\section*{Abstract}
We introduce a framework that uses simulation-based calibration to assess data contribution in multi-source inference. The framework systematically evaluates what each data source contributes to parameter estimation, enabled by computational efficiency such that when inference is fast enough, calibration becomes a practical diagnostic tool.

We demonstrate the framework using Bayesian optimization likelihood-free inference (BOLFI) applied to a target cell limited model of influenza A virus dynamics, combining RNA measurements and endpoint dilution assay data from both single-cycle and multiple-cycle experiments, with an ordinary differential equation model for RNA dynamics and a stochastic simulator for the endpoint dilution assay. BOLFI achieves a 20-fold speed-up over traditional MCMC, illustrating the substantial efficiency gain despite differences in computing architecture, and requires no likelihood knowledge. A Gaussian process classifier handles simulation failures, and a discrepancy design integrates heterogeneous data types. Using 100 synthetic datasets per configuration, we compare five data configurations. 

The results reveal a counter-intuitive finding: combining all data sources does not always improve inference. The single-cycle experiment alone is most reliable for estimating the virion entry rate ($\beta$), while the ED-only data is surprisingly best for the total RNA production rate ($P_{\text{RNA}}$), outperforming the combined data. ED data, when combined with RNA data, improves the calibration of the infectious virion production rate ($P$) and the duration of the infectious phase ($\tau_I$). In contrast, for the eclipse phase duration ($\tau_E$), the multiple-cycle experiment alone (combining both its RNA and ED measurements) is more reliable than the combined data, indicating that adding single-cycle data introduces inconsistency. For the infection establishment rate ($\gamma$), all configurations yield well-centered posteriors with wide tails, reflecting its partial identifiability. This reveals a trade-off between precision (combining all data) and calibration (using the most reliable source for each parameter).
Based on these findings, we explore a composite diagnostic that selects the best-calibrated data type for each parameter. This approach serves as a sensitivity check rather than a formal replacement for combined inference. The inference also reveals a 35-fold disparity between RNA and infectious virus production, highlighting a key bottleneck in viral replication. 

This work demonstrates that computational efficiency enables a new level of methodological rigor: systematic calibration-guided data fusion. The framework is general and applicable to other multi-source inference problems where simulation-based models are used.

\section*{Author summary}
Traditional inference methods for viral dynamics typically require explicit likelihood functions and substantial computation, often on the order of tens of hours, limiting the analyses researchers can perform. We present a framework that combines computational efficiency with systematic analysis of data contribution. Our method, Bayesian optimization likelihood-free inference (BOLFI), completes inference in approximately 2 hours, substantially faster than traditional MCMC. The inference pipeline itself is virus-agnostic and can be applied to other viral systems with similar experimental data.

This efficiency enables us to introduce a calibration-based framework to assess what each data source actually contributes to parameter estimation. By comparing five data configurations, we reveal a counter-intuitive finding: combining all data sources does not always improve inference. For example, data on infectious virus (ED) alone outperforms combined data to estimate the total RNA production rate, while data from multiple-cycle experiments alone is more reliable for estimating the duration of the eclipse phase. Based on this, we explore a composite diagnostic that selects the best-calibrated data type for each parameter, serving as a sensitivity check on combined inference. This framework is general and applicable to other multi-source inference problems where simulation-based models are used, regardless of the scientific domain.

Keywords: calibration, Bayesian optimization, likelihood-free inference, virus dynamics, multi-source data, data fusion


\section{Introduction}
\label{sec:introduction}

Combining multiple data sources is common practice in data-driven science. The underlying assumption is intuitive: more data leads to better inference.

\subsection*{Data Contribution Analysis: Current Approaches and Gaps}
There is a rich literature on quantifying the contribution of different data sources to an analysis. In machine learning and statistics, methods such as Shapley values \cite{shapley1953value}, feature importance \cite{breiman2001random}, and Data Shapley \cite{ghorbani2019data} have been developed to quantify the contribution of individual features or data points to predictive performance. In machine learning, the Shapley value is applied predominantly to Explainable AI through feature attribution \cite{lundberg2017unified}. In multimodal learning, occlusion-based techniques and ablation studies are commonly used to assess modality importance \cite{liang2022multimodal}. In medical imaging, classification rate-based approaches measure the additive value of different imaging modalities \cite{calhoun2009feature, calhoun2017fusion}. In official statistics, quality frameworks assess data sources based on relevance, coverage, and validity \cite{eurostat2023quality}.

Recent extensions have incorporated uncertainty into these frameworks. Data Shapley has been extended with confidence intervals to quantify the uncertainty of contribution estimates \cite{chavda2024uncertainty}. Shapley values have been adapted for uncertain data, measuring expected contribution and variance from random distributions \cite{han2024shapley}. InfoSHAP quantifies contribution to predictive uncertainty rather than point estimates \cite{watson2023explaining}. These developments represent important progress in understanding data contribution for predictive tasks. However, a fundamental distinction remains: these methods assess contribution to prediction performance or model uncertainty, not inference reliability.
They ask: ``Does this data source improve predictions?'' or ``How much does this data source reduce predictive uncertainty?'' They do not assess contribution to inference reliability, which is whether the credible intervals produced by the model are correctly calibrated. A model can produce accurate point estimates while still having poorly calibrated uncertainty --- for example, a $90\%$ credible interval that contain the true value only $70\%$ of the time. This is a critical distinction. A data source can improve point estimates while still producing overconfident intervals --- intervals that are systematically too narrow and contain the true value less often than their stated coverage level would imply. Conversely, a data source may produce underconfident intervals that are unnecessarily wide, or may not improve point estimates but still provide valuable information for correctly quantifying uncertainty, helping to avoid miscalibration. This distinction between point estimate accuracy and uncertainty calibration is often overlooked in practice \cite{gelman2013bayesian}. For scientific inference, the reliability of credible intervals is as important as the accuracy of point estimates.

This is the gap that we address: using simulation-based calibration to assess what each data source contributes to the reliability of parameter-specific uncertainty quantification. To our knowledge, this is the first framework to do this. We introduce a calibration-guided data fusion framework that evaluates data contribution through the lens of inference reliability rather than prediction performance.

\subsection*{The Limits of Point Estimates and Interval Widths}

A common misconception is that accurate point estimates and correctly sized credible intervals guarantee reliable inference. Yet these measures alone do not ensure reliability.

\begin{table}[ht]
\centering
\caption{\bf What different diagnostic measures tell us about inference quality.}
\label{tab:diagnostic_measures}
\begin{tabular}{>{\raggedright\arraybackslash}p{3.9cm}p{3.9cm}p{4.1cm}}
\hline
\textbf{What is checked} & \textbf{What it tells us} & \textbf{What it does not tell us} \\
\hline
Point estimate (mean, median) & Where the distribution is centered & Whether the uncertainty is correctly sized or placed \\
Interval width (variance, CI length) & How spread out the distribution is & Whether the interval is correctly placed \\
Simulation-based calibration & Whether the credible interval contains the true value at the claimed frequency &  \\
\hline
\end{tabular}
\end{table}

For example, consider a right-skewed distribution. The median may be 10 and the $90\%$ credible interval may be [8, 16] --- correct placement and correct width. However, because the distribution is skewed, the true value may fall in this interval only 80\% of the time. The point estimate and width cannot detect this. The only way to check whether credible intervals are correctly calibrated is simulation-based calibration (SBC) \cite{talts2018validating}.

\subsection*{The Role of Computational Efficiency}

Simulation-based calibration has traditionally been computationally expensive. Running hundreds of calibration cases requires hundreds of inference runs --- a prohibitive cost when each inference takes days or weeks.

Recent advances in likelihood-free inference (LFI) have changed this landscape. LFI techniques enable inference without explicit likelihood functions, but their value extends beyond this: they can also be highly sample-efficient when implemented with modern algorithms. However, without enough prior knowledge, inference can require a large number of simulations, which is infeasible if the simulator is slow to execute. To address this, a wide range of efficient algorithms have been developed, including sequential Monte Carlo approaches and surrogate-based methods \cite{Lintusaari2017, Price2017, Simola2021, Ikonomov2020, beaumont2002approximate, marjoram2003markov, sisson2007sequential, csillery2010approximate, drovandi2018likelihood}.

Among these, Bayesian optimization likelihood-free inference (BOLFI) \cite{gutmann2016bayesian} has emerged as a powerful tool for accelerating inference. It uses a Gaussian process surrogate model for the discrepancy between simulated and observed data, with active learning to iteratively propose where to sample next based on learned uncertainty. Based on the learned surrogate, the likelihood function is approximated and posterior distributions are sampled. BOLFI has been extended with various acquisition functions \cite{Jarvenpaa2019, jarvenpaa2021parallel}.
Furthermore, BOLFI has been applied to computationally heavy parameter estimation problems in a range of domains, including precision cosmology \cite{leclercq2018bayesian}, plasma physics simulations \cite{honkonen2025}, and cognitive decision-making models \cite{chen2025mobolfi}.

The key insight for our work is that when inference is fast enough, calibration becomes a practical diagnostic tool rather than a computational burden. Computational efficiency is not just about speed --- it enables a new level of methodological rigor.

\subsection*{Our Contributions and Application}
We demonstrate the data contribution framework using BOLFI applied to influenza A virus dynamics. We infer the six parameters of a target cell limited ordinary differential equation model \cite{Sachak-Patwa2023} describing influenza A virus dynamics using experimental data: the virion entry rate ($\beta$), the infection establishment rate per virion ($\gamma$), the infectious virion production rate ($P$), the total RNA production rate ($P_{\text{RNA}}$), and the durations of the eclipse and infectious phases ($\tau_E$ and $\tau_I$). Traditionally, this problem has been addressed with comprehensive MCMC-based approaches, requiring extensive computation time, approximately 43~hours on a cluster to obtain 500,000 parameter sets after burn-in \cite{quirouette2024does}. Moreover, these methods are typically applied to small-scale datasets. We apply the BOLFI algorithm to the same datasets and simulation models. Remarkably, BOLFI completes the inference process including surrogate model learning and sampling, in roughly 2~hours using only 1,000 simulation runs, a 20-fold reduction in computational effort. The actual times depend on the computing architecture; nevertheless, the comparison illustrates the substantial efficiency gain.

Our contributions are threefold:
\begin{enumerate}
\item \textbf{Efficient likelihood-free inference for virus dynamics.} We apply BOLFI to influenza A virus dynamics, achieving a 20-fold speed-up over traditional MCMC. We introduce a GP-based classifier for handling simulation failures and design a discrepancy that integrates heterogeneous data types (RNA and ED measurements from single and multiple cycle experiments).
\item \textbf{A general framework for data contribution analysis.} We use simulation-based calibration (SBC) to systematically evaluate what each data source contributes to parameter estimation. To our knowledge, this is the first time SBC has been used to identify which data sources are most reliable for each parameter.
\item \textbf{Composite diagnostic as an exploratory tool.} We explore a composite diagnostic that selects the best-calibrated data type for each parameter, illustrating how SBC can serve as a sensitivity check on combined inference and revealing potential trade-offs between precision and bias in multi-source data fusion.
\end{enumerate}

Our results reveal a counter-intuitive finding: combining all available data sources does not monotonically improve inference. RNA measurements introduce noise that degrades calibration for the total RNA production rate ($P_{\text{RNA}}$), while single-cycle data introduces inconsistency for the eclipse phase duration ($\tau_E$). ED data, when combined with RNA data, improves calibration for the infectious virion production rate ($P$) and infectious phase duration ($\tau_I$). This finding --- that data sources can be beneficial for some parameters and detrimental for others --- has implications beyond virus dynamics, offering a cautionary tale for any domain relying on multi-source data integration.

\subsection*{Multi-Source Likelihood-Free Inference Methods}
Several recent methods have been developed for likelihood-free inference with multi-source data, utilizing different underlying surrogate models. Neural network architectures, such as neural posterior estimation (NPE) \cite{papamakarios2016fast} and its multi-round variant, sequential neural posterior estimation (SNPE) \cite{greenberg2019automatic}, learn conditional density estimators to approximate target distributions directly. Similarly, sequential neural likelihood (SNL) \cite{papamakarios2019sequential} uses deep surrogates to model the synthetic likelihood function, allowing for amortized or highly localized inference across disparate datasets, though these neural approaches typically require 10,000--100,000 simulations for training. Alternatively, MOBOLFI \cite{chen2025mobolfi} extends the classic Gaussian process-based BOLFI framework to handle multi-source data using multi-objective Bayesian optimization; however, this method focuses on algorithmic extensions rather than systematic data contribution analysis. 

The remainder of this paper is organized as follows. The Materials and methods section describes the virus dynamics simulator, the experimental data, and the target parameters to be inferred. It then explains the BOLFI algorithm, the classifier integration for handling simulation failures, and the discrepancy design for combining data from four different experiments, followed by a theoretical justification of the discrepancy design and the simulation-based calibration procedure. The Results section presents parameter inference results from both synthetic and experimental data, including convergence diagnostics and calibration results for the discrepancy design. It then introduces the data contribution analysis across five data configurations, explores a composite diagnostic for parameter-specific data selection, and concludes with a summary of the calibration-guided data fusion framework. Finally, the Discussion and Conclusion sections summarize the contributions, discuss the insights and broader implications of the framework, and address limitations and future work.

\section{Materials and methods}
\label{sec:problem}

\subsection{Model parameters and experimental data}

\begin{figure}[!ht]
\centering
\includegraphics[width=0.8\textwidth]{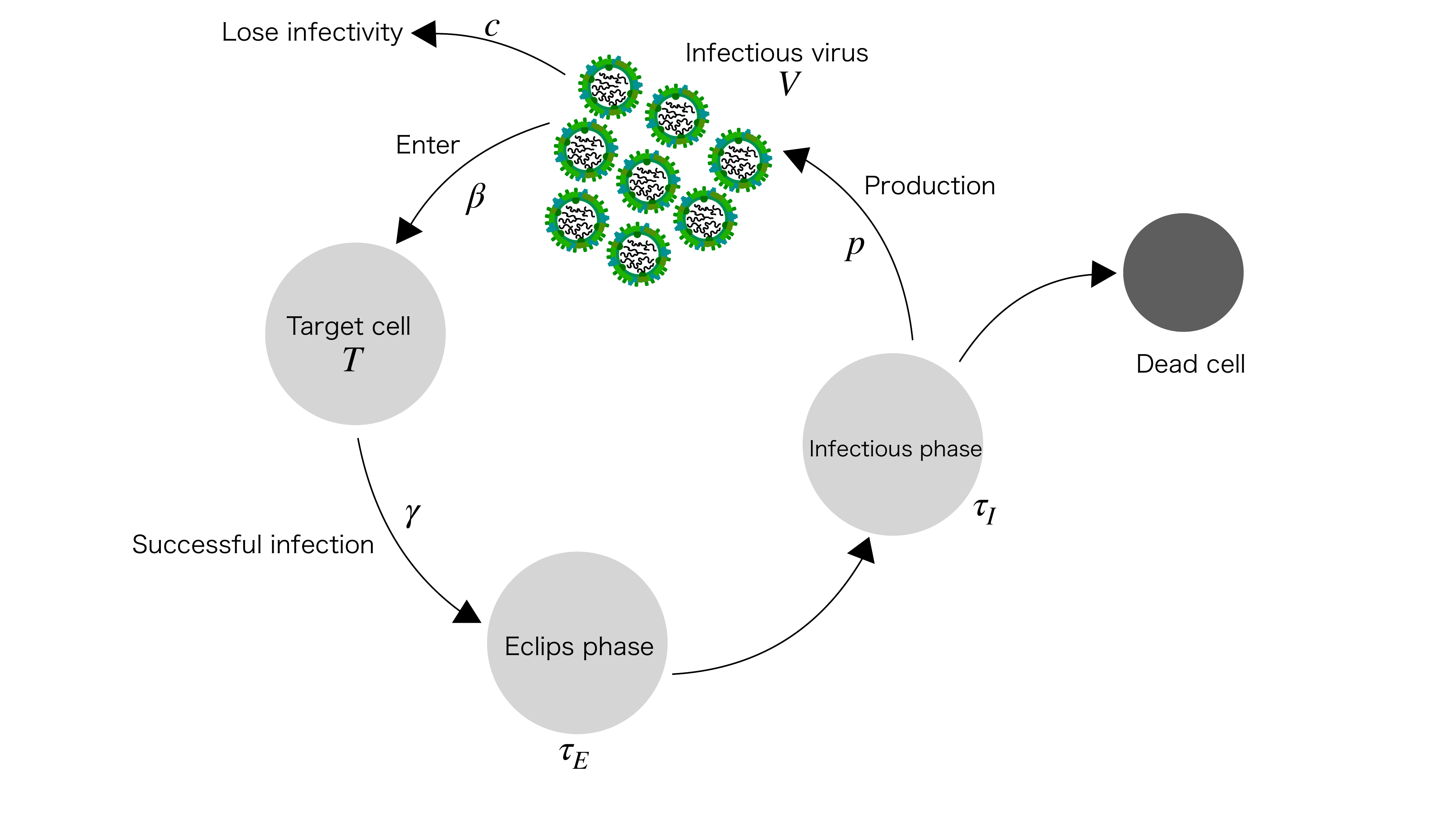}
\caption{\textbf{Virus cell infection process.} Schematic representation of the influenza A virus replication cycle showing target cells (T), eclipse phase (E), infectious phase (I), infectious virus (V), and total viral RNA ($V_{\text{RNA}}$). Key rate parameters are indicated at each step.}
\label{fig:virus_cell_infection_process}
\end{figure}

We study the dynamics of virus-cell infection, which involves several key steps: entry into the host cell, establishment of infection, an eclipse phase, viral replication and production of new virions, and eventual cell death. Released virus particles may lose infectivity or infect other cells, sustaining the infection cycle. Fig.~\ref{fig:virus_cell_infection_process} illustrates this process, with the corresponding rate parameters listed in Table~\ref{table_parameters}. 
We use a target cell limited model governed by a system of ordinary differential equations (ODEs) \cite{Sachak-Patwa2023}, with six unknown parameters: the virion entry rate $\beta$, the infection establishment rate per virion $\gamma$, the production rate of total viral RNA $P_{\text{RNA}}$, the production rate of infectious virions $P$, and the durations of the eclipse phase $\tau_E$ and infectious phase $\tau_I$. The full system of ODEs is provided in Section~\ref{sec:simulator}. Additional model parameters and symbols are listed in Table~\ref{table_parameters2}.

\begin{table}[!ht]
\small
\centering
\caption{\textbf{Parameters to be inferred.}}
\begin{tabular}{lll}
\toprule
Symbol & Parameter & Unit \\
\midrule
$\beta$  & rate of irreversible virion entry into a target cell & ml/(cell$\cdot$h) \\
$\gamma$ & successful infection establishment rate per virion entry & cell/[V] \\
$P$ & production rate for infectious virions & [V]/(cell$\cdot$h) \\
$P_{\text{RNA}}$ & production rate for total RNA virions & vRNA/(cell$\cdot$h) \\
$\tau_{E}$  & mean duration for the eclipse phase & h \\
$\tau_{I}$ & mean duration for the infectious phase & h \\
\bottomrule
\end{tabular}
\label{table_parameters}
\end{table}

\begin{table}[!ht]
\centering
\caption{\textbf{Additional model parameters.} These symbols are used in the ODE system presented in Section~\ref{sec:simulator}.}
\begin{tabular}{ll}
\toprule
Symbol & Parameter \\ 
\midrule
$T$   & target cells \\ 
$V$   & infectious virus \\ 
$V_{\text{RNA}}$   & total viral RNA \\ 
$E_{i}, i=1,2,..,n_{E}$  & the number of cells in the $i$th eclipse compartment \\ 
$I_{i}, i=1,2,..,n_{I}$  & the number of cells in the $i$th infectious compartment \\ 
$n_{E}$   & total number of eclipse compartments \\ 
$n_{I}$   & total number of infectious compartments \\ 
$c$   & infectious virus clearance rate \\ 
$c_{\text{RNA}}$   & total virus clearance rate \\
\bottomrule
\end{tabular}
\label{table_parameters2}
\end{table}

\begin{figure}[!ht]
\centering
\includegraphics[width=0.6\textwidth]{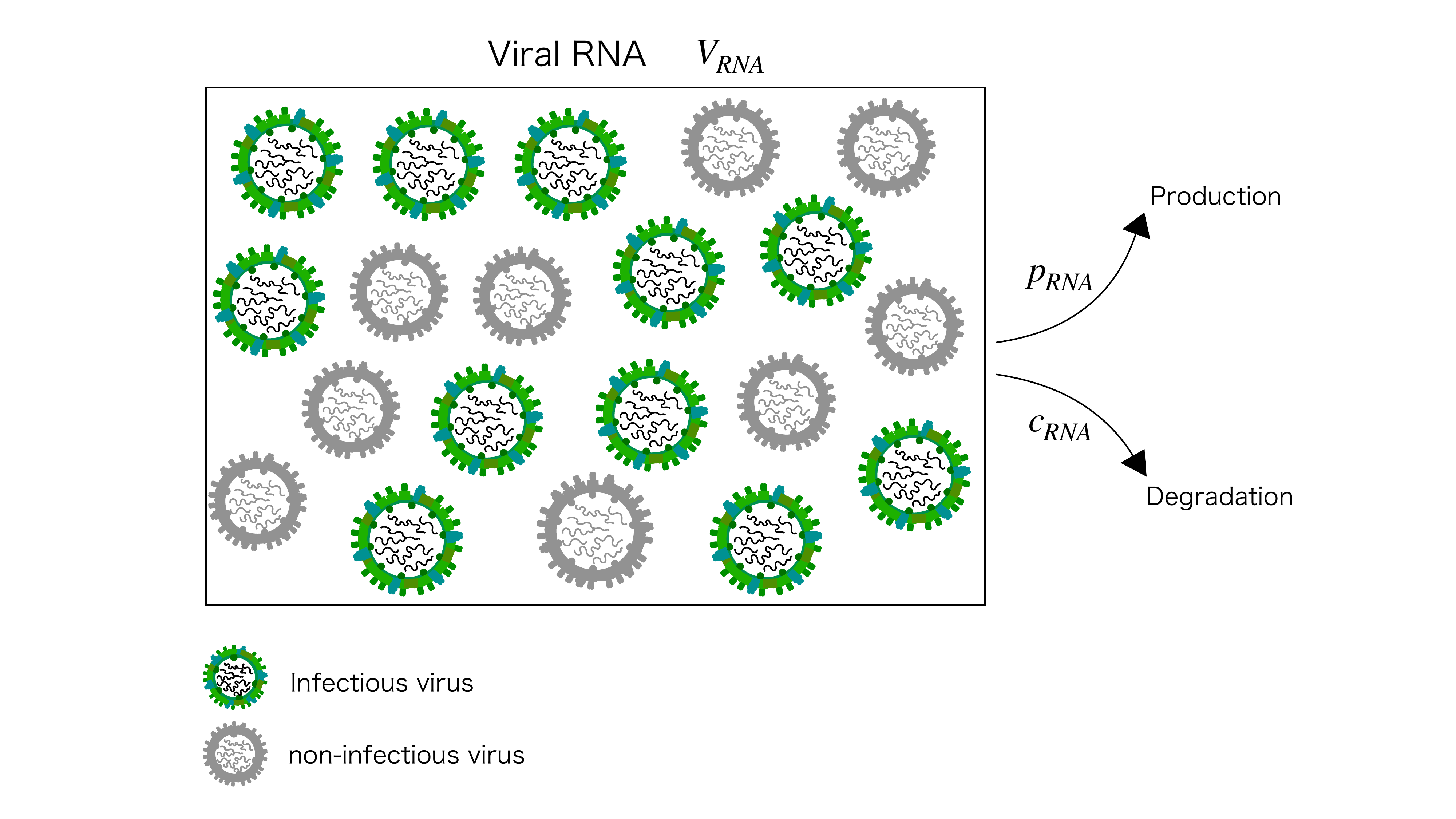}
\caption{\textbf{Hierarchical relationship between $V_{\text{RNA}}$ and $V$.} Viral RNA ($V_{\text{RNA}}$) includes infectious virus ($V$) and non-infectious virus RNA. This hierarchical relationship is fundamental to understanding the two types of experimental measurements.}
\label{fig:viral_RNA}
\end{figure}

In the modeling framework, the observable variable $V_{\text{RNA}}$ (viral RNA) includes all RNA species measured from the experiment, which encompasses both infectious virions (denoted as $V$) and non-infectious viral RNA strands. Therefore, $V$ is considered a subset of $V_{\text{RNA}}$. The production of $V_{\text{RNA}}$ is governed by parameters such as $P_{\text{RNA}}$ (total RNA production rate) and $c_{\text{RNA}}$ (degradation factor), which appear in the ODE system presented in Section~\ref{sec:simulator} and are listed in Table\ref{table_parameters2}. These parameters indirectly influence $V$ through shared replication mechanisms. This distinction is important because the RNA measurements from experiments quantify total viral RNA regardless of infectivity, while the infectious virion data ($V$) correspond only to a functionally active subset. See Fig~\ref{fig:viral_RNA} for visual explanation.

We analyze the data from \cite{Simon2016}. We have two experimental settings: the single-cycle (SC) infections and the multiple-cycle (MC) infections. In the SC experiment, a high multiplicity of infection (MOI = 3), which means that on average three infectious virions are added per target cell, ensures that all target cells are infected simultaneously within one round of replication. The virus was inoculated onto cells at \( t = -1 \, \text{h} \) to achieve a MOI of 3 within one hour, and the cell culture was rinsed at time \( t = 0 \, \text{h} \) to remove any virus still present from the high inoculum.
In the MC experiment, a low initial virus titer allows the infection to spread over multiple rounds, with the newly produced virus infecting the remaining susceptible cells. Infection is initiated at time \( t = 0 \, \text{h} \) and is not rinsed afterward. During the course of both the SC and the MC experiment, at each sampling time, 0.5 ml of the 10 ml supernatant is removed for virus quantification and replaced with 0.5 ml of fresh media.

For both Single Cycle and Multiple Cycle infection experiments, we have time steps samples of both total viral RNA concentration and Endpoint Dilution (ED) assay experiments. The viral RNA concentration data are the measurements that quantify the amount of viral RNA present in a sample. This data is often obtained through techniques like quantitative PCR (qPCR) or digital PCR. Viral RNA concentration data is a real number with unit vRNA/ml.
ED assay experiment is used to measure the amount of infectious virus in a sample \cite{Reed1938, Cresta2021}. In the experiment, the sample was serially diluted and cell cultures were prepared with each dilution. After incubation, assess whether the cells show signs of infection. They determine the highest dilution of a sample at which the virus can still infect cells and produce a detectable effect. This method allows researchers to calculate the $50\%$ infectious dose (TCID50), which is a standard measure of viral infectivity. In our study, we are not using TCID50; instead, we use the infectious numbers from the ED assay outcome count directly as data for parameter inference. In our case, each ED assay experiment output is an 8-digit vector with integers (0,1,2,3,4) as elements, corresponding to the number of infected wells in each of 8 dilution columns, where each column has 4 replicate wells.

\begin{figure}[!ht]
\centering
\includegraphics[width=.8\textwidth]{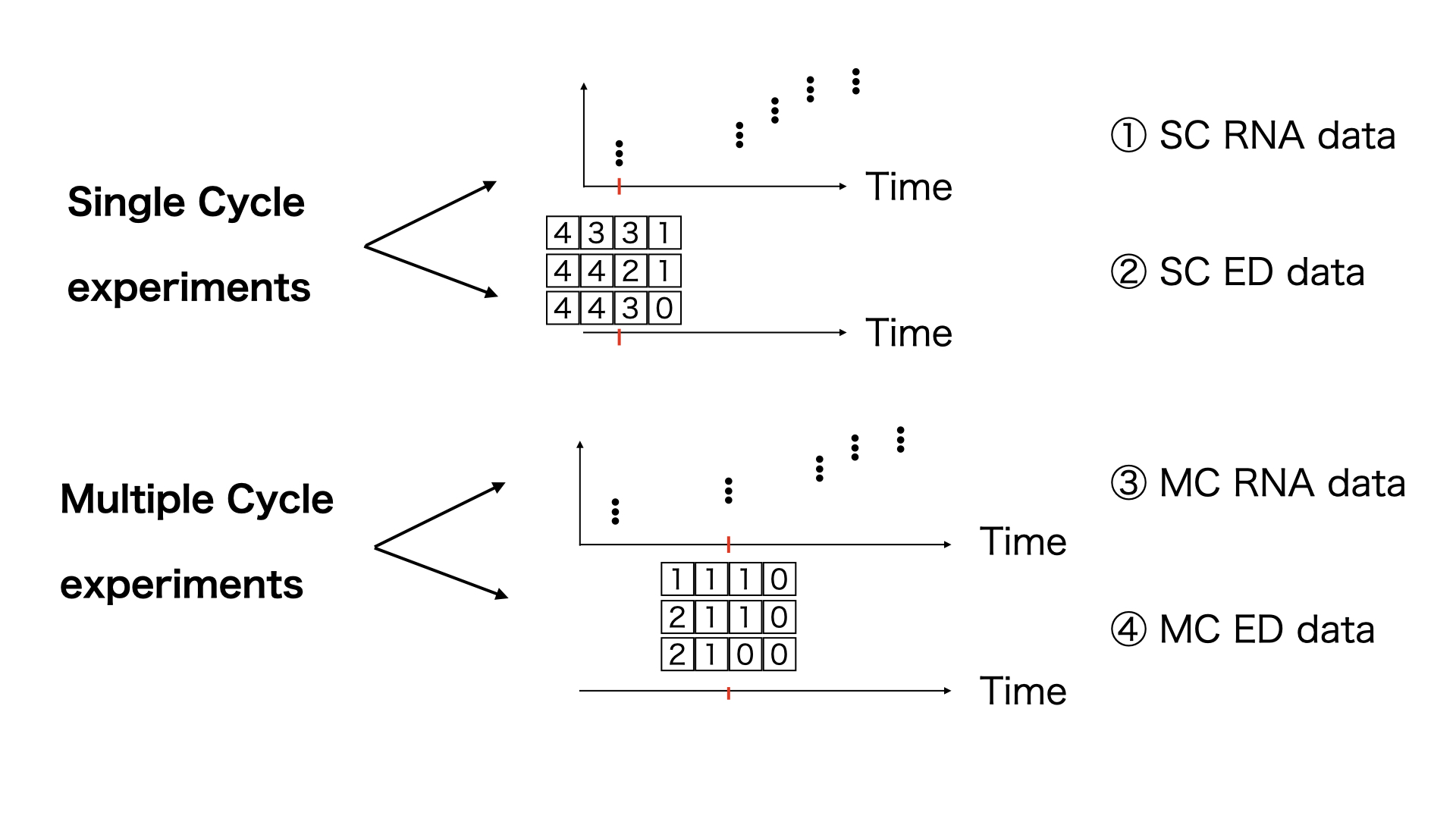}
\caption{\textbf{Overview of experimental data sets.} Four data sets are available: single cycle RNA data, single cycle ED data, multiple cycle RNA data, and multiple cycle ED data. For both Single Cycle and Multiple Cycle infection experiments, there are 12 time steps measurements. Each time step measurement includes both total viral RNA concentration and ED assay outcomes.}
\label{experiment data image}
\end{figure}

\begin{figure}[!ht]
\centering
\includegraphics[width=.8\textwidth]{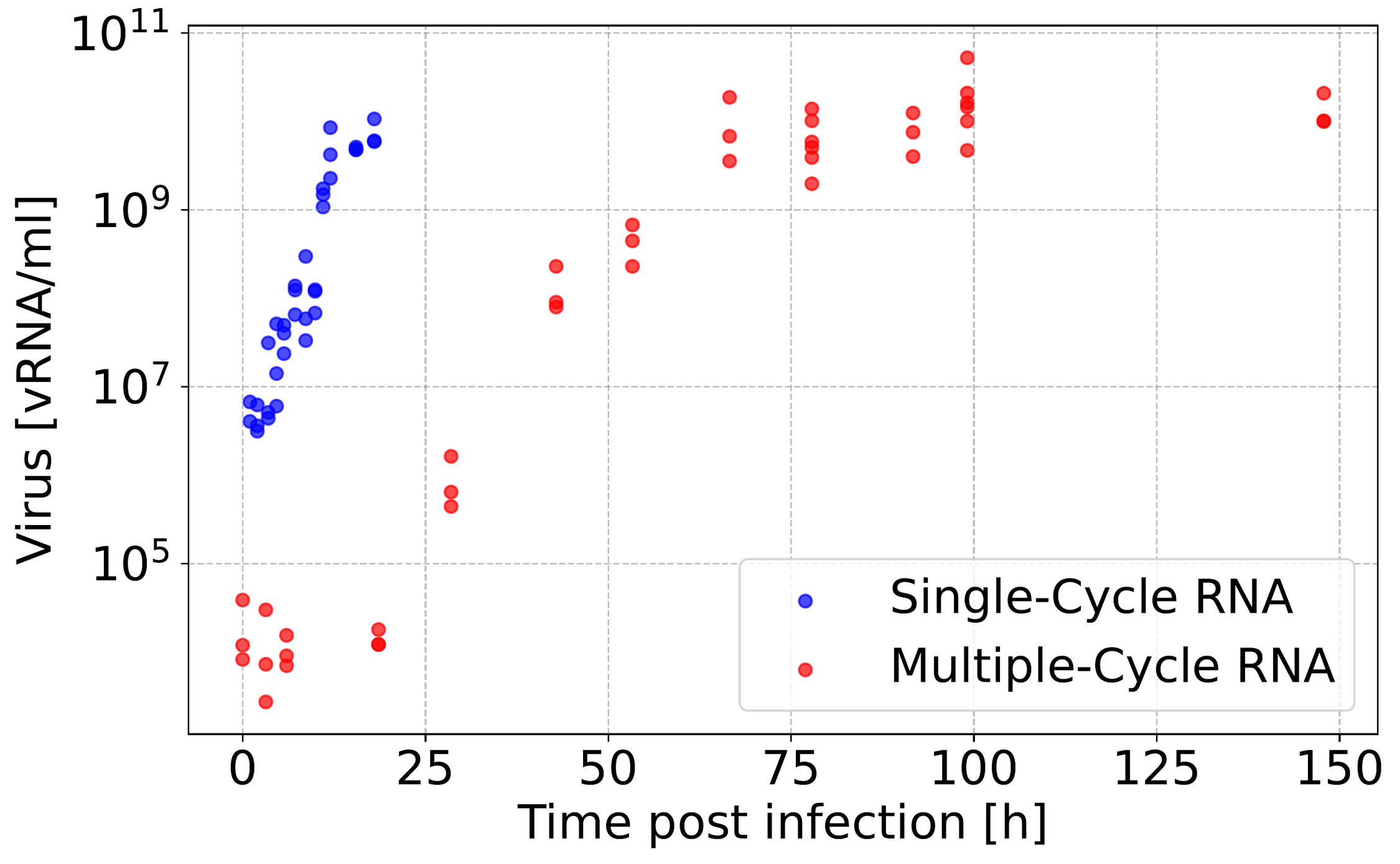}
\caption{\textbf{Single Cycle and Multiple Cycle virus RNA data together.} In SC experiment cells reach a fully infected saturation state in a short time compared to MC experiment, because the starting amount of viruses are much more than in MC experiments. Data is from \cite{Simon2016}. The raw RNA data are provided in Table~\ref{tab:SC RNA data} and Table~\ref{tab:MC RNA data}.}
\label{SC_MC_RNA_data_together_plot}
\end{figure}

\begin{figure}[ht]
\centering
\includegraphics[width=\textwidth]{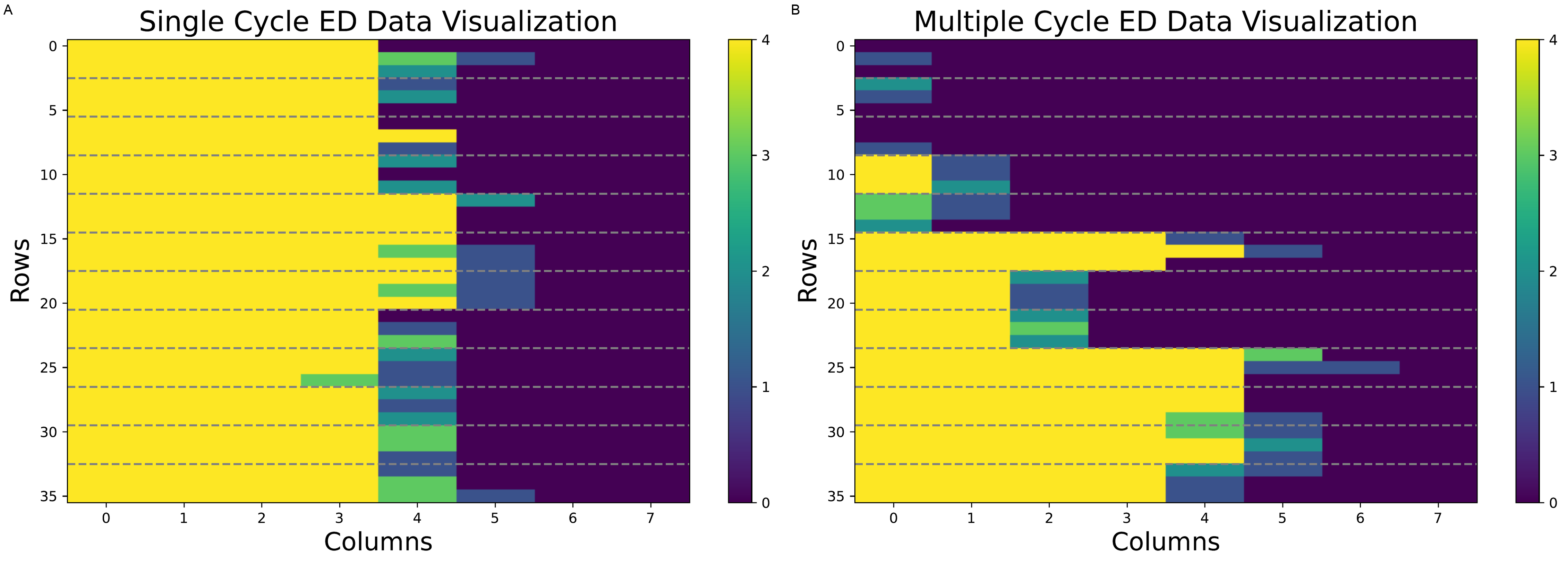}
\caption{\textbf{ED assay data visualization for single and multiple cycle experiments.} 
(A) Single cycle ED data: Color map showing ED assay outcomes over time. Most columns are fully infected even at early time points due to high initial virus dose, with variation primarily in the most diluted columns. Dilution factor shifts from $10^{-1}$--$10^{-8}$ (times 1--8.62 h) to $10^{-2}$--$10^{-9}$ (times 9.9--18 h). 
(B) Multiple cycle ED data: Infection spreads progressively over time, consistent with multiple rounds of viral replication. Dilution factors increase by 2--3 orders at later time points (28.5, 53.3, 66.6 h) to accommodate increasing viral titers. 
For both, each column has 4 replicates (value 4 = all wells infected). Data is from \cite{Simon2016}. The raw ED data are provided in Table~\ref{tab:SC_ED_data} and Table~\ref{tab:MC_ED_data}.}
\label{fig:SC_MC_ED_data_colormap}
\end{figure}

Therefore, we have four data sets in total: SC RNA data, MC RNA data, SC ED data, MC ED data. Fig~\ref{experiment data image} shows an overview of the four data sets and the relationship between the two experimental settings. The data are listed in Table~\ref{tab:SC RNA data}, Table~\ref{tab:MC RNA data}, Table~\ref{tab:SC_ED_data}, Table~\ref{tab:MC_ED_data} and visualized in Fig~\ref{SC_MC_RNA_data_together_plot}, Fig~\ref{fig:SC_MC_ED_data_colormap}. The ED data show distinct patterns between experiment types. In single cycle experiments (Fig~\ref{fig:SC_MC_ED_data_colormap} A), most columns are fully infected even at early time points due to the high initial virus dose, with variation primarily in the most diluted columns. 
In MC experiments (Fig~\ref{fig:SC_MC_ED_data_colormap} B), the infection spreads progressively over time, consistent with multiple rounds of viral replication. The gradual increase in infected wells across columns reflects the multi-cycle infection dynamics.
For both experiment types, there are multiple replicates: generally 3 replicates for ED data, while RNA data has 2-6 replicates per time point.
Both experiments have 12 time steps, though the measurement times differ between SC and MC.

\subsection{The simulator}
\label{sec:simulator}

We developed a custom simulator that models virus dynamics and experimental assay processes, based on two simulators described in \cite{Quirouette2023}: one capturing the virus dynamics over time via an ordinary differential equation (ODE), and the other simulating the ED assay experiment. The simulator outputs four synthetic data sets (SC RNA data, SC ED data, MC RNA data, and MC ED data) that have the same format as the experimental data that we are going to analyze.

The ordinary differential equation (ODE) model used herein is given by
\begin{align}
\frac{dT}{dt} &= -\gamma\beta TV/s \\
\frac{dE_1}{dt} &= \gamma\beta TV/s - \frac{E_1}{\tau_E} \\
\frac{dE_i}{dt} &= \frac{E_{i-1}}{\tau_E} - \frac{E_i}{\tau_E}, \quad i = 2, 3, \dots, n_E \\
\frac{dI_1}{dt} &= \frac{E_{n_E}}{\tau_E} - \frac{I_1}{\tau_I} \\
\frac{dI_j}{dt} &= \frac{I_{j-1}}{\tau_I} - \frac{I_j}{\tau_I}, \quad j = 2, 3, \dots, n_I \\
\frac{dV}{dt} &= P \sum_{j=1}^{n_I} I_j - cV - \beta TV/s \\
\frac{dV_{\text{RNA}}}{dt} &= P_{\text{RNA}} \sum_{j=1}^{n_I} I_j - c_{\text{RNA}}V_{\text{RNA}} - \beta TV/s
\end{align}
The eclipse and infectious phase durations are modeled using a chain of $n_E = 60$ and $n_I = 60$ compartments, respectively, following \cite{Quirouette2023}. This gives an Erlang distribution for each phase duration, parameterized by the mean durations $\tau_E$ and $\tau_I$, which are the parameters we infer.

The ED assay simulator is stochastic, whereas the ODE itself is deterministic. To introduce stochasticity, additive Gaussian noise is added on the $\log{10}$ scale to the ODE-predicted RNA concentrations, with noise variances estimated from the experimental data. Let $Y_{\text{RNA}}$ denote the simulated RNA measurement after noise addition, and $V_{\text{RNA}}$ is the deterministic ODE-predicted value. The noise model is:
\begin{align}
\log_{10}(Y_{\text{RNA}}) = \log_{10}(V_{\text{RNA}}) + \varepsilon_{\text{RNA}}, \quad \varepsilon_{\text{RNA}} \sim \mathcal{N}(0, \sigma^2)
\end{align}
Equivalently, on the linear scale:
\begin{align}
Y_{\text{RNA}} = V_{\text{RNA}} \times 10^{\varepsilon_{\text{RNA}}}, \quad \varepsilon_{\text{RNA}} \sim \mathcal{N}(0, \sigma^2)
\end{align}

The standard deviation $\sigma$ values are adopted from \cite{Quirouette2023}, where they were calculated as the pooled standard deviation of the residuals between the $\log_{10}$ transformed experimental measurements and their time-point-specific means, aggregated across all time points for each experiment. We tested the normality of the pooled $\log_{10}$-transformed residuals using the Shapiro-Wilk test (where $W$ is the test statistic and $p$ is the p-value). The Shapiro-Wilk test showed no significant deviation from normality for the MC RNA data ($W = 0.982$, $p = 0.78$) or the SC RNA data ($W = 0.976$, $p = 0.54$). Skewness and kurtosis values were within acceptable ranges for approximate normality (MC: skewness $= -0.12$, kurtosis $= 0.34$; SC: skewness $= -0.08$, kurtosis $= 0.21$), indicating no severe departures from normality. Visual inspection of Q-Q plots and histograms (Supplementary Figures~\ref{fig:MC_normality_check} and~\ref{fig:SC_normality_check}) confirmed these findings, supporting that it is reasonable to add Gaussian noise on the $\log_{10}$ scale. 
However, it is important to note that with $35$ (SC) or $42$ (MC) residuals per dataset, even the Shapiro-Wilk test has limited power. Consequently, p-values $>0.05$ do not constitute proof of normality, but rather indicate a failure to detect significant deviation. The normality assumption should therefore be viewed as a practical approximation that facilitates parameter estimation, rather than a statistically validated property of the data.

The resulting noise standard deviations are as follows:
\begin{itemize}
\item For multi-cycle RNA data: $\sigma_{\text{MC RNA}} = 0.258$
\item For single-cycle RNA data: $\sigma_{\text{SC RNA}} = 0.237$
\end{itemize}
The minor differences between our calculated values ($\sigma_{\text{MC RNA}} = 0.261$, $\sigma_{\text{SC RNA}} = 0.233$) and the previously reported values ($0.258$ and $0.237$, respectively) are within rounding error of the published data and do not impact inference. We use the $\sigma$ value consistently from previous reported ones in our simulator.

The ED assay simulator is stochastic and models the experimental endpoint dilution assay process. For a given viral sample, the simulator deposits virions into 8 dilution columns with 4 replicate wells per column. The number of virions deposited in each well follows a Binomial distribution, and a well is classified as infected if at least one deposited virion successfully establishes infection. The extinction probability depends on the model parameters through the burst size and the probability of successful cell entry. The simulator outputs the number of infected wells per dilution column, matching the format of the experimental ED data. Full details, including the extinction probability expression, are provided in the Supporting Information (S1 Appendix).

For the ED assay, an explicit likelihood function has been derived in previous work \cite{Cresta2021, Quirouette2023} based on the binomial process underlying the assay. However, we deliberately chose not to use this likelihood in our inference. Instead, we adopt a likelihood-free approach using the $L_1$ discrepancy between simulated and observed ED counts. This choice reflects the broader goal of our framework: to develop a methodology that does not rely on explicitly derived likelihoods, making it applicable to more complex, uncontrolled systems where such likelihoods are difficult to obtain. The binomial distribution underlying the ED assay is used in our theoretical justification (Section~\ref{sec:discrepancy}) to prove the sufficient statistic for the ED data.

To mimic repeated experimental measures, our combined simulator ($\mathcal{S}(\boldsymbol{\theta})$) repeats the simulation of each parameter set six times (matching the maximum number of replicates observed in the data). For each time point, we randomly select the same number of replicates as observed in the experimental data, so the simulator generates data that match the experimental replicating proportions.

The combined simulator can be viewed as a single process with two interconnected compartments:
\begin{align}
\mathcal{S}(\boldsymbol{\theta}) = 
\begin{cases}
\text{Virus Dynamics (ODE)}  \\
\text{Assay Simulation} 
\end{cases}
\end{align}
where $\boldsymbol{\theta} = (\beta, \gamma, P, P_{\text{RNA}}, \tau_I, \tau_E)$ represents the parameters to be inferred. In the Supporting Information (S1 Appendix), we provide detailed descriptions of the ODE-based virus dynamics model and the stochastic ED assay simulator. 

\subsubsection*{Units}

In the simulation in this study, we use [IV] unit for \( V_o \). There are two units in previous study \cite{quirouette2024does} --- [SIN] unit and [IV] unit. The initial [SIN] unit is converted to [IV] unit using 
\begin{equation}
C_{\text{measured}} = C_{\text{actual}} \cdot P_{V \rightarrow \text{Establishment}}
\label{eq:actual to measured}
\end{equation}
in both SC and MC infections as in previous study \cite{quirouette2024does}. This equation relates the experimentally measured concentration of infection-causing units (SIN/ml) to the actual infectious virion concentration (IV/ml) by considering the probability of a successful infection establishment \cite{quirouette2024does}.

\subsection{The BOLFI algorithm}
\label{sec:bolfi}

When prior information about the parameter values is limited, exploring the parameter space requires many simulations. Although our virus dynamics simulator is relatively efficient, traditional MCMC methods still require substantial computational time \cite{quirouette2024does}.

We use BOLFI \cite{gutmann2016bayesian}, a method that addresses this challenge by using Bayesian optimization to learn a Gaussian process (GP) surrogate model of the discrepancy surface --- the relationship between parameters and model-data fit. This surrogate is then used to guide the selection of new parameter combinations, balancing exploration of uncertain regions and exploitation of promising areas, making the inference highly sample-efficient. In practice, this means BOLFI can obtain posterior estimates with only a few hundred simulations, rather than the thousands or millions required by other methods.

The algorithm works as follows. For any parameter set $\theta$, we define a discrepancy $d(\theta)$ that measures how well the simulated data match the real observations. This discrepancy is computed as $d(\theta) = d(s(y), s(y_{\text{obs}}))$, where $s(\cdot)$ extracts summary statistics from the data, and $d(\cdot,\cdot)$ is a distance measure (e.g., Euclidean distance). As BOLFI runs simulations, it builds an evidence set $\mathcal{E}_t = \{(\theta_i, d(\theta_i))\}_{i=1}^t$ of parameter-discrepancy pairs.

A Gaussian process serves as a probabilistic surrogate model for the discrepancy function. After $t$ simulations, the GP provides both a predicted mean $\mu_t(\theta)$ and uncertainty $v_t(\theta)$ for any unsimulated parameter set:
\begin{align}
d(\theta) \mid \mathcal{E}_t \sim \text{GP} \left(\mu_t(\theta), v_t(\theta) + \eta^2 \right),
\label{GP_d}
\end{align}

The mean and variance functions are given by:
\begin{align}
\label{pred_mean}
\mu_t(\theta) &= \mathbf{k}_t(\theta)^T \mathbf{K}_t^{-1} \mathbf{d}_t, \\
\label{pred_variance}
v_t(\theta) + \eta^2 &= k(\theta, \theta) - \mathbf{k}_t(\theta)^T \mathbf{K}_t^{-1} \mathbf{k}_t(\theta) + \eta^2,
\end{align}
where $\mathbf{k}_t(\theta)$ captures correlations between $\theta$ and previously simulated points, and $\mathbf{K}_t$ captures correlations among the simulated points themselves. These correlations are defined by a covariance function $k(\cdot,\cdot)$. A common choice is the squared exponential covariance function:
\begin{align}
k(\theta', \theta'') = \sigma_f^2 \exp \left( -\sum_{j=1}^d \frac{(\theta'_j - \theta''_j)^2}{\lambda_j^2} \right),
\end{align}
where $\lambda_j$ are length scale parameters that determine how smoothly the discrepancy changes with each parameter. The GP hyperparameters ($\eta$, $\sigma_f$, $\lambda_j$) are optimized as evidence accumulates \cite{rasmussen2006gaussian}.

From the GP surrogate, we can approximate the likelihood function pointwise as:
\begin{align}
L(\theta) \approx \Phi \left( \frac{h - \mu_t(\theta)}{\sqrt{v_t(\theta) + \eta^2}} \right),
\label{likelihood_function}
\end{align}
where $\Phi(\cdot)$ is the Gaussian cumulative distribution function, and $h$ is a threshold (typically the minimum observed discrepancy).

An acquisition function determines where to sample next by balancing exploration (sampling where uncertainty is high) and exploitation (sampling where discrepancy is low). This study uses the maximum variance acquisition method \cite{Jarvenpaa2019}, which selects the next point by maximizing the variance of the unnormalized posterior:
\begin{align}
\theta_{t+1} = \arg \max \text{Var}(p(\theta) \cdot p_a(\theta)),
\end{align}
where $p(\theta)$ is the prior and $p_a(\theta)$ is an approximate likelihood derived from the GP (Eq.~\ref{likelihood_function}). For details, see \cite{Jarvenpaa2019}.

\subsection{Handling simulation failures with a classifier}
\label{sec:classifier}

Our simulator can fail for certain parameter combinations that are physically impossible or violate internal consistency checks. These failures act as hidden constraints that are not known in advance. While explicit constraints like $P_{\text{RNA}}>P$ can be encoded in the prior, hidden constraints cannot.

Failed simulations pose a problem for BOLFI because we cannot compute a discrepancy $d(\theta')$ without simulator output. Simply ignoring failures is problematic, without updating the model, BOLFI may repeatedly select invalid parameters, wasting computational resources.

To address this, we extend BOLFI with a classifier that learns which parameter regions lead to simulation failures. This approach incorporates the classifier-based failure handling mechanism from \cite{elgammal2023fast}, but replaces their SVM-based classifier with a GP classifier, so we use Gaussian processes for both regression and classification. The regression model learns the discrepancy surface from successful simulations, while the classifier learns to distinguish valid from invalid parameter combinations. The classifier updates with every simulation attempt (successful or failed), while the regression model updates only with successes.

When predicting for a new parameter set $\theta$:
\begin{itemize}
    \item If classified as valid: predictions use Eqs.~\eqref{pred_mean} and \eqref{pred_variance}.
    \item If classified as invalid: we set GP mean to $\infty$ and the variance to $0$ in Eqs.~\eqref{GP_d}, making the point unappealing to the acquisition function and guiding it away from invalid regions.
\end{itemize}

The classifier learns to identify invalid parameter combinations during the inference process. As the evidence set grows, failures become progressively less frequent, and the acquisition function successfully guides sampling toward valid regions. Figure~\ref{fig:bolfi_convergence} confirms convergence of the GP surrogate model after approximately 220 simulations in this application.

Algorithm~\ref{BOLFI_with_classifier_algorithm} outlines the complete procedure. We implement it using the ELFI software package \cite{lintusaari2018elfi}. A demo notebook  illustrating the failure-robust BOLFI approach on a 2D example is available online \cite{BOLFIrobust2024}.

\NoHyper
\begin{algorithm}[tbp]
\caption{BOLFI with classifier}
\begin{algorithmic}[1]
\State Given simulator $\mathcal{S}$ that generates data $y \sim \mathcal{S}(\theta)$ 
\State Define discrepancy $d$ and prior $p(\theta)$
\State Initialization:
\For{$t = 1, \ldots, N_{\text{init}}$}
    \State Sample $\theta_t \sim p(\theta)$
    \State Generate synthetic data $y' \sim \mathcal{S}(\theta_t)$
    \If{simulation succeeds}
        \State Update GP classifier (valid)
        \State Compute $d_t = d(s(y_{\text{obs}}), s(y'))$
    \Else
        \State Update GP classifier (invalid)
    \EndIf    
\EndFor
\State Build evidence set $\mathcal{E}_{N_{\text{init}}} = \{(\theta_t, d_t)\}_{t=1}^{N_{\text{init}}}$
\State Fit GP surrogate $d(\theta) \mid \mathcal{E}_{N_{\text{init}}}$
\State
\State Active learning loop:
\For{$t = N_{\text{init}} + 1, \ldots, N_E$}
    \If{$t$ multiple of $T_{\text{update}}$}
        \State Optimize GP hyperparameters
    \EndIf
    \State Select $\theta_t$ via acquisition function
    \State Generate synthetic data $y' \sim \mathcal{S}(\theta_t)$
    \If{simulation succeeds}
        \State Update GP classifier (valid)
        \State Compute $d_t$
        \State Update $\mathcal{E}_t$ and refit GP
    \Else
        \State Update GP classifier (invalid)
    \EndIf    
\EndFor
\State

\State Compute likelihood $L(\theta)$ via Eq~(\ref{likelihood_function})
\State Obtain posterior $\pi(\theta | y_{\text{obs}}) \propto p(\theta) L(\theta)$
\State Sample from posterior using sampling algorithm (e.g., Metropolis sampler)
\State \textbf{Output:}
\State \quad Posterior samples $\{\theta^{(i)}\}_{i=1}^{M}$
\State \quad Point estimates and credible intervals for each parameter

\end{algorithmic}
\label{BOLFI_with_classifier_algorithm}
\end{algorithm}
\endNoHyper

\subsection{Parameter settings and priors}

Following previous studies, we transform some parameters to the logarithmic scale:
$\hat{\gamma}=\log_{10}{\gamma}$, $\hat{P}=\log_{10}{P}$, $\hat{\beta}=\log_{10}{\beta}$, $\hat{P}_{\text{RNA}}=\log_{10}{P_{\text{RNA}}}$. The duration parameters $\tau_{E}$ and $\tau_{I}$ remain on the linear scale.

\begin{table}[!ht]
\centering
\caption{\textbf{Prior distributions for model parameters.}}
\begin{tabular}{llll}
\toprule
\textbf{Parameter} & \textbf{Distribution} & \textbf{Range} & \textbf{Constraints} \\
\midrule
\(\hat{P}\) & Uniform (log scale) & (\(0.5\), \(3\)) & Subject to \(\hat{P} < \hat{P}_{\text{RNA}}\) \\
\(\hat{P}_{\text{RNA}}\) & Uniform (log scale) & (\(2\), \(4.8\)) & Subject to \(\hat{P} < \hat{P}_{\text{RNA}}\) \\
\(\hat{\gamma}\) & Uniform (log scale) & (\(-1\), \(0.5\)) & --- \\
\(\hat{\beta}\) & Uniform (log scale) & (\(-8\), \(-6\)) & --- \\
\(\tau_E\) & Uniform (linear scale) & (\(2\), \(15\)) & --- \\
\(\tau_I\) & Uniform (linear scale) & (\(2\), \(55\)) & --- \\
\bottomrule
\end{tabular}
\label{tab:prior_settings}
\end{table}

Table~\ref{tab:prior_settings} shows the prior distributions. 
The ranges are based on biological knowledge of influenza A virus dynamics: $\gamma$ ranges from very low to near-certain infection establishment\cite{Simon2016}; the range for $\beta$ is narrow because values below $10^{-8}$ led to near-certain simulation failure due to insufficient infection, and $\beta$ is expected to be on the order of $10^{-7}$ ml/(cell$\cdot$h) due to the rarity of successful virus-cell encounters\cite{Simon2016}; and production rates span orders of magnitude to cover plausible RNA and infectious virion production\cite{Simon2016, Quirouette2023}. The duration parameters $\tau_E$ and $\tau_I$ cover the known ranges for eclipse and infectious phase durations at the cellular level \cite{Simon2016, Quirouette2023}. Note that we use the [IV] unit for $V$; the parameter ranges account for the conversion from [SIN] to [IV] units as described in \cite{quirouette2024does}. 

We enforce $\hat{P} < \hat{P}_{\text{RNA}}$ based on the biological constraint that infectious virions ($V$) are a subset of total viral RNA ($V_{\text{RNA}}$) (Fig~\ref{fig:viral_RNA}). During inference, we adaptively update bounds: the upper bound for $\hat{P}$ becomes the minimum observed $\hat{P}_{\text{RNA}}$, focusing sampling in plausible regions. 

The log-transformation for $\beta$, $\gamma$, $P$ and $P_{\text{RNA}}$ follows the approach of \cite{quirouette2024does}, who found that the parameter scales are better behaved on the logarithmic scale for inference. The duration parameters $\tau_E$ and $\tau_I$ remain on the linear scale, as their biological ranges are well-established in the literature \cite{Simon2016, Quirouette2023}.

The BOLFI algorithm requires bounded parameter domains for constructing the Gaussian process surrogate \cite{lintusaari2018elfi, gutmann2016bayesian}. Our ranges were initially set to be sufficiently wide to cover the biologically plausible parameter space, and then refined through an iterative process of running preliminary inference, examining posterior shapes, and adjusting ranges to ensure adequate coverage of the posterior without excessive simulation failures. At the current range, approximately $35\%$ of simulations fail; setting the range wider would increase the failure rate further. This type of iterative refinement of the GP domain bounds is standard practice in Bayesian inference, in simulation-based inference where prior bounds influence the efficiency of surrogate modeling \cite{lintusaari2018elfi, elgammal2023fast}. This iterative prior refinement is conceptually similar to prior conditioning approaches used in neural posterior estimation and related methods \cite{cranmer2020frontier, papamakarios2019sequential}, where prior distributions are refined based on initial inference results to improve sampling efficiency. However, these methods learn the shape of the prior distribution, whereas we only adjusted the range while keeping the uniform distribution unchanged.

\subsection{Discrepancy design}
\label{sec:discrepancy}

In likelihood-free inference, the discrepancy function quantifies how well simulated data match observations. Since we cannot compute the true likelihood directly, discrepancy serves as a proxy: smaller discrepancy means better fit. Typically:
\begin{align}
d(\theta) = \| s(y_{\text{sim}}(\theta)) - s(y_{\text{obs}}) \|
\end{align}
where $s(\cdot)$ extracts summary statistics, and $\|\cdot\|$ denotes a norm.

Our discrepancy combines information from four datasets: MC ED, MC RNA, SC ED, and SC RNA. The key components are:

\begin{itemize}
    \item \textbf{ED data discrepancies ($d_1$, $d_3$):} $L_1$ distances normalized by the number of data points (36 = 12 time points $\times$ 3 replicates):
    \begin{align*}
    d_1 &= \frac{1}{36}\sum_{i=1}^{36} |y_{\text{sim,MC\_ED},i} - y_{\text{obs,MC\_ED},i}| \\
    d_3 &= \frac{1}{36}\sum_{i=1}^{36} |y_{\text{sim,SC\_ED},i} - y_{\text{obs,SC\_ED},i}|
    \end{align*}
    
    \item \textbf{RNA data discrepancies ($d_2$, $d_4$):} Euclidean distances on log-transformed data:
    \begin{align*}
    d_2 &= \|\log_{10}(y_{\text{sim,MC\_RNA}}) - \log_{10}(y_{\text{obs,MC\_RNA}})\|_2 \\
    d_4 &= \|\log_{10}(y_{\text{sim,SC\_RNA}}) - \log_{10}(y_{\text{obs,SC\_RNA}})\|_2
    \end{align*}
\end{itemize}

These individual discrepancies are combined additively: $d_{13} = d_1 + d_3$, $d_{24} = d_2 + d_4$, and finally $d = d_{13} + d_{24}$. To ensure consistent scaling, we normalize $d$ to $d\_\text{norm} = (d - \text{mean})/\text{std}$, where mean=20 and std=10 were estimated from preliminary simulations. For failed simulations, we set them as $d\_\text{masked} = \infty$, effectively excluding them from consideration.

\subsection{Theoretical Justification of the Discrepancy Measure}
\label{sec:discrepancy_justification}

The choice of discrepancy function critically determines the quality of the posterior approximation in likelihood-free inference. Here we justify our discrepancy design for RNA and ED data, with detailed proofs are provided in the Supporting Information (S1 Appendix).

\subsubsection[RNA Data: Euclidean Distance on log10 Transformed Data]{RNA Data: Euclidean Distance on $\log_{10}$-Transformed Data}

Our RNA discrepancy is the Euclidean distance on $\log_{10}$-transformed, pooled data across replicates:
\begin{equation}
d_{\text{RNA}}(D, D_{\text{obs}}) = \left[ \sum_{i=1}^m \sum_{j=1}^{n_i} (z_{ij} - z_{ij}^{\text{obs}})^2 \right]^{1/2}, \quad z_{ij} = \log_{10}(y_{ij}).
\end{equation}
Here, $D = \{z_{ij}\}$ denotes a simulated data set and $D_{\text{obs}} = \{z_{ij}^{\text{obs}}\}$ denotes the observed data. The quantity $y_{ij}$ is the RNA measurement for replicate $j$ at time point $i$, and $z_{ij}$ is its $\log_{10}$-transformed value; $m$ is the number of time points and $n_i$ is the number of replicates at time $i$.

\textbf{Key theoretical properties:}
\begin{enumerate}
\item \textbf{Connection to sufficient statistics.} Under the log-normal noise model validated in Section 2.2, the time-point-specific sample means $\bar{z}_i = \frac{1}{n_i}\sum_j z_{ij}$ are sufficient statistics for the ODE parameters $\boldsymbol{\theta}$. The expected squared discrepancy satisfies:
\begin{equation}
\mathbb{E}[d_{\text{RNA}}^2] = C + \sum_{i=1}^m n_i(\bar{z}_i^{\text{obs}} - \mu_i(\theta))^2
\end{equation}
where $C$ is a constant independent of $\theta$ and $\mu_i(\theta) = \log_{10}(V(t_i, \theta))$, where $V(t_i, \theta)$ is the ODE model prediction at time $t_i$. Thus, ranking parameters by expected discrepancy is equivalent to ranking them by distance between the sufficient statistic and model predictions.

\item \textbf{Robustness to non-Gaussian noise.} Unlike parametric MCMC with Gaussian likelihood, our discrepancy-based approach does not require the normality assumption to be exact. It remains valid for any noise distribution with finite variance. This is critical given the limited power of normality tests with small sample sizes ($n=35\text{--}42$ residuals).

\item \textbf{Optimality for small replicates.} With $n_i = 2\text{--}6$ replicates, using all data directly (rather than first computing noisy sample means) provides more stable inference. Time points with more replicates naturally receive higher weight, which is appropriate as they provide more precise information.
\end{enumerate}

The detailed derivation of these properties is provided in S3 Appendix.

\subsubsection[ED Data: L1 Distance on Binomial Counts]{ED Data: $L_1$ Distance on Binomial Counts}

For ED data, we use the $L_1$ distance normalized by the number of observations:
\begin{equation}
d_{\text{ED}} = \frac{1}{R \times K \times T} \sum_{r=1}^R \sum_{k=1}^K \sum_{i=1}^T |c_{ik}^{(r),\text{sim}} - c_{ik}^{(r),\text{obs}}|
\end{equation}
where $R=3$ replicates, $K=8$ dilutions, and $T=12$ is the number of time points, and $c_{ik}^{(r)}$ is the number of infected wells (out of 4) for replicate $r$, dilution $k$, and time point $i$.

\textbf{Key theoretical properties:}

\begin{enumerate}
\item \textbf{Sufficiency preservation.} For each dilution $k$ and time $i$, the sufficient statistic for the infection probability $p_{ik}(\theta)$ is $s_{ik} = \sum_r c_{ik}^{(r)}$, the total infected wells across replicates (S1 Appendix). The $L_1$ discrepancy is minimized when the model-predicted probability matches the empirical proportion $s_{ik}/12$, preserving the information in the sufficient statistic.

\item \textbf{Robustness to saturation.} Unlike $L_2$ distance, which is dominated by saturated wells ($c=4$), the $L_1$ distance treats all dilution factors equitably.

\item \textbf{No parametric assumptions.} The $L_1$ distance requires no distributional assumptions beyond finite moments, complementing the non-parametric nature of our LFI approach.
\end{enumerate}

The detailed derivation of these properties and a comparison of alternative discrepancies ($L_2$, Poisson deviance, Hamming distance) are provided in S3 Appendix.

\subsubsection{Additive Combination of Data Sources}

We combine four data sources additively:
\begin{equation}
d(\theta) = d_{\text{MC,RNA}}(\theta) + d_{\text{MC,ED}}(\theta) + d_{\text{SC,RNA}}(\theta) + d_{\text{SC,ED}}(\theta)
\end{equation}

This additive design is grounded in recent methodological developments \cite{Thomas2025SplitBOLFI}. When the discrepancy is additive, the exponentiated loss becomes multiplicative, corresponding to an approximate joint likelihood as a product over conditionally independent data sources --- a standard and interpretable assumption for our independently conducted experiments.

We assessed the discrepancy design through synthetic data recovery (Fig.~\ref{synthetic_plot}) and posterior predictive checks (Fig.~\ref{fig:model_validation}). We also performed simulation-based calibration for both the original and weighted discrepancy designs (Section~\ref{sec:calibration_results}). The normalization (mean=20, std=10) serves only to improve GP numerical stability, not to re-weight data sources.
Detailed proofs and derivations for all theoretical results are provided in S3 Appendix.

\subsection{Simulation-Based Calibration}

To assess the statistical reliability of our inference, we performed 
simulation-based calibration (SBC) \cite{talts2018validating} using 100 synthetic datasets generated from the prior predictive distribution in each test. For each dataset, we ran the full BOLFI inference pipeline and recorded whether the true parameter value fell within the 50\%, 90\%, and 95\% credible intervals. We applied SBC in two distinct contexts:

\textbf{Data contribution analysis.} Our primary application of SBC was to 
systematically evaluate what each data source contributes to parameter 
estimation. We compared five data configurations (RNA-only, ED-only, 
MC-only, SC-only, and combined; see Section~\ref{sec:data_contribution_analysis}), 
using a fixed acceptable range of $\pm 5\%$ around the nominal coverage 
level. This criterion facilitated comparison across configurations and 
highlighted systematic patterns of over- or under-coverage, revealing 
which data sources are most reliable for each parameter.

\textbf{Calibration of the discrepancy design.} In Section~\ref{sec:calibration_results}, we evaluated the original 
and weighted discrepancy designs. The weighted discrepancy was 
constructed by weighting ED and RNA components by the inverse of their 
estimated variances, following the principle that each data source should 
contribute in proportion to its information content \cite{chen2025mobolfi}. We considered coverage within $\pm 2$ standard errors of the nominal level as well-calibrated. For $N=100$ calibration replicates, the standard error 
for a nominal coverage level $p$ is $\text{SE}(p) = \sqrt{p(1-p)/N}$. 
This yields bands of $\pm 2 \times 5.0\% = \pm 10\%$ for 50\% nominal 
coverage, $\pm 2 \times 3.0\% = \pm 6\%$ for 90\% nominal coverage, and 
$\pm 2 \times 2.2\% = \pm 4.4\%$ for 95\% nominal coverage. These bands 
correspond to approximately a 95\% confidence interval for the estimated 
coverage and are shown as the shaded region in the calibration figures 
(Fig.~\ref{fig:calibration_comparison}).

\section{Results}
\label{sec:results}

\subsection{Test result with synthetic data}

\begin{figure}[!ht]
\centering
\includegraphics[width=\textwidth]{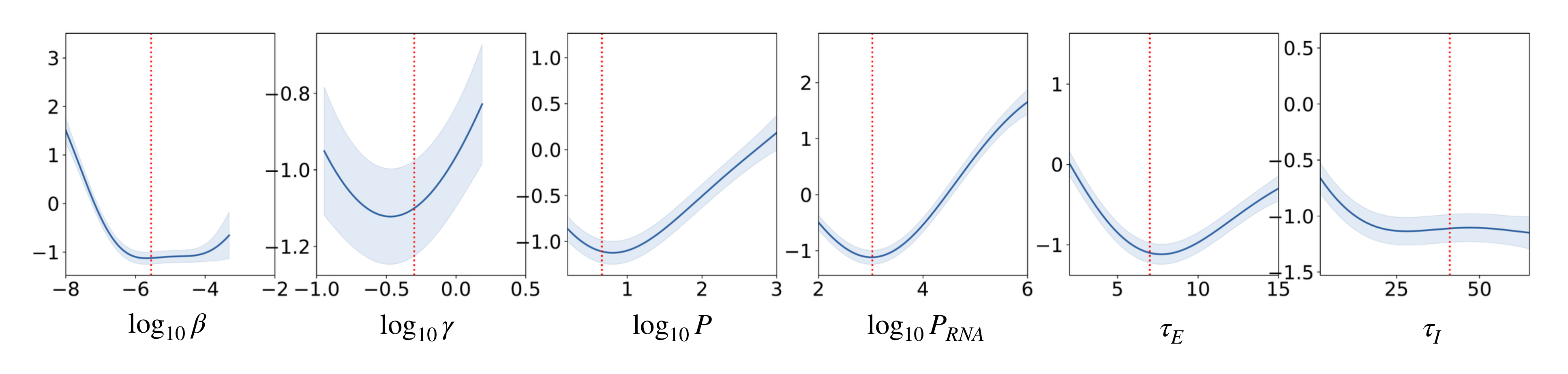}
\caption{\textbf{One-dimensional slices of the learned multidimensional GP surrogate model for synthetic data.} A series of one-dimensional slices of the learned multidimensional Gaussian Process (GP) surrogate model (the target model in BOLFI) is shown for each parameter dimension, generated using synthetic data produced with default parameter values (the red dashed vertical lines). The solid lines represent the mean estimates, while the shaded regions indicate the confidence intervals around these estimates. The fixed point used for plotting is posterior sample mean estimated from 50,000 MCMC samples. Values for $\beta$, $\gamma$, $P$, and $P_{\text{RNA}}$ are shown on the $\log_{10}$ scale; their original units are ml/(cell$\cdot$h), cell/[V], [V]/(cell$\cdot$h), and vRNA/(cell$\cdot$h), respectively. $\tau_E$ and $\tau_I$ are on a linear scale with unit [h].}
\label{synthetic_plot}
\end{figure}

We tested the proposed inference procedure using synthetic data generated with the simulator and the fixed set of parameters as observed data. The result is shown in Fig~\ref{synthetic_plot}. The lowest values of the surrogate model along each one-dimensional slice of the parameters are close to or correspond to the true parameter values used to generate the synthetic data. This suggests that our inference method is effective and capable of accurately recovering the true parameters. 

\subsection{Learning results from experimental data}

The inference was performed using 1,000 simulations, resulting in 759 valid evidence points and 241 failed outputs. The BOLFI algorithm took approximately 1 hour and 26 minutes to learn the target model on the CSC Puhti supercomputing infrastructure. Sampling from the posterior distribution using a Metropolis sampler required about 25 minutes for 100,000 samples.

\begin{figure}[!ht]
\centering
\includegraphics[width=\textwidth]{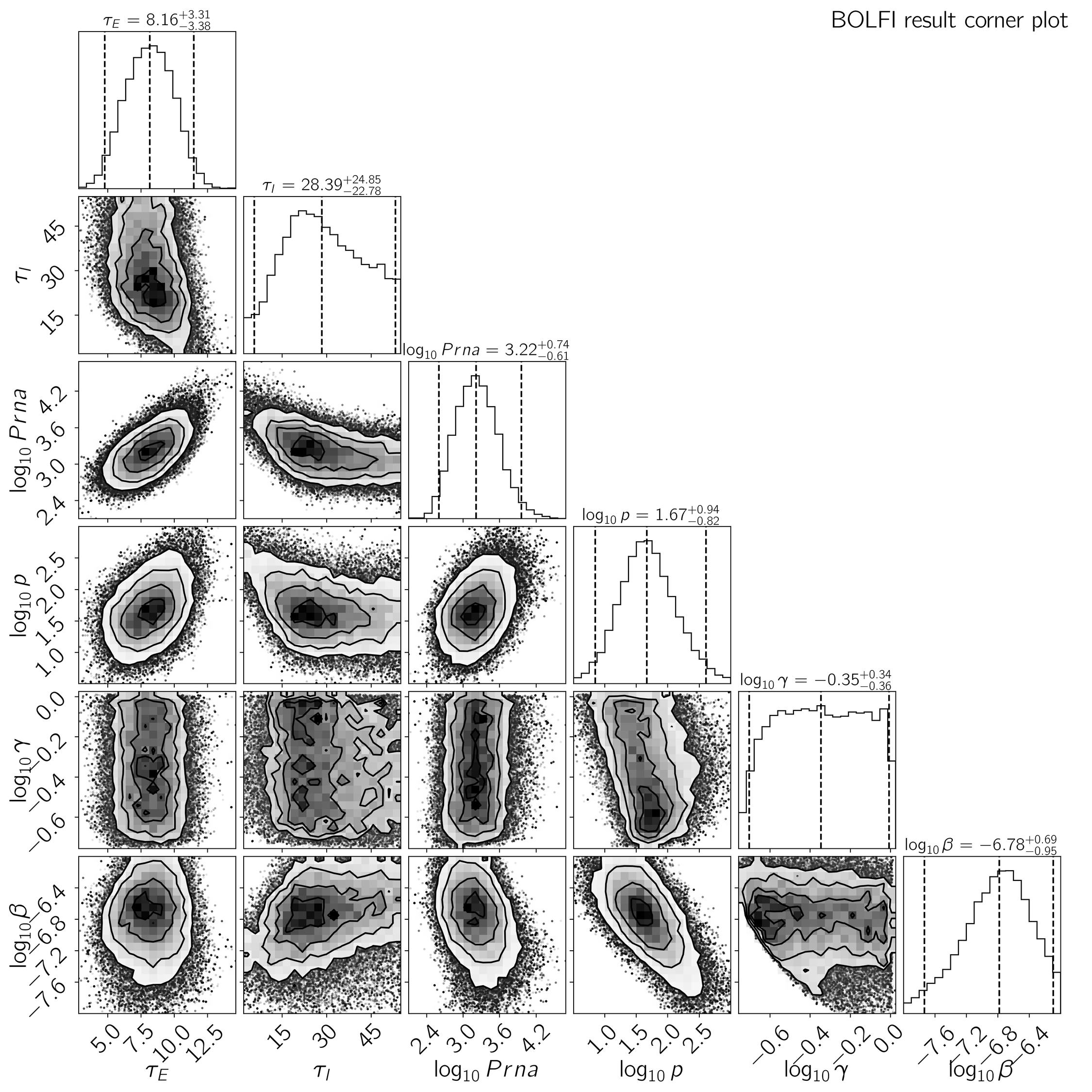}
\caption{\textbf{Pairwise posterior samples from BOLFI.} Pairwise posterior samples from the BOLFI learned distribution. Vertical lines are quantiles [0.025, 0.5, 0.975]. The units are $\beta$ (ml/(cell$\cdot$h)), $\gamma$ (cell/[V]), $P$ ([V]/(cell$\cdot$h)), $P_{\text{RNA}}$ (vRNA/(cell$\cdot$h)), $\tau_E$ (h), and $\tau_I$ (h). This plot is based on 100,000 samples from the BOLFI learned posterior distribution using a Metropolis sampler. Four chains of 100,000 iterations were acquired.}
\label{BOLFI_conerReorder}
\end{figure}

\begin{figure}[!ht]
\centering
\includegraphics[width=\textwidth]{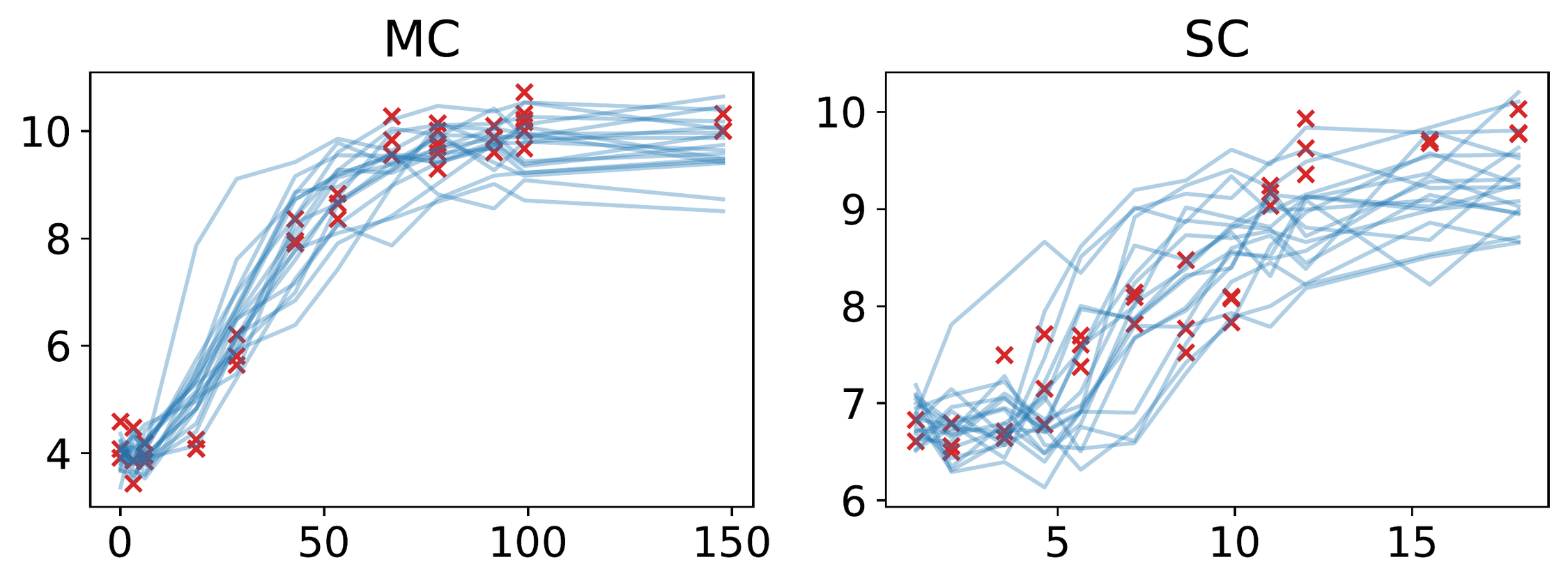}
\caption{\textbf{Model validation against experimental RNA data.} Comparison of model predictions (blue lines) with observed RNA data (red crosses) for multiple-cycle (left) and single-cycle (right) experiments. The predictions are generated using the inferred posterior parameter distributions and successfully capture the range and dynamics of the experimental measurements.}
\label{fig:model_validation}
\end{figure}

Table~\ref{tab:BOLFI inferred parameters} presents the estimated parameter values with their 95\% credible intervals. The complete posterior distributions, including pairwise correlations and convergence diagnostics, are shown in Fig~\ref{BOLFI_conerReorder}. Fig~\ref{fig:model_validation} demonstrates that these inferred parameters successfully reproduce the experimental RNA data, with model predictions closely matching observations for both multiple and single cycle experiments. 
Detailed diagnostics of the GP surrogate model are provided in S4 Appendix (Figs~\ref{fig:gp_pairwise}, \ref{fig:gp_1d_slices}, and \ref{fig:posterior_2d_slices}).

\begin{table}[!ht]
\centering
\caption{\textbf{Parameters estimated by the BOLFI algorithm.}}
\begin{tabular}{lc}
\toprule
Parameter (unit) & Mean [95\% CI] \\
\midrule
$\beta$ (ml/(cell$\cdot$h)) &  $ 1.52\times 10^{-7} [1.85\times 10^{-8}, 8.05\times 10^{-7}]$  \\
$\gamma$ (cell/[V]) & $0.45 [0.20, 0.99]$   \\
$P$ ([V]/(cell$\cdot$h)) &  $ 48.98 [7.08, 408.32]$   \\
$P_{\text{RNA}}$ (vRNA/(cell$\cdot$h)) &  $1702.16 [401.79, 9141.13]$   \\
$\tau_E$ (h) &  8.14 [4.78, 11.47]   \\
$\tau_I$ (h) &   29.35 [5.67, 53.24]  \\
\bottomrule
\end{tabular}
\label{tab:BOLFI inferred parameters}
\begin{flushleft} 
Estimates are provided as: Mean [95\% CI] where the 95\% credible interval (CI) corresponds to the marginal posterior distribution (MPD) for each parameter, marginalized over all others. 95\% CI is represented as [0.025 0.975] quantiles. The medians (0.5 quantiles) of each parameter are $\beta = 10^{-6.78}$ (ml/(cell$\cdot$h)), $\gamma = 10^{-0.35}$ (cell/[V]), $P = 10^{1.67}$ ([V]/(cell$\cdot$h)), $P_{\text{RNA}} = 10^{3.22}$ (vRNA/(cell$\cdot$h)), $\tau_E = 8.16$ (h), $\tau_I = 28.39$ (h). For the Maximum A Posteriori (MAP), which corresponds to the parameter set with the highest likelihood in the joint, multi-dimensional posterior $P_{\text{post}}$(data), the learned result is $\beta = 10^{-6.958}$ (ml/(cell$\cdot$h)), $\gamma = 10^{0.007}$ (cell/[V]), $P = 10^{1.414}$ ([V]/(cell$\cdot$h)), $P_{\text{RNA}} = 10^{3.309}$ (vRNA/(cell$\cdot$h)), $\tau_E = 7.898$ (h), $\tau_I = 19.109$ (h).
\end{flushleft}
\end{table}

\subsection{Interpretation of the results}

\paragraph{Virion entry rate ($\beta$).} 
The estimated rate of infectious virion entry into cells is $\beta = 1.52\times 10^{-7}$ ml/(cell$\cdot$h) (95\% CI: $[1.85\times 10^{-8}, 8.05\times 10^{-7}]$). This extremely small value indicates that successful cell entry is a rare event, consistent with the biological reality that most virions never encounter a susceptible cell. The posterior distribution for $\beta$ (Fig~\ref{BOLFI_conerReorder}) is slightly right-skewed but has a well-defined peak, suggesting good identifiability despite the small magnitude.

\paragraph{Infection establishment rate per virion entry ($\gamma$).} 
\label{gamma challenge}
After entering a cell, the rate at which an infectious virion successfully establishes infection is estimated as $\gamma = 0.45$ (cell/[V]) with a wide 95\% credible interval of $[0.20, 0.99]$. Biologically, this parameter encapsulates multiple factors: cellular defense mechanisms, stochastic replication outcomes, and potential virion defects. The marginal posterior distribution (Fig~\ref{BOLFI_conerReorder}) is notably flat, with the lower bound constrained by the data and the upper bound constrained by the valid parameter region of the simulator — values of $\log_{10} \gamma$ above approximately 0 are excluded by the classifier (Figs.~\ref{fig:gp_pairwise} and~\ref{fig:posterior_2d_slices}). This indicates that while we can establish a minimum infectivity threshold, the data provide limited information to pinpoint an upper bound or a most probable value --- a finding consistent with previous work \cite{quirouette2024does}. A uniform prior was placed over the logarithmic scale of $\gamma$, which, when transformed back to linear scale, inherently gives more weight to smaller values and flattens the posterior unless the data are sufficiently informative.
 The shallow discrepancy surface for $\gamma$ further confirms weak parameter sensitivity within biologically plausible ranges. From a modeling point of view, caution should be taken when interpreting the point estimates of $\gamma$; the inferred range should be viewed as a conservative estimate indicating that a minimum level of infectivity can be established. These findings highlight the need for improved experimental resolution — such as single-cell infection assays — to reduce uncertainty in this biologically significant parameter.

\paragraph{Viral production rates ($P_{\text{RNA}}$ and $P$).} 
The total RNA virion production rate is estimated at $P_{\text{RNA}} = 1702$ vRNA/(cell$\cdot$h) (95\% CI: $[402, 9141]$), meaning each infected cell produces thousands of viral RNA copies per hour. This high rate reflects the explosive growth potential of influenza A virus populations. In contrast, the infectious virion production rate is substantially lower: $P = 49$ [V]/(cell$\cdot$h) (95\% CI: $[7, 408]$) -- approximately 35-fold less than total RNA production. This ~35-fold disparity provides a quantitative estimate of the well-known inefficiency in influenza virus production, highlighting that virion assembly and packaging represent a major post-transcriptional bottleneck. As only a small fraction of viral RNA is packaged into infectious particles, this bottleneck may present a promising target for antiviral strategies aimed at disrupting assembly or packaging. Despite being lower, a production rate of 49 infectious virions per cell per hour remains remarkably high and represents a critical parameter for predicting viral spread and evaluating antiviral strategies.

\paragraph{Eclipse phase duration ($\tau_E$).} 
The estimated mean eclipse phase duration --- the period between cell infection and the onset of viral production --- is $\tau_E = 8.14$ hours (95\% CI: $[4.78, 11.47]$ h). This aligns closely with the established biological range of 8--10 hours for influenza A, providing validation that our inference captures realistic infection dynamics.

\paragraph{Infectious phase duration ($\tau_I$).} 
The estimated mean infectious phase duration is $\tau_I = 29.35$ hours (95\% CI: $[5.67, 53.24]$ h). The posterior distribution exhibits a long flat tail (Fig~\ref{fig:gp_1d_slices}), indicating that longer durations remain plausible. This pattern, also observed in synthetic data tests (Fig~\ref{synthetic_plot}), suggests that the data constrain a minimum infectious period more strongly than a precise maximum. This extended tail may also reflect biological heterogeneity in cell survival, where some infected cells could enter a non-productive, persistent state, making it difficult for the assay to precisely define the end of the infectious phase. While human influenza infectious periods are typically reported as 96--168 hours, our estimate of $\sim$29 hours represents the cell-level infectious phase within a single replication cycle, not the duration of host-level shedding. The broad credible interval captures variability across cells and experimental conditions, consistent with the stochastic nature of viral infection.

In summary, our inference yields the following parameter estimates:
\begin{itemize}
    \item \textbf{Virion entry rate ($\beta$):} $1.52\times10^{-7}$ ml/(cell$\cdot$h) --- extremely small, reflecting rare cell entry events; well-identified posterior.
    \item \textbf{Infection establishment rate ($\gamma$):} 0.45 cell/[V] --- the data only inform the lower bound ($\gamma > 0.20$); the upper bound is determined by the boundary of the feasible parameter space, with weak parameter sensitivity.
    \item \textbf{Total RNA production rate ($P_{\text{RNA}}$):} 1702 vRNA/(cell$\cdot$h) --- high RNA production reflecting explosive viral growth potential.
    \item \textbf{Infectious virion production rate ($P$):} 49 [V]/(cell$\cdot$h) --- approximately 35-fold lower than RNA production, highlighting the assembly and packaging bottleneck; this rate remains critically high for viral spread and antiviral evaluation.
    \item \textbf{Eclipse phase duration ($\tau_E$):} 8.14 hours --- aligns with the established biological range.
    \item \textbf{Infectious phase duration ($\tau_I$):} 29.35 hours --- broad credible interval and long posterior tail; reflects cell-level variability.
\end{itemize}

Having established the biological interpretation of the parameter estimates, we next assess the reliability of the inference procedure itself.
\subsection{BOLFI Convergence}
\begin{figure}[ht]
\centering
\includegraphics[width=0.8\textwidth]{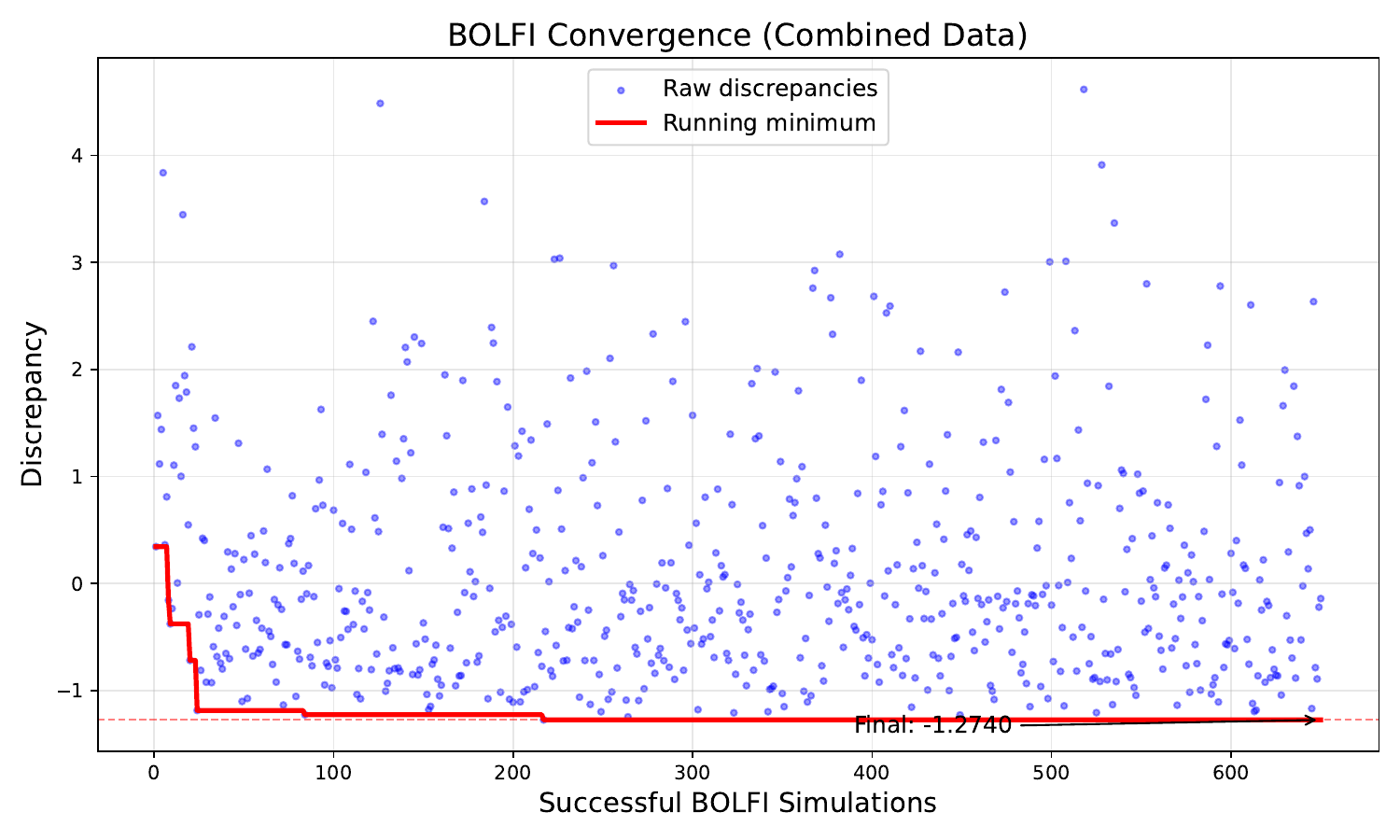}
\caption{\textbf{BOLFI convergence trajectory} for the combined discrepancy using synthetic data generated from the prior distribution.
The blue points show the raw discrepancy values for each successful simulation; the red line shows the running minimum. The discrepancy stabilized after approximately 220 simulations, indicating that the best-fitting region had been identified.}
\label{fig:bolfi_convergence}
\end{figure}

To verify that 1,000 BOLFI simulations were sufficient for convergence, 
we tracked the minimum discrepancy as a function of the successful simulation count using synthetic data generated from the prior
(Fig.~\ref{fig:bolfi_convergence}). The minimum discrepancy decreased 
rapidly during the first 200 simulations and stabilized after approximately 
220 successful simulations at a value of -1.274. No further decrease was observed beyond this point, indicating that BOLFI had identified the region of the best-fitting parameters. The same convergence pattern was observed in the inference on the real experimental data.

Although the minimum discrepancy stabilizes early, BOLFI's acquisition function continues to balance exploration and exploitation, refining the GP surrogate model in regions of high uncertainty. In practice, monitoring the stabilization of the minimum discrepancy and the GP hyperparameters is the standard way to assess BOLFI convergence, with the maximum number of simulations set conservatively to ensure adequate exploration \cite{gutmann2016bayesian, lintusaari2018elfi}. The original BOLFI paper \cite{gutmann2016bayesian} and the ELFI implementation \cite{lintusaari2018elfi} report that a few hundred simulations are typically sufficient for convergence on standard problems, and our 1,000 simulations provide a conservative margin. The convergence plot (Fig.~\ref{fig:bolfi_convergence}) confirms that the surrogate model is reliable and that additional simulations beyond this point would not meaningfully improve the inference.

We assessed convergence of the MCMC sampling in the last step of BOLFI inference using the Gelman-Rubin R-hat statistic. Using $100,000$ posterior samples, the R-hat values were below 1.01 for all parameters ($\beta$: $1.001$, $\gamma$: $1.001$, $P$: $1.001$, $P_{\text{RNA}}$: $1.001$, $\tau_E$: $1.000$, $\tau_I$: $1.001$), indicating convergence. 
\textbf{For the simulation-based calibration study}, we used $1{,}000$ posterior samples per case (in contrast to the $100,000$ samples used for the full inference), which provides sufficient precision to estimate 50\%, 90\%, and 95\% credible intervals.

\subsection{Calibration Results for the Discrepancy Design}
\label{sec:calibration_results}

The calibration results for the original discrepancy design stated in Section~\ref{sec:discrepancy} are summarized 
in Table~\ref{tab:calibration_original} and visualized in Figure~\ref{fig:calibration_comparison} (left). Four parameters $P$, $\gamma$, $\beta$, and $\tau_I$ showed excellent calibration, with 50\% credible interval coverage close to the nominal level (49.0\%, 55.0\%, 53.0\%, and 49.0\%, respectively).
The 90\% and 95\% credible intervals were also close to nominal for these parameters (91.0--97.0\% and 95.0--100.0\%, respectively). For $P_{\text{RNA}}$ and $\tau_E$, the 50\% coverage was 81.0\% and 77.0\%, respectively, indicating more conservative behavior.

\begin{table}[ht]
\centering
\caption{\textbf{Calibration results for the original discrepancy design.}}
\label{tab:calibration_original}
\begin{tabular}{lccc}
\hline
Parameter & 50\% CI & 90\% CI & 95\% CI \\
\hline
$\beta$   & 53.0\% & 94.0\% & 99.0\%  \\
$\gamma$ & 55.0\% & 97.0\% & 100.0\% \\
$P$       & 49.0\% & 94.0\% & 98.0\%  \\
$P_{\text{RNA}}$ & 81.0\% & 98.0\% & 99.0\%  \\
$\tau_E$  & 77.0\% & 94.0\% & 95.0\%  \\
$\tau_I$  & 49.0\% & 91.0\% & 95.0\%  \\
\hline
\end{tabular}
\end{table}

\subsubsection{Sensitivity to Discrepancy Weighting}
\label{sec:weighted_discrepancy_calibration_results}
\begin{figure}[ht]
\centering
\includegraphics[width=0.9\textwidth]{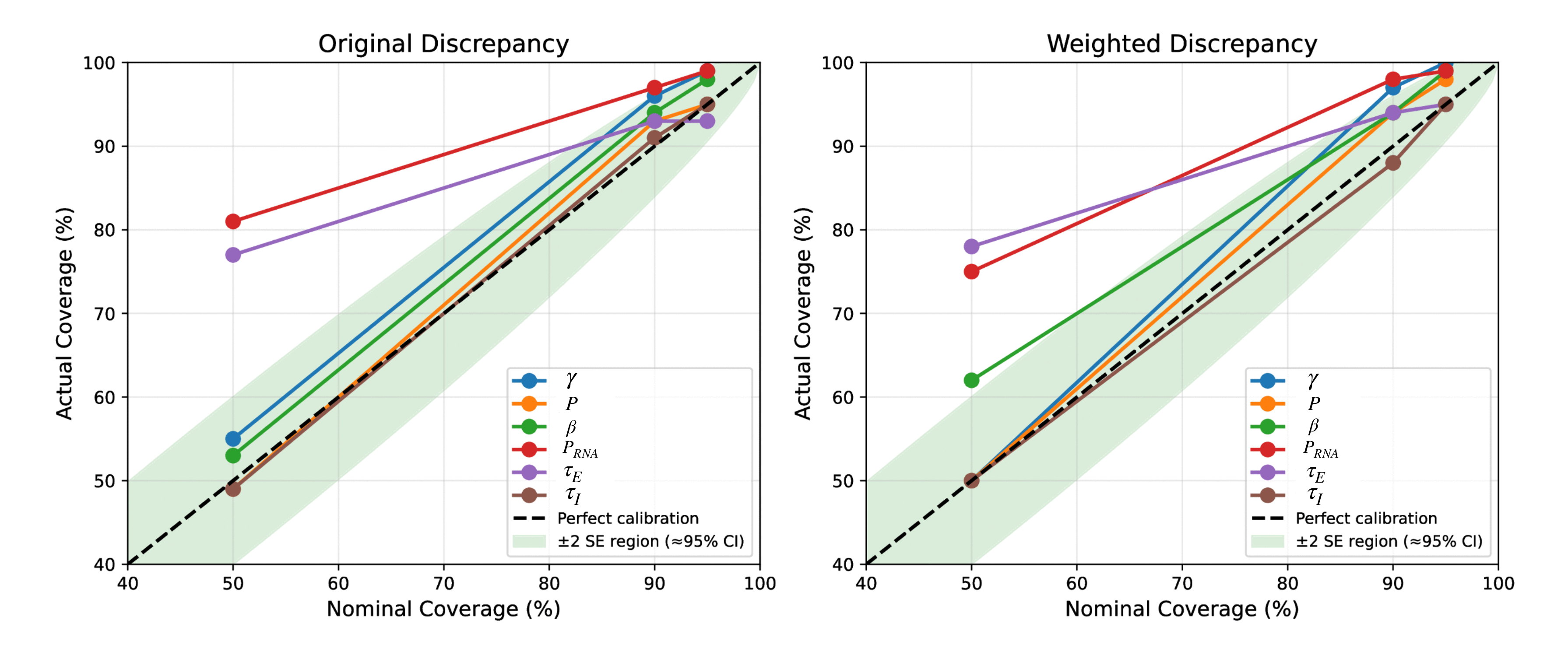}
\caption{\textbf{Comparison of calibration performance between the original (left) and weighted (right) discrepancy designs.} The dashed black line represents perfect calibration. The green shaded region indicates $\pm 2$ standard errors 
($\approx 95\%$ confidence interval) for the estimated coverage, calculated 
as $2\sqrt{p(1-p)/100}$ for nominal coverage level $p$. This corresponds to 
$\pm 10\%$ at 50\% nominal, $\pm 6\%$ at 90\% nominal, and $\pm 4.4\%$ at 
$95\%$ nominal coverage. Coverage values falling within this shaded region 
are considered statistically consistent with well-calibrated inference.}
\label{fig:calibration_comparison}
\end{figure}

To assess the sensitivity of our results to the discrepancy design, we also explored a weighted discrepancy where each component was multiplied by a weight inversely proportional to its variance. The weights were estimated empirically: for 500 parameter sets sampled from the prior, we generated pairs of independent simulations, computed the four component discrepancies between each pair, and estimated the variance of each component. The weight for component $k$ was then set to $w_k = 1/\text{Var}(d_k)$, normalized so that $\sum w_k = 4$. This ensures that noisier data sources (with higher variance) contribute less to the total discrepancy, following the principle that each data source should contribute in proportion to its information content. The weighted discrepancy was used as a sensitivity check on the original design; the results are presented in Table~\ref{tab:calibration_weighted} and visualized in Figure~\ref{fig:calibration_comparison} (right).

\begin{table}[ht]
\centering
\caption{\textbf{Calibration results for the weighted discrepancy design}, where each component is weighted by the inverse of its estimated variance. The ``Change (50\% CI)'' column shows the difference in 50\% coverage relative to the original discrepancy design (Table~\ref{tab:calibration_original}).}
\label{tab:calibration_weighted}
\begin{tabular}{lcccc}
\hline
Parameter & 50\% CI & 90\% CI & 95\% CI & Change (50\% CI) \\
\hline
$\beta$   & 62.0\% & 94.0\% & 99.0\%  & $+9.0\%$ \\
$\gamma$ & 50.0\% & 97.0\% & 100.0\% & $-5.0\%$ \\
$P$       & 50.0\% & 94.0\% & 98.0\%  & $+1.0\%$ \\
$P_{\text{RNA}}$ & 75.0\% & 98.0\% & 99.0\%  & $-6.0\%$ \\
$\tau_E$  & 78.0\% & 94.0\% & 95.0\%  & $+1.0\%$ \\
$\tau_I$  & 50.0\% & 88.0\% & 95.0\%  & $+1.0\%$ \\
\hline
\end{tabular}
\end{table}

The weighted approach improved the 50\% CI coverage for $P_{\text{RNA}}$ 
from 81.0\% to 75.0\%, moving it closer to the nominal level and confirming that the discrepancy weighting partially addresses the conservative behavior. This improvement likely arises because ED data, which receives higher weight, provides a cleaner indirect signal for total RNA production than the noisier RNA measurements themselves. This is consistent with the data contribution analysis in Section~\ref{sec:data_contribution_analysis} for parameter $P_{\text{RNA}}$.
However, $\tau_E$ remained conservative (77.0\% to 78.0\%), suggesting that 
more sophisticated data fusion approaches may be needed for this parameter. 
For $\beta$, the original 50\% coverage was 53.0\% (well-calibrated), but 
the weighted design increased this to 62.0\% (slightly conservative), 
indicating that ED data is not the optimal source for estimating the virion 
entry rate. The remaining parameters ($P$, $\gamma$, $\tau_I$) remained 
stable across both designs, with coverage values changing by at most 1 percentage point, confirming that our method's core performance is robust to the choice of discrepancy design.

Figure~\ref{fig:calibration_comparison} visualizes these results. The green shaded region indicates that the formal $\pm 2$ standard error range, within which coverage is statistically consistent with well-calibrated inference. For both designs, the parameters outside the shaded region at the 50\% level are the same ones identified above ($P_{\text{RNA}}$ and $\tau_E$ for the original; $P_{\text{RNA}}$, $\beta$, and $\tau_E$ for the weighted). The 90\% and 95\% coverage for all parameters fall within or close to the acceptable ranges. In summary, although neither design achieves perfect calibration across all parameters, both produce coverage consistent with nominal levels for most parameters, with the original design demonstrating robustness for the majority.

\subsection{Data Contribution Analysis}
\label{sec:data_contribution_analysis}
To understand the contribution of each data source to parameter estimation, we performed a systematic data contribution analysis. We ran simulation-based 
calibration with five different data configurations: 
\begin{itemize}
\item \textbf{RNA-only:} Using only viral RNA measurements (both MC and SC).
\item \textbf{ED-only:} Using only infectious virus (ED) measurements (both MC and SC).
\item \textbf{MC-only:} Using only multiple-cycle experiment data (both RNA and ED).
\item \textbf{SC-only:} Using only single-cycle experiment data (both RNA and ED).
\item \textbf{Combined:} Using all data sources (original method).
\end{itemize} 
Each configuration was evaluated using $100$ synthetic datasets.

This calibration study required 500 BOLFI inference runs (100 cases $\times$ 5 data configurations), each using 1,000 simulations. When run in parallel, the total wall time was approximately 20 hours, demonstrating the practical feasibility of this calibration-based framework enabled by BOLFI's efficiency.

We focus on the 50\% credible interval coverage as the primary metric for assessing data contribution, as it tests whether the posterior is correctly centered on the true parameter value. The 90\% and 95\% coverage provide complementary information about whether the uncertainty spread is correctly sized. For example, a parameter may have correct 50\% coverage (well-centered) but elevated 90--95\% coverage (overly wide tails), indicating partial identifiability rather than centering issues. We therefore examine all three levels together to characterize the full calibration pattern for each parameter.

\begin{figure}[ht]
\centering
\includegraphics[width=0.9\textwidth]{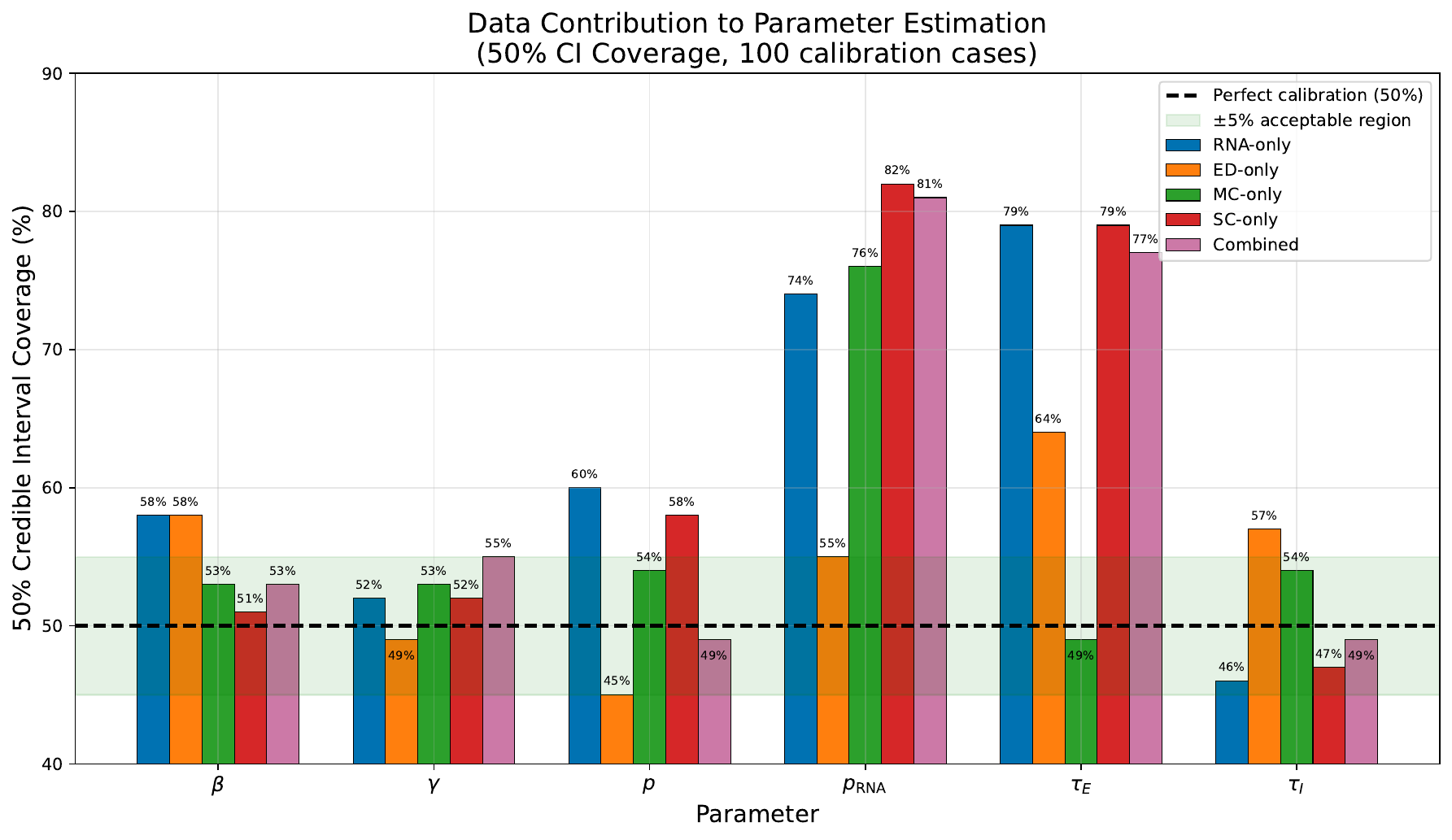}
\caption{\textbf{Comparison of 50\% credible interval coverage across five data configurations.} The dashed line at 50\% indicates perfect calibration. The light green shaded region in the figure shows the $\pm 5\%$ acceptable range, within which coverage is statistically consistent with well-calibrated inference. The combined configuration performs best for $P$ and $\tau_I$, while MC-only performs best for $\tau_E$, ED-only for $\gamma$ and $P_{\text{RNA}}$, and SC-only for $\beta$.}
\label{fig:data_contribution_50}
\end{figure}

The calibration results for all five data configurations are summarized in Table~\ref{tab:data_contribution} and visualized in Figs.~\ref{fig:data_contribution_50}, \ref{fig:data_contribution_90}, and \ref{fig:data_contribution_95}. The complete results at all three credible levels are provided in Table~\ref{tab:calibration_all_levels} in the Supporting Information.

\begin{table}[ht]
\centering
\caption{\textbf{50\% credible interval coverage for each data configuration.} Bold values indicate the best performance for each parameter.}
\label{tab:data_contribution}
\begin{tabular}{lcccccc}
\hline
Data Type & $\beta$ & $\gamma$ & $P$ & $P_{\text{RNA}}$ & $\tau_E$ & $\tau_I$ \\
\hline
RNA-only & 58.0\% & 52.0\% & 60.0\% & 74.0\% & 79.0\% & 46.0\% \\
ED-only  & 58.0\% & \textbf{49.0\%} & 45.0\% & \textbf{55.0\%} & 64.0\% & 57.0\% \\
MC-only  & 53.0\% & 53.0\% & 54.0\% & 76.0\% & \textbf{49.0\%} & 54.0\% \\
SC-only  & \textbf{51.0\%} & 52.0\% & 58.0\% & 82.0\% & 79.0\% & 47.0\% \\
Combined & 53.0\% & 55.0\% & \textbf{49.0\%} & 81.0\% & 77.0\% & \textbf{49.0\%} \\
\hline
\end{tabular}
\end{table}

Data contribution analysis reveals several important patterns in how different data sources inform parameter estimation. Below, we examine each parameter in detail, considering both the calibration coverage (focusing primarily on 50\% coverage in Table~\ref{tab:data_contribution}) and the biological relationships between parameters. The best-calibrated data type for each parameter is summarized in 
Table~\ref{tab:data_contribution_summary}.

\paragraph{$\beta$ (Rate of infectious virion entry)}
SC-only data gives the best calibration for $\beta$ (51.0\%), with 
MC-only (53.0\%) and Combined (53.0\%) also performing well. ED-only 
(58.0\%) and RNA-only (58.0\%) are more conservative. The calibration 
ranges from 51.0\% to 58.0\% -- a spread of 7 percentage points. The 
narrow spread shows that parameter $\beta$ is rather robust to data configuration. And, the fact that SC-only achieves near-perfect calibration suggests that the single-cycle experiment captures the essential dynamics for estimating the rate of virion entry. Biologically, this result is intuitive: the single-cycle experiment 
starts with a large, synchronized inoculum of viruses. This provides 
a clear signal for the initial entry rate ($\beta$), as the dynamics 
are dominated by the early infection events. In contrast, the 
multiple-cycle experiment involves ongoing rounds of infection, 
which may obscure the initial entry dynamics. 
Notably, both ED-only and RNA-only are more conservative (58.0\%), 
indicating that relying solely on one measurement type (either ED or 
RNA) introduces additional uncertainty. The combined approach improves the calibration from 58.0\% 
to 53.0\% -- a meaningful improvement of 5 percentage points -- 
demonstrating that combining complementary data sources enhances 
the estimation of $\beta$. While SC-only remains the best single data configuration, the combined approach offers a robust alternative with only a slight conservative bias (53.0\% vs 51.0\%).

\paragraph{$\gamma$ (Rate of successful infection per virion entry)}
ED-only data gives the best calibration for $\gamma$ (50\% CI: 49.0\%), 
followed closely by SC-only (52.0\%) and RNA-only (52.0\%). MC-only 
(53.0\%) and Combined (55.0\%) are slightly more conservative. The 
50\% calibration ranges from 49.0\% to 55.0\% across all configurations 
--- a narrow spread of only 6 percentage points, suggesting that all 
data configurations yield well-centered posterior estimates for 
$\gamma$.

However, examining the 90\% and 95\% coverage reveals a different 
picture: all configurations show elevated coverage at 90\% (95--97\% 
vs nominal 90\%) and near-perfect coverage at 95\% (99--100\% vs 
nominal 95\%). This pattern --- correct 50\% coverage but overly wide 
tails --- is consistent with the biological and statistical challenges 
of estimating $\gamma$ described in Section~\ref{gamma challenge}.
As discussed in Section~\ref{gamma challenge}, $\gamma$ encapsulates multiple biological processes: cellular defense mechanisms, stochastic replication outcomes, and potential virion defects. The data can clearly 
establish a lower bound ($\gamma > 0.20$), ruling out very low infection establishment rates. However, values up to the classifier boundary remain plausible, indicating that the data provide almost no information on the upper end of the infection establishment rate. This results in a posterior for $\gamma$ that is constrained at the lower bound but otherwise flat, which translates into correct centering (50\% coverage) but overly wide credible intervals (90--95\% coverage).

The ED-only configuration achieves the best 50\% coverage (49.0\%), 
suggesting that ED data alone provides sufficient information for 
centering the posterior. The slightly conservative behavior of the 
combined configuration (55.0\%) indicates that adding RNA data does 
not significantly improve the estimate.

\paragraph{$P$ (Infectious production rate)}
Combined all data gives the best calibration for $P$ (49.0\%), followed 
by MC-only (54.0\%) and ED-only (45.0\%). RNA-only (60.0\%) and SC-only 
(58.0\%) are noticeably more conservative. The calibration ranges 
from 45.0\% to 60.0\% --- a spread of $15$ percentage points. This wide 
spread indicates that $P$ is sensitive to the data configuration. The fact that MC-only performs better than SC-only (54.0\% vs 58.0\%) is consistent with the nature of infectious 
production: the multiple-cycle experiment captures the accumulation 
of infectious virus over multiple rounds of replication, providing 
more information about the production rate ($P$). The single-cycle 
experiment, which only captures one round of infection, provides 
a more limited view of infectious production.
Notably, RNA-only performs poorly, suggesting that RNA 
data alone (without ED data from the same experiment) leads to 
overly wide credible intervals. The combined approach brings the 
calibration to the ideal level, demonstrating that ED data provides 
essential complementary information for $P$. This makes biological 
sense: $P$ describes infectious virion production, and ED data 
directly measures infectious virus. Overall, the approach of combined all data type achieves the best balance.

\paragraph{$P_{\text{RNA}}$ (Total RNA virion production rate)}
This parameter shows the most striking variation. ED-only gives the 
best calibration (55.0\%), followed by RNA-only (74.0\%), MC-only 
(76.0\%), Combined (81.0\%), and SC-only (82.0\%). The calibration 
ranges from 55.0\% to 82.0\% --- a dramatic spread of 27 percentage 
points. The fact that ED-only gives the best calibration is 
counter-intuitive: $P_{\text{RNA}}$ is the total RNA production rate, 
so RNA data should be directly informative. Yet the direct RNA measurements are outperformed by indirect ED measurements. 
A possible explanation for this surprising result lies in the 
distinction between how $V_{\text{RNA}}$ is defined in the model 
and what the RNA measurement actually captures. In the model, 
$V_{\text{RNA}}$ represents extracellular viral RNA from intact 
virions. However, RT-qPCR-based RNA quantification measures total 
extracellular viral RNA, which may include RNA from damaged or 
degraded virions as well as free-floating RNA fragments --- components 
that are not explicitly represented in the model. The assay cannot 
distinguish between RNA that is packaged in infectious virions and 
RNA from non-infectious or degraded particles, and is also subject 
to technical variability including RNA degradation during sample 
handling, reverse transcription biases, and amplification 
inefficiencies \cite{Lamb2024, Mohorianu2023, GenomicsInsights2019}.
ED data, by contrast, measures only infectious virus ($V$) through a functional assay that detects productive infection, yielding integer counts that are more stable 
and less prone to technical noise. Although ED data does not directly measure total RNA, the strong biological correlation between infectious virus production and total RNA production makes it an effective indirect proxy for $P_{\text{RNA}}$ --- one that avoids the measurement complications inherent in RNA quantification. We emphasize that this interpretation is speculative; other factors, such as the temporal dynamics of RNA production versus virion assembly, could also contribute to the observed pattern.

This contrasts with the results for $P$, where combined data 
outperforms ED-only. For $P$ (infectious production), RNA data 
provides useful correlated information that supplements the ED 
signal, improving estimation. For $P_{\text{RNA}}$ (total RNA 
production), the direct RNA signal appears less reliable for inference, and adding it to the ED data degrades the calibration. This distinction highlights that the same data source can be beneficial for one parameter and detrimental for another, depending on the biological interpretation of the measurement, underscoring the importance of understanding both the biological relationships and the measurement properties of different data sources.

SC-only gives the poorest calibration, consistent with the limited information 
provided by a single round of infection for estimating ongoing 
RNA production. For estimating infectious production rate, it is better to start with a low virus titer and observe the infectious cycle. This observation is consistent with calibration result for parameter $P$. 

\paragraph{$\tau_E$ (Mean eclipse phase duration)}
MC-only gives perfect calibration for $\tau_E$ (49.0\%), while all 
other configurations perform poorly: ED-only (64.0\%), Combined (77.0\%), 
RNA-only (79.0\%), and SC-only (79.0\%). The calibration ranges from 
49.0\% to 79.0\% -- a spread of 30 percentage points, the largest of 
all parameters. 

The superiority of MC-only can be understood by considering the 
experimental design. The eclipse phase is the time between initial 
infection and the onset of virus production. The multiple-cycle 
experiment captures this process over several rounds of replication, 
providing rich temporal information about when infectious virus 
first appears. Critically, MC-only combines both RNA and ED data 
from the same experiment, ensuring internal consistency between 
the two measurement types.

The poor performance of SC-only (79.0\%) and RNA-only (79.0\%) 
indicates that neither single-cycle data nor RNA data alone is 
sufficient for estimating $\tau_E$. The single-cycle experiment 
uses a high multiplicity of infection (MOI) to synchronize infection, 
which may obscure the eclipse phase dynamics by overwhelming the 
cells with virus. The RNA data alone lacks the infectious virus 
information needed to pinpoint when new infectious particles are 
produced.

Strikingly, Combined (77.0\%) performs worse than MC-only (49.0\%). 
This indicates that adding SC data (both RNA and ED) to the MC data 
introduces inconsistency for this parameter. The SC experiment captures a different 
infection regime (high MOI, synchronized infection), and its data 
are not fully compatible with the MC experiment. When combined, 
these inconsistencies widen the posterior distribution, leading 
to conservative credible intervals. This demonstrates that, for certain parameters such as $\tau_E$, combining 
data from different experimental designs is not always beneficial 
--- internal consistency within a single experiment can be more 
valuable than adding data from a different experimental regime.

Biologically, this result suggests that the eclipse phase is best 
captured by the multiple-cycle experimental design, which follows 
the natural course of infection over multiple rounds of replication, 
rather than the synchronized single-cycle design.

\paragraph{$\tau_I$ (Mean infectious phase duration)}
Combined data gives the best calibration for $\tau_I$ (49.0\%), 
followed by RNA-only (46.0\%), SC-only (47.0\%), MC-only (54.0\%), 
and ED-only (57.0\%). The calibration ranges from 46.0\% to 57.0\% --- 
a spread of 11 percentage points. The combined approach achieves perfect calibration, demonstrating that integrating both RNA and ED measurements from both single-cycle and multiple-cycle experiments provides complementary information. This integration is essential for accurately estimating the infectious phase duration. 
The slightly low coverage for RNA-only (46.0\%) and SC-only (47.0\%) suggests that RNA data alone, particularly from the single-cycle experiment, may under-estimate uncertainty for $\tau_I$. Notably, MC-only (54.0\%) and ED-only (57.0\%) are more conservative, indicating that each of these data sources alone introduces some uncertainty. However, when combined, the complementary information from all four data sources yields well-calibrated inference for this parameter. This demonstrates that the infectious phase duration is best estimated when both viral RNA and infectious virus measurements are integrated across both experimental designs.

\paragraph{Contrasting Behavior: $\tau_E$ vs $\tau_I$}
A particularly revealing finding emerges from comparing the calibration results for $\tau_E$ (eclipse phase duration) and $\tau_I$ (infectious phase duration). Both parameters describe temporal aspects of the infection cycle, yet they show opposite responses to the inclusion of single-cycle (SC) data. MC-only gives perfect calibration for $\tau_E$ (49.0\%), while Combined (which adds SC data) performs poorly (77.0\%). In contrast, Combined gives the best calibration for $\tau_I$ (49.0\%), while MC-only is more conservative (54.0\%). This difference can be understood biologically: the eclipse phase ($\tau_E$) is sensitive to initial infection conditions --- the high MOI used in the single-cycle experiment synchronizes infection and artificially alters the timing of the eclipse phase, making SC data inconsistent with MC data for estimating $\tau_E$, and adding it worsens the calibration. In contrast, the infectious phase duration ($\tau_I$) measures how long virus production persists. This parameter is less sensitive to initial conditions and more dependent on the overall duration of the infection cycle. SC data, which captures later stages of infection, 
provides complementary information that improves $\tau_I$ estimation. Figure~\ref{fig:tau_comparison} is a conceptual illustration of how SC data affect $\tau_E$ and $\tau_I$ differently. 

\begin{figure}[ht]
\centering
\begin{tikzpicture}[scale=0.8]
\draw[->, thick] (0,0) -- (9,0) node[right] {Time};

\draw[fill=black] (0,0) circle (2pt) node[below=6pt] {Infection};
\draw[fill=black] (3.5,0) circle (2pt) node[below=6pt] {Production starts};
\draw[fill=black] (7,0) circle (2pt) node[below=6pt] {Production ends};

\draw[decorate,decoration={brace,amplitude=6pt},xshift=0pt,yshift=2pt] 
    (0,0.1) -- (3.5,0.1) node[midway,above=8pt] {$\tau_E$};
\draw[decorate,decoration={brace,amplitude=6pt},xshift=0pt,yshift=2pt] 
    (3.5,0.1) -- (7,0.1) node[midway,above=8pt] {$\tau_I$};

\node at (4.5,-1.5) {\footnotesize SC: high MOI (synchronized)};
\node at (4.5,-2.0) {\footnotesize MC: low MOI (natural)};

\node[align=center, blue] at (1.3,-3.0) {SC data affects $\tau_E$\\ (inconsistent)};
\node[align=center, red] at (6.0,-3.0) {SC data complements $\tau_I$\\ (consistent)};

\end{tikzpicture}
\caption{\textbf{Conceptual illustration of the contrasting effects of SC data on $\tau_E$ and $\tau_I$.} The eclipse phase ($\tau_E$) is sensitive to initial infection conditions, making SC data inconsistent with MC data. The infectious phase ($\tau_I$) is less sensitive to initial conditions, allowing SC data to provide complementary information.}
\label{fig:tau_comparison}
\end{figure}

This contrast demonstrates that combining data from different 
experimental designs is not universally beneficial. The value of additional data depends on the biological parameter being estimated 
and the compatibility of the experimental conditions. This finding 
has important implications for experimental design: researchers 
should consider which parameters they aim to estimate when deciding 
whether to combine data from different experimental setups.

\begin{table}[ht]
\centering
\caption{{\bf{Best-calibrated data type}} and 50\% credible interval coverage for each parameter. See Section~\ref{sec:data_contribution_analysis} for data type definitions.}
\label{tab:data_contribution_summary}
\begin{tabular}{lccc}
\hline
Parameter & Best Data Type & 50\% Coverage \\
\hline
$\beta$ & SC-only & 51.0\% \\
$\gamma$ & ED-only & 49.0\% \\
$P$ & Combined & 49.0\% \\
$P_{\text{RNA}}$ & ED-only & 55.0\% \\
$\tau_E$ & MC-only & 49.0\% \\
$\tau_I$ & Combined & 49.0\% \\
\hline
\end{tabular}
\end{table}

In summary, the data contribution analysis reveals that no single data configuration is optimal for all parameters (Table~\ref{tab:data_contribution_summary}). The single-cycle experiment alone is most reliable for estimating the virion entry rate ($\beta$), while ED-only data is best for the infection establishment rate ($\gamma$) and, surprisingly, the total RNA production rate ($P_{\text{RNA}}$). The multiple-cycle experiment alone provides the best calibration for the eclipse phase duration ($\tau_E$), whereas combining all data sources is optimal for the infectious virion production rate ($P$) and the infectious phase duration ($\tau_I$). These findings underscore that the choice of data configuration should be guided by the parameter of interest, and that combining all available data is not universally beneficial.

\paragraph{Patterns across credible levels.}

While the 50\% coverage assesses centering, the 90\% and 95\% coverage 
reveal whether the uncertainty spread is correctly sized (see the S5 Appendix, Figs~\ref{fig:data_contribution_90}, \ref{fig:data_contribution_95} 
and Table~\ref{tab:calibration_all_levels}). Examining all three levels together reveals three patterns: \textbf{Well-calibrated} parameters 
($P$ and $\tau_I$ with combined data) show near-nominal coverage across 
all levels; \textbf{conservative} parameters ($\gamma$ across all 
configurations, and $P_{\text{RNA}}$ and $\tau_E$ with RNA- or SC-inclusive 
data) have correct 50\% coverage but elevated 90--95\% tails, reflecting 
partial identifiability or data inconsistency; and \textbf{robust} 
parameters ($\beta$) maintain near-nominal coverage across all data 
configurations.

\subsubsection{Implications for Experimental Design}

Our data contribution analysis has meaningful implications for experimental 
design in virus dynamics studies:

\begin{itemize}
\item If the goal is to estimate $\tau_E$, the multiple-cycle experiment alone is sufficient and more reliable than the combined data, as adding single-cycle data introduces inconsistency.
\item If the goal is to estimate $P$ or $\tau_I$, combining RNA and ED data 
is recommended.
\item The single-cycle experiment alone is sufficient for $\beta$ estimation, as it provides a clear signal for the initial entry dynamics.
\end{itemize}

These findings caution against the common practice of combining all available data sources without evaluating their individual contributions.
Data contribution analysis can guide more efficient experimental design 
and improve parameter estimation accuracy.

\subsection{Composite diagnostic: parameter-specific data selection}
\label{sec:composite_diagnostic}

The data contribution analysis above reveals that different data types yield the best calibration for different parameters (Table~\ref{tab:data_contribution_summary}). This raises a natural question: if one were to select the best-calibrated data type for each parameter, how would the resulting estimates compare to those from the combined inference?

\begin{figure}[ht]
\centering
\includegraphics[width=0.9\textwidth]{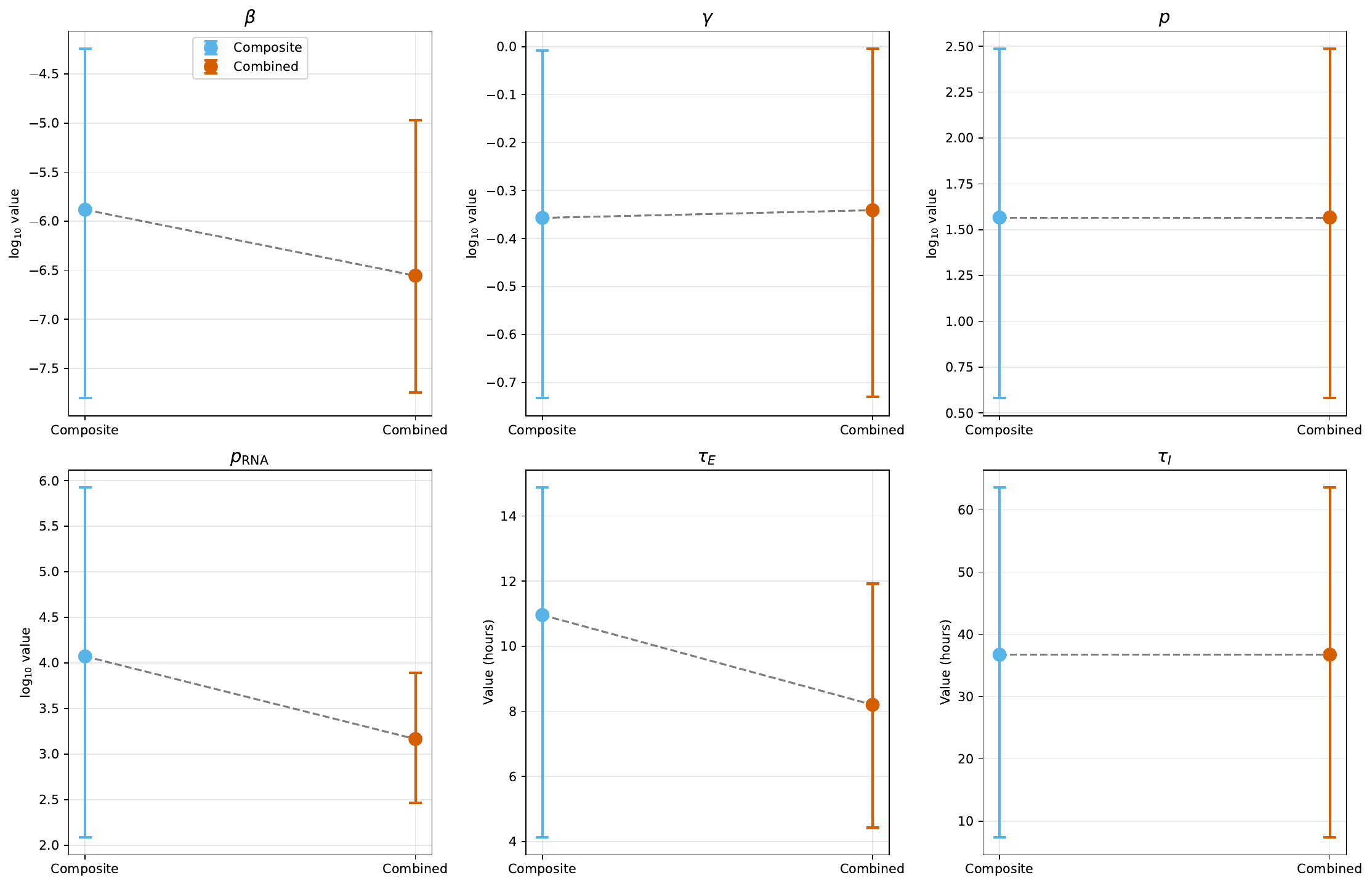}
\caption{\textbf{Comparison of composite diagnostic and combined inference results.} 
Composite estimates (blue) use the best-calibrated data type for each 
parameter, while combined estimates (orange) use all data sources. 
Error bars represent 95\% credible intervals. For parameters in $\log_{10}$ 
scale ($\beta$, $\gamma$, $P$, $P_{\text{RNA}}$), the composite $\beta$ 
estimate is approximately $4.7$ times larger than the combined estimate in linear scale, and the composite $P_{\text{RNA}}$ estimate is approximately $8.1$ times 
larger in linear scale. These differences highlight the trade-off between precision (narrower combined intervals) and potential bias (composite estimates 
may be more accurate for parameters where specific data types are 
identified as more reliable).}
\label{fig:composite_vs_combined}
\end{figure}

To explore this question, we constructed a \textbf{composite diagnostic} that selects, for each parameter, the posterior from the data configuration that yielded the best calibration (closest to 50\% coverage). Table~\ref{tab:composite_vs_combined} and Fig~\ref{fig:composite_vs_combined} compare these composite estimates with the combined inference results. 

\begin{table}[ht]
\centering
\caption{Comparison of composite diagnostic and combined inference results. 
Composite estimates are selected from the best-calibrated data 
configuration for each parameter. All values for $\beta$, $\gamma$, $P$ and $P_{\text{RNA}}$ are in $\log_{10}$ scale; $\tau_E$ and $\tau_I$ are in 
linear scale. The units in linear scale are the same as in Table~\ref{tab:BOLFI inferred parameters}.}
\label{tab:composite_vs_combined}
\begin{tabular}{lcccccc}
\hline
\multirow{2}{*}{Parameter} & \multicolumn{3}{c}{Composite} & \multicolumn{3}{c}{Combined} \\
\cline{2-7}
 & Mean & \multicolumn{2}{c}{95\% CI} & Mean & \multicolumn{2}{c}{95\% CI} \\
\hline
$\beta(\log_{10})$ & -5.883 & -7.804 & -4.241 & -6.555 & -7.746 & -4.973 \\
$\gamma(\log_{10})$ & -0.357 & -0.733 & -0.008 & -0.341 & -0.730 & -0.004 \\
$P(\log_{10})$ & 1.566 & 0.580 & 2.488 & 1.566 & 0.580 & 2.488 \\
$P_{\text{RNA}}(\log_{10})$ & 4.072 & 2.087 & 5.924 & 3.165 & 2.464 & 3.894 \\
$\tau_E$ & 10.958 & 4.132 & 14.872 & 8.202 & 4.423 & 11.919 \\
$\tau_I$ & 36.739 & 7.402 & 63.569 & 36.739 & 7.402 & 63.569 \\
\hline
\end{tabular}
\end{table}

Several patterns emerge from this comparison. For $\log_{10}{P}$ and $\tau_I$, the composite and combined estimates are identical, as the combined data was identified as the best data type for both parameters. For $\log_{10}{\gamma}$, the composite estimate ($-0.357$, 95\% CI: $[-0.733, -0.008]$) is nearly identical to the combined estimate ($-0.341$, 95\% CI: $[-0.730, -0.004]$), confirming that ED-only data provides essentially the same information for this parameter as the full combined data.
More substantial differences appear for the remaining parameters. The composite estimate for $\log_{10}{\beta}$ ($-5.883$, 95\% CI: $[-7.804, -4.241]$) is notably higher than the combined estimate ($-6.555$, 95\% CI: $[-7.746, -4.973]$), suggesting that SC-only data yields a higher infection rate estimate. In linear scale, this corresponds to a factor of approximately $4.7$. Similarly, the composite estimate for $\log_{10}{P_{\text{RNA}}}$ ($4.072$, 95\% CI: $[2.087, 5.924]$) is substantially higher than the combined estimate ($3.165$, 95\% CI: $[2.464, 3.894]$), corresponding to a factor of approximately $8.1$ in linear scale. For $\tau_E$, the composite estimate ($10.958$ h, 95\% CI: $[4.132, 14.872]$) is notably higher than the combined estimate ($8.202$ h, 95\% CI: $[4.423, 11.919]$), a difference of $2.76$ hours in the mean.

A notable pattern emerges: the combined estimates consistently fall within the credible intervals of the composite estimates, and the combined credible intervals are generally narrower. This is particularly evident for $\log_{10}{P_{\text{RNA}}}$ (composite CI width: $3.8$ vs combined: $1.4$) and $\log_{10}{\beta}$ (composite: $3.6$ vs combined: $2.8$). This indicates that while combined data provides more precise estimates (narrower intervals), the composite diagnostic suggests that for some parameters, this precision may come at the cost of potential conservative bias.

\paragraph{Caveats and interpretation.}

We emphasize several important caveats regarding this composite diagnostic. First, it is \textbf{not a formal statistical estimator}. The composite approach does not preserve joint correlations between parameters, nor does it account for differences in information content across data configurations. For applications requiring joint inference, the combined approach, which preserves the full posterior covariance structure, remains the appropriate choice.

Second, the term ``best-calibrated'' in this context refers specifically to the 50\% credible interval coverage. The 90\% and 95\% coverage provide additional information about the tails of the posterior distribution, which we report in Table~\ref{tab:data_contribution}. The selection of 50\% coverage as the primary criterion reflects our focus on whether the posterior is correctly centered, rather than whether the tails are correctly sized.

Third, the composite diagnostic should be understood as an \textbf{exploratory tool} for checking the combined inference, not as a replacement for it. Its value lies in revealing which parameters are well-constrained by each data source and identifying cases where combining data may introduce conservative bias.

Despite these caveats, the composite diagnostic provides a useful check on the combined inference. The differences observed for $\beta$, $P_{\text{RNA}}$, and $\tau_E$ suggest that these parameters warrant closer scrutiny when interpreting combined inference results. The composite diagnostic thus serves as a complement to the combined approach, offering additional insight into the reliability of parameter-specific uncertainty quantification.

\textbf{In practice, for this application, we recommend:} (i) using combined inference for final parameter estimates. Although combined data can introduce conservative bias for some parameters (notably $P_{\text{RNA}}$ and $\tau_E$), it preserves the full joint correlation structure between parameters, which is essential for any application requiring coherent multivariate inference. The combined posterior also provides narrower credible intervals, reflecting the additional information from multiple data sources; 
(ii) using the composite diagnostic as a sensitivity check to identify parameters where combining data sources may introduce bias. The composite diagnostic is not a formal estimator --- it is an exploratory tool that reveals which parameters are well-constrained by each data source and flags cases where the combined inference may be overly conservative; 
and (iii) if combined inference gives poor calibration for specific parameters, reporting this finding with appropriate caveats and, where scientifically appropriate, considering whether a single data source provides more reliable marginal credible intervals for that parameter. However, any such parameter-specific selection should be understood as an exploratory exercise rather than a replacement for joint inference.

\subsection{Summary of the calibration-guided data fusion framework}
\label{sec:framework_summary}
The framework above provides a general approach that can be applied to other multi-source inference problems where simulation-based models are used. The key insight is that computational efficiency (here enabled by BOLFI) makes it practical to run SBC across multiple data configurations, revealing how each data source contributes to parameter estimation.
The complete framework is summarized in the following workflow (Algorithm~\ref{alg:framework}):

\begin{algorithm}[H]
\caption{\textbf{Workflow for calibration-guided data fusion}}
\label{alg:framework}
\begin{algorithmic}[1]
\State \textbf{Step 1: Define inference target}
\State \quad Identify model parameters $\boldsymbol{\theta}$ and prior distributions $p(\boldsymbol{\theta})$
\State \quad Specify simulator $\mathcal{S}(\boldsymbol{\theta})$ that generates synthetic data

\State
\State \textbf{Step 2: Perform inference for each data configuration}
\State \quad For each data source or combination of sources $k = 1, \ldots, K$:
\State \quad \quad Run BOLFI (or other inference algorithm) to obtain posterior $\pi_k(\boldsymbol{\theta} \mid D_k)$
\State \quad \quad Store posterior samples and credible intervals
\State
\State \textbf{Step 3: Simulation-based calibration (SBC)}
\State \quad For each data configuration $k$ and $N$ synthetic datasets generated from the prior:
\State \quad \quad Run inference on each synthetic dataset
\State \quad \quad Record whether true parameter $\theta_j$ falls within credible intervals
\State \quad Compute coverage for each parameter and credible level
\State
\State \textbf{Step 4: Data contribution analysis}
\State \quad For each parameter $\theta_j$, examine calibration patterns across data configurations
\State \quad Identify which data sources yield well-centered posteriors, which introduce conservative bias, and which lead to overconfidence
\State \quad Interpret relationships: which data sources are informative, redundant, or contradictory for each parameter
\State \quad Derive insights for experimental design --- which experiments are most valuable for which parameters
\State
\State \textbf{Step 5: Sensitivity check and decision}
\State \quad Compare combined inference results with estimates from best-calibrated data type (optional)
\State \quad Based on Step 4 insights, decide whether combined inference is appropriate, or whether parameter-specific data selection is warranted
\State \quad If new experimental designs are suggested, consider updating the inference with new data
\State \quad When computational efficiency is a priority, approximate algorithms (e.g., BOLFI) can be used for the calibration tests; once the data configuration is decided, more computationally intensive but potentially more accurate algorithms can be applied for final inference
\State
\State \textbf{Output:}
\State \quad Understanding of how each data source contributes to each parameter (informativeness, redundancy, contradiction, bias)
\State \quad Guidance for experimental design
\State \quad Informed decision on whether to use combined inference, single-source inference, or a composite diagnostic
\State \quad Composite diagnostic as a sensitivity check
\end{algorithmic}
\end{algorithm}

The framework is modular: inference can be performed with any algorithm (not just BOLFI), data configurations can be tailored to the problem, and the sensitivity check in Step 5 is optional. This framework is most valuable when (i) multiple data sources are available, (ii) the simulator is computationally tractable for repeated runs, and (iii) parameter-specific inference reliability is a concern.

\section{Discussion and Conclusion}

\subsection{Summary of Contributions}

This work makes three interconnected contributions, each building on the previous.

\paragraph{1. Efficient likelihood-free inference for virus dynamics.}
First, we demonstrated that BOLFI provides a computationally efficient alternative to traditional MCMC for virus dynamics parameter estimation. With only 1,000 simulations, BOLFI achieved a 20-fold speed-up compared to the previous approach, while maintaining well-calibrated inference for most parameters. The minimum discrepancy stabilized after approximately 220 successful simulations, confirming BOLFI's sample efficiency. Methodologically, we introduced a GP-based classifier that automatically learns to avoid simulator failures, a common challenge in complex simulation-based models across many scientific domains. We also designed a discrepancy that integrates RNA and ED data from both single and multiple cycle experiments. We have discussed the reason why we set the discrepancy in such a way and the close relation with sufficient statistics for these data. 

This establishes BOLFI as a powerful, scalable, and interpretable tool for parameter inference in biological systems modeling and beyond. The substantial computational gains we achieved --- reducing runtime from days to hours --- remove a key barrier to routine application of efficient inference methods. As experimental data become increasingly rich and heterogeneous, approaches like ours will be essential for extracting maximum insight from available measurements while honestly representing remaining uncertainties.

Our parameter estimates yield several biologically meaningful insights into influenza A dynamics. The extremely low virion entry rate ($\beta \approx 1.5\times10^{-7}$ ml/(cell$\cdot$h)) reflects the stochastic nature of virus-cell encounters. The infection probability $\gamma$ has a well-defined lower bound but uncertain upper bound, indicating that current experimental designs provide limited resolution for this parameter --- a finding confirmed across independent inference strategies. High production rates for both total RNA ($P_{\text{RNA}} \approx 1700$ vRNA/(cell$\cdot$h)) and infectious virions ($P \approx 49$ [V]/(cell$\cdot$h)) reveal a $\sim$35-fold disparity between RNA transcription and infectious particle assembly. Eclipse phase estimates ($\tau_E \approx 8.1$ h) align precisely with known biological ranges, while the infectious phase ($\tau_I \approx 29.4$ h) captures cell-level variability consistent with stochastic infection dynamics.

\paragraph{2. A general systematic framework for data contribution analysis.}
Second, we introduced a general framework for assessing data contribution in multi-source inference (Algorithm~\ref{alg:framework}). The idea is simple yet powerful: use simulation-based calibration to evaluate what each data source actually contributes to parameter estimation. This framework is not limited to likelihood-free inference, to virus dynamics, or to biological data; it applies to any multi-source inference problem where simulation-based models are used.

The key insight is that computational efficiency enables methodological rigor. When inference is fast enough, calibration becomes a practical diagnostic tool. With BOLFI completing inference in approximately 2 hours, we can afford to run hundreds of calibration cases across multiple data configurations in acceptable time. Beyond the computational improvement, it opens the door to systematic analyses --- such as calibration-based data contribution assessment --- that were previously impractical.

To our knowledge, this is the first time that simulation-based calibration has been used not only to validate an inference method but to systematically identify which data sources are most reliable for each parameter. While existing approaches to data contribution analysis, such as Shapley values, feature importance, and ablation studies, focus on prediction performance, our framework focuses on \textit{inference reliability}. We are answering the question: does a data source provide correct credible intervals for each parameter?

\paragraph{3. Composite inference as a diagnostic and practical tool.}
Third, we explored a composite diagnostic that selects the best-calibrated data type for each parameter, illustrating how SBC can be used as a sensitivity check on combined inference. The comparison reveals that for some parameters ($\beta$, $P_{\text{RNA}}$, $\tau_E$), the combined estimates differ substantially from those obtained from the best-calibrated single source, highlighting the trade-off between precision and potential conservative bias in multi-source inference. Researchers should be aware that combining all available data is not always optimal.

\subsection{Insights, Recommendations, and Broader Implications}

Our application reveals a counter-intuitive finding: combining all available data sources does not always improve inference. ED data is beneficial for estimating the infectious virion production rate ($P$) and infectious phase duration ($\tau_I$), where combined data yields the best calibration. For the total RNA production rate ($P_{\text{RNA}}$), ED-only data surprisingly outperforms combined data (55\% vs 81\% 50\% coverage), indicating that RNA measurements introduce noise rather than information for this parameter. For the eclipse phase duration ($\tau_E$), MC-only data (49\% coverage) is more reliable than combined data (77\% coverage), indicating that adding single-cycle data introduces inconsistency.

This finding is distinct from existing work on conflicting data sources (e.g., MOBOLFI \cite{chen2025mobolfi}), which focuses on detecting conflict between data sources. Our work addresses a different question: \textit{even without obvious conflict, does combining data sources improve inference for each parameter?} We found that a data source can be beneficial for some parameters and detrimental for others, which is a parameter-specific contribution that global conflict detection alone cannot capture. 
More broadly, this framework can guide experimental design and applies to any domain where multiple data sources are combined in simulation-based inference.

Based on our analysis for this specific application, we recommend using combined inference for final parameter estimates for this case, as it leverages all available data and provides the most precise results. The composite diagnostic serves as a sensitivity check to identify parameters where combining data sources may introduce bias. For parameters where combined inference gives poor calibration (in our study, $P_{\text{RNA}}$ and $\tau_E$), considering a single data source may provide more reliable credible intervals. Importantly, the combined estimates consistently fall within the composite credible intervals, indicating that combined inference is not \textit{wrong} but rather a more precise estimate consistent with the composite uncertainty range.

\subsection{Limitations and Future Work}

Several limitations should be acknowledged. First, the simulator has a failure rate of approximately 35\%, which affects efficiency. 

Second, the discrepancy design could be further optimized. The conservative calibration for $P_{\text{RNA}}$ and $\tau_E$ suggests that RNA measurements (for $P_{\text{RNA}}$) and single-cycle data (for $\tau_E$) may not be optimally weighted relative to other data sources. While the discrepancy design is theoretically justified through its connection to sufficient statistics, this justification holds in expectation and does not guarantee perfect calibration in finite samples. The improvement from the weighted discrepancy suggests that a more principled approach to weighting --- such as learning weights from data or using a hierarchical model — could further improve calibration. This highlights the distinction between theoretical validity (preserving information) and practical calibration; empirical calibration checks remain essential for assessing inference reliability. Despite this, our method provides well-calibrated uncertainty estimates for the key parameters $P$, $\gamma$, and $\tau_I$; the conservative behavior for $P_{\text{RNA}}$ and $\tau_E$ produces wider intervals, which is preferable to overconfident inference.

Third, the composite diagnostic is exploratory and does not preserve joint correlation structures. It should be understood as a sensitivity check on combined inference, not a formal method for posterior combination.

Fourth, BOLFI relies on a Gaussian process surrogate model, which introduces approximation error \cite{helin2024introduction}. While SBC confirms this does not compromise reliability for key parameters, the surrogate's accuracy depends on tuning of GP hyperparameters and the acquisition function. We followed standard practices \cite{lintusaari2018elfi, gutmann2016bayesian} and monitored convergence via stabilization of GP parameters and minimum discrepancy. More generally, BOLFI results may be sensitive to kernel choice, length scales, and acquisition function; users should assess surrogate quality through diagnostics.

The computational efficiency achieved here opens new possibilities for iterative analyses (multiple datasets, prior sensitivity, model variations) that were previously prohibitive. Future work will extend this framework to other domains, explore more sophisticated data fusion methods such as MOBOLFI \cite{chen2025mobolfi}, and develop automated tools for data contribution analysis. Additionally, the framework could use approximate algorithms for calibration tests, followed by more accurate algorithms for final inference once the data configuration is decided.

 In summary, we have shown that computational efficiency enables a new level of methodological rigor in multi-source inference, revealing that combining all available data is not monotonically optimal and providing a framework for systematically assessing what each data source contributes to parameter estimation.

\section*{Acknowledgments}
This work was initiated while the author (Y.X.) was supported by the RIKEN Special Postdoctoral Researcher (SPDR) program at the RIKEN Interdisciplinary Theoretical and Mathematical Sciences (iTHEMS) Center, Japan, and continued at the University of Helsinki, Finland, where the author (Y.X.) was supported by the Finnish Center for Artificial Intelligence (FCAI). The funders had no role in study design, data collection and analysis, decision to publish, or preparation of the manuscript. The author thanks Catherine Beauchemin for sharing the experimental data, Ulpu Remes for discussions during the early stages of this work, and Jukka Corander for hosting at the University of Helsinki. Computational resources were provided by CSC-IT Center for Science, Finland, via the Puhti supercomputing infrastructure. The author thanks the reviewers for their constructive comments, which helped to improve this manuscript. 

The code for this project is available on GitHub: \url{https://github.com/yingyingxu-kyo/virology_likelihood_free_inference_BOLFI/tree/main}

\nolinenumbers
\bibliography{Bibliography}

@article{Quirouette2023,
  author = {Quirouette, Christian and Cresta, Daniel and Li, Jizhou and Wilkie, Kathleen P. and Liang, Haozhao and Beauchemin, Catherine A. A.},
  title     = {The effect of random virus failure following cell entry on infection outcome and the success of antiviral therapy},
  journal   = {Scientific Reports},
  year      = {2023},
  volume    = {13},
  number    = {1},
  pages     = {17243},
  doi       = {10.1038/s41598-023-44180-w},
  pmid      = {37821517},
  pmcid     = {PMC10567758}
}

@article{quirouette2024does,
  title = {Does the random nature of cell-virus interactions during in vitro infections affect TCID50 measurements and parameter estimation by mathematical models?},
  author = {Quirouette, Christian and Thevakumaran, Risavarshni and Adachi, Kyosuke and Beauchemin, Catherine A. A.},
  year = {2024},
  eprint = {2412.12960},
  archivePrefix = {arXiv},
  primaryClass = {physics.bio-ph},
  url = {https://doi.org/10.48550/arXiv.2412.12960},
  note = {arXiv preprint},
  publisher = {arXiv},
  journal = {arXiv preprint}
}

@article{Simon2016,
  author = {Simon, P. F. and de La Vega, M.-A. and Paradis, Éric and Mendoza, E. and Coombs, K. M. and Kobasa, D. and Beauchemin, C. A. A.},
  title = {Avian influenza viruses that cause highly virulent infections in humans exhibit distinct replicative properties in contrast to human H1N1 viruses},
  journal = {Scientific Reports},
  volume = {6},
  pages = {24154},
  year = {2016},
  doi = {10.1038/srep24154},
  url = {https://www.nature.com/articles/srep24154}
}

@article{gutmann2016bayesian,
  title={Bayesian optimization for likelihood-free inference of simulator-based statistical models},
  author={Gutmann, Michael U and Corander, Jukka},
  journal={Journal of Machine Learning Research},
  volume={17},
  number={1},
  pages={47},
  year={2016},
  publisher={JMLR.org},
  url={https://jmlr.org/papers/volume17/15-017/15-017.pdf}
}

@book{rasmussen2006gaussian,
  title={Gaussian processes for machine learning},
  author={Rasmussen, Carl Edward and Williams, Christopher KI},
  year={2006},
  publisher={MIT press}
}

@article{lintusaari2018elfi,
  title={ELFI: Engine for Likelihood-Free Inference},
  author={Lintusaari, J. and Vuollekoski, H. and Kangasraasi, A. and Skytten, K. and J{\"a}rvenp{\"a}{\"a}, M. and Marttinen, P. and Gutmann, Michael U. and Vehtari, A. and Corander, J. and Kaski, S.},
  journal={Journal of Machine Learning Research},
  volume={19},
  number={16},
  pages={1--7},
  year={2018}
}

@article{Reed1938,
  author    = {L. J. Reed and H. Muench},
  title     = {A simple method of estimating fifty percent endpoints},
  journal   = {American Journal of Epidemiology},
  volume    = {27},
  number    = {3},
  pages     = {493--497},
  year      = {1938},
  doi       = {10.1093/oxfordjournals.aje.a118408}
}

@article{Cresta2021,
  author    = {D. Cresta and D. C. Warren and C. Quirouette and A. P. Smith and L. C. Lane and A. M. Smith and C. A. A. Beauchemin},
  title     = {Time to revisit the endpoint dilution assay and to replace the TCID50 as a measure of a virus sample's infection concentration},
  journal   = {PLoS Computational Biology},
  volume    = {17},
  number    = {10},
  pages     = {e1009480},
  year      = {2021},
  month     = {10},
  doi       = {10.1371/journal.pcbi.1009480},
  note      = {Erratum in: PLoS Comput Biol. 2023 Jan 31;19(1):e1010877. doi: 10.1371/journal.pcbi.1010877},
  pmid      = {34662338},
  pmcid     = {PMC8553042}
}

@techreport{BOLFIrobust2024,
  author    = {Remes, Ulpu},
  title     = {ELFI Notebooks: Failure-robust BOLFI},
  year      = {2024},
  howpublished = {GitHub repository},
  url       = {https://github.com/uremes/elfi-notebooks/blob/extend_model/failure_robust_BOLFI.ipynb},
  note      = {Accessed: 2024-02-20},
  publisher = {GitHub},
  institution = {GitHub}
}

@article{Lintusaari2017,
  author = {Lintusaari, Jarno and Gutmann, Michael U. and Dutta, Ritabrata and Kaski, Samuel and Corander, Jukka},
  title = {Fundamentals and Recent Developments in Approximate Bayesian Computation},
  journal = {Systematic Biology},
  volume = {66},
  number = {1},
  pages = {e66--e82},
  year = {2017},
  doi = {10.1093/sysbio/syw077},
  url = {https://doi.org/10.1093/sysbio/syw077}
}

@article{Price2017,
  author    = {Fishman Price and C. C. Drovandi and A. Lee and D. J. Nott},
  title     = {Bayesian Synthetic Likelihood},
  journal   = {Journal of Computational and Graphical Statistics},
  volume    = {27},
  number    = {1},
  pages     = {1--11},
  year      = {2017},
  doi       = {10.1080/10618600.2017.1302882},
  url       = {https://doi.org/10.1080/10618600.2017.1302882}
}

@article{Simola2021,
  author = {Simola, Umberto and Cisewski-Kehe, Jessi and Gutmann, Michael U. and Corander, Jukka},
  title = {Adaptive Approximate Bayesian Computation Tolerance Selection},
  journal = {Bayesian Analysis},
  volume = {16},
  number = {2},
  pages = {397--423},
  year = {2021},
  doi = {10.1214/20-BA1211},
  url = {https://doi.org/10.1214/20-BA1211}
}

@InProceedings{Ikonomov2020,
  title     = {Robust Optimisation Monte Carlo},
  author    = {Borislav Ikonomov and Michael U. Gutmann},
  booktitle = {Proceedings of the Twenty Third International Conference on Artificial Intelligence and Statistics},
  pages     = {2819--2829},
  year      = {2020},
  editor    = {Silvia Chiappa and Roberto Calandra},
  volume    = {108},
  series    = {Proceedings of Machine Learning Research},
  month     = {8},
  publisher = {PMLR},
  url       = {https://proceedings.mlr.press/v108/ikonomov20a.html}
}

@article{Jarvenpaa2019,
  author       = {J{\"a}rvenp{\"a}{\"a}, Marko and Gutmann, Michael U. and Pleska, Arijus and Vehtari, Aki and Marttinen, Pekka},
  title        = {Efficient Acquisition Rules for Model-Based Approximate Bayesian Computation},
  journal      = {Bayesian Analysis},
  volume       = {14},
  number       = {2},
  pages        = {595--622},
  year         = {2019},
  publisher = {International Society for Bayesian Analysis},
  doi       = {10.1214/18-BA1121},
  url          = {https://projecteuclid.org/euclid.ba/1537258134}
}

@article{jarvenpaa2021parallel,
  author    = {J{\"a}rvenp{\"a}{\"a}, Marko and Gutmann, Michael U. and Vehtari, Aki Caps and Marttinen, Pekka},
  title     = {Parallel {Gaussian} process surrogate {Bayesian} inference with noisy likelihood evaluations},
  journal   = {Bayesian Analysis},
  volume    = {16},
  number    = {1},
  pages     = {147--176},
  year      = {2021},
  publisher = {International Society for Bayesian Analysis},
  doi       = {10.1214/20-BA1200},
  url       = {https://projecteuclid.org}
}

@article{leclercq2018bayesian,
  author    = {Leclercq, Florent},
  title     = {Bayesian optimisation for likelihood-free cosmological inference},
  journal   = {Physical Review D},
  volume    = {98},
  number    = {6},
  pages     = {063511},
  year      = {2018},
  publisher = {APS}
}

@phdthesis{honkonen2025,
  author    = {Honkonen, I.},
  title     = {Bayesian likelihood-free inference for effective uncertainty quantification in plasma physics simulations},
  school    = {University of Helsinki},
  year      = {2025},
  type      = {{Ph.D.} dissertation}
}

@article{beaumont2002approximate,
  title={Approximate Bayesian computation in population genetics},
  author={Beaumont, Mark A and Zhang, Wenyang and Balding, David J},
  journal={Genetics},
  volume={162},
  number={4},
  pages={2025--2035},
  year={2002},
  publisher={Oxford University Press}
}

@article{marjoram2003markov,
  title={Markov chain Monte Carlo without likelihoods},
  author={Marjoram, Paul and Molitor, John and Plagnol, Vincent and Tavar{\'e}, Simon},
  journal={Proceedings of the National Academy of Sciences},
  volume={100},
  number={26},
  pages={15324--15328},
  year={2003},
  publisher={National Acad Sciences}
}

@article{sisson2007sequential,
  title={Sequential Monte Carlo without likelihoods},
  author={Sisson, Scott A and Fan, Yanan and Tanaka, Mark M},
  journal={Proceedings of the National Academy of Sciences},
  volume={104},
  number={6},
  pages={1760--1765},
  year={2007},
  publisher={National Acad Sciences}
}

@article{drovandi2018likelihood,
  title={Likelihood-free Bayesian estimation of multivariate quantile distributions},
  author={Drovandi, Christopher C and Pettitt, Anthony N},
  journal={Computational Statistics \& Data Analysis},
  volume={117},
  pages={180--196},
  year={2018},
  publisher={Elsevier}
}

@article{csillery2010approximate,
  title={Approximate Bayesian computation (ABC) in practice},
  author={Csill{\'e}ry, Katalin and Blum, Michael GB and Gaggiotti, Oscar E and Fran{\c{c}}ois, Olivier},
  journal={Trends in Ecology \& Evolution},
  volume={25},
  number={7},
  pages={410--418},
  year={2010},
  publisher={Elsevier}
}

@article{Thomas2025SplitBOLFI,
  title={Misspecification-robust likelihood-free inference in high dimensions},
  author={Thomas, Owen and others},
  journal={Computational Statistics},
  volume={40},
  pages={4399--4439},
  year={2025},
  doi={10.1007/s00180-025-01607-4}
}

@article{chen2025mobolfi,
  title={Multi-objective Bayesian optimization for Likelihood-Free inference in sequential sampling models of decision making},
  author={Chen, David and Li, Xinwei and Kim, Eui-Jin and Bansal, Prateek and Nott, David J.},
  journal={Transactions on Machine Learning Research},
  volume={2025},
  year={2025},
  url={https://openreview.net/forum?id=hQjwDqfSzj}
}

@article{talts2018validating,
  author    = {Talts, Sean and Betancourt, Michael and Simpson, Daniel and Vehtari, Aki and Gelman, Andrew},
  title     = {Validating Bayesian Inference Algorithms with Simulation-Based Calibration},
  year      = {2018},
  eprint = {1804.06788},
  archivePrefix = {arXiv},
  primaryClass = {stat.ME},
  url       = {https://arxiv.org/abs/1804.06788},
  journal = {arXiv preprint}
}

@article{GenomicsInsights2019,
  title={Inter-Laboratory Variability in Array-Based RNA Quantification Methods},
  author={Wong, V.Y.},
  journal={Genomics Insights},
  year={2019},
  doi={10.4137/GEI.S11909}
}

@article{Lamb2024,
  title={Quantification of influenza virus mini viral RNA dynamics using Cas13},
  author={Lamb, C.H. and others},
  journal={bioRxiv},
  year={2024},
  doi={10.1101/2023.11.03.565460}
}

@article{Mohorianu2023,
  title={noisyR: A comprehensive noise filter for RNA sequencing data},
  author={Mohorianu, I.I. and others},
  journal={bioRxiv},
  year={2023}
}

@book{shapley1953value,
  author = {Shapley, Lloyd S.},
  title = {A Value for n-Person Games},
  book = {Contributions to the Theory of Games, Vol. II},
  series = {Annals of Mathematics Studies},
  volume = {28},
  pages = {307--317},
  year = {1953},
  publisher = {Princeton University Press}
}

@article{lundberg2017unified,
  title={A Unified Approach to Interpreting Model Predictions},
  author={Lundberg, Scott M and Lee, Su-In},
  journal={Advances in Neural Information Processing Systems 30},
  pages={4765--4774},
  year={2017},
  publisher={Curran Associates, Inc.}
}

@article{breiman2001random,
  title={Random forests},
  author={Breiman, Leo},
  journal={Machine Learning},
  volume={45},
  pages={5--32},
  year={2001}
}

@article{ghorbani2019data,
  title={Data Shapley: Equitable valuation of data for machine learning},
  author={Ghorbani, Amirata and Zou, James},
  journal={International Conference on Machine Learning},
  year={2019}
}

@article{liang2022multimodal,
  title={Multimodal data fusion and analysis},
  author={Liang, Xiao and others},
  journal={IEEE Signal Processing Magazine},
  year={2022}
}

@techreport{eurostat2023quality,
  author      = {{Eurostat}},
  title       = {Quality Framework for Multi-Source Statistics},
  institution = {Eurostat},
  year        = {2023},
  type        = {Guidelines},
  url         = {https://ec.europa.eu/eurostat/web/quality/european-quality-standards}
}

@article{calhoun2009feature,
  author  = {Calhoun, Vince D. and Adal{\i}, T{\"u}lay},
  title   = {Feature-Based Fusion of Medical Imaging Data},
  journal = {IEEE Transactions on Information Technology in Biomedicine},
  year    = {2009},
  volume  = {13},
  number  = {5},
  pages   = {711--720},
  doi     = {10.1109/TITB.2008.923773}
}

@article{calhoun2017fusion,
  author  = {Calhoun, Vince D. and Adal{\i}, T{\"u}lay},
  title   = {Fusion Analysis of Functional {MRI} Data for Classification of Individuals with Schizophrenia},
  journal = {IEEE Transactions on Medical Imaging},
  year    = {2017},
  volume  = {36},
  number  = {5},
  pages   = {1088--1097},
  doi     = {10.1109/TMI.2016.2642125}
}

@article{chavda2024uncertainty,
  title={Uncertainty Quantification of Data Shapley via Statistical Inference},
  author={Chavda, B. and others},
  journal={arXiv preprint},
  year={2024}
}

@article{han2024shapley,
  title={Shapley Value on Uncertain Data},
  author={Han, Y. and others},
  journal={VLDB Journal},
  year={2024}
}

@inproceedings{watson2023explaining,
  title     = {Explaining Predictive Uncertainty with Information Theoretic Shapley Values},
  author    = {Watson, David S. and O'Hara, Joshua and Tax, Niek and Mudd, Richard and Guy, Ido},
  booktitle = {Advances in Neural Information Processing Systems},
  volume    = {36},
  pages     = {72215--72238},
  year      = {2023},
  publisher = {Curran Associates, Inc.},
  url       = {https://proceedings.neurips.cc/paper_files/paper/2023/file/16e4be78e61a3897665fa01504e9f452-Paper-Conference.pdf}
}

@inproceedings{papamakarios2019sequential,
  title     = {{Sequential Neural Likelihood}: {Fast Likelihood-Free Inference} with {Autoregressive Flows}},
  author    = {Papamakarios, George and Sterratt, David and Murray, Iain},
  booktitle = {International Conference on Artificial Intelligence and Statistics},
  pages     = {837--848},
  year      = {2019},
  organization={PMLR}
}

@inproceedings{papamakarios2016fast,
  title     = {{Fast $\epsilon$-free Inference} of {Simulation Models} with {Normalizing Flows}},
  author    = {Papamakarios, George and Murray, Iain},
  booktitle = {Advances in Neural Information Processing Systems},
  volume    = {29},
  pages     = {4373--4381},
  year      = {2016},
  publisher = {Curran Associates, Inc.}
}

@inproceedings{greenberg2019automatic,
  title     = {{Automatic Posterior Transformation} for {Likelihood-Free Inference}},
  author    = {Greenberg, David and Nonnenmacher, Marcel and Macke, Jakob},
  booktitle = {International Conference on Machine Learning},
  pages     = {2404--2414},
  year      = {2019},
  organization={PMLR}
}

@article{elgammal2023fast,
  title     = {Fast and robust {Bayesian} inference using {Gaussian} processes with {GPry}},
  author    = {El Gammal, Jonas and Sch{\"o}neberg, Nils and Torrado, Jes{\'u}s and Fidler, Christian},
  journal   = {Journal of Cosmology and Astroparticle Physics},
  volume    = {2023},
  number    = {10},
  pages     = {021},
  year      = {2023},
  publisher = {IOP Publishing}
}

@article{cranmer2020frontier,
  title={The frontier of simulation-based inference},
  author={Cranmer, Kyle and Brehmer, Johann and Louppe, Gilles},
  journal={Proceedings of the National Academy of Sciences},
  volume={117},
  number={48},
  pages={30055--30062},
  year={2020},
  publisher={National Acad Sciences},
  doi={10.1073/pnas.1912789117}
}

@article{Sachak-Patwa2023,
  title = {A target-cell limited model can reproduce influenza infection dynamics in hosts with differing immune responses},
  author = {Sachak-Patwa, Rahil and Lafferty, Erin I. and Schmit, Claude J. and Thompson, Robin N. and Byrne, Helen M.},
  journal = {Journal of Theoretical Biology},
  volume = {568},
  pages = {111490},
  year = {2023},
  doi = {10.1016/j.jtbi.2023.111490},
  url = {https://www.sciencedirect.com/science/article/pii/S0022519323000875}
}

@book{gelman2013bayesian,
  title={Bayesian Data Analysis},
  author={Gelman, Andrew and Carlin, John B. and Stern, Hal S. and Dunson, David B. and Vehtari, Aki and Rubin, Donald B.},
  edition={3},
  year={2013},
  publisher={CRC Press},
  address={Boca Raton, FL}
}

@incollection{helin2024introduction,
  author    = {Helin, Tapio and Stuart, Andrew M. and Teckentrup, Aretha L. and Zygalakis, Konstantinos C.},
  title     = {Introduction to Gaussian Process Regression in Bayesian Inverse Problems, with New Results on Experimental Design for Weighted Error Measures},
  booktitle = {Monte Carlo and Quasi-Monte Carlo Methods - MCQMC 2022},
  pages     = {49--79},
  year      = {2024},
  publisher = {Springer Cham},
  doi       = {10.1007/978-3-031-59762-6\_3}
}

\newpage
\section*{Supporting Information}

The following supporting information is provided:
\begin{itemize}
\item \textbf{S1 Appendix: Simulator Implementation} --- Detailed description of the ODE-based virus dynamics model and the stochastic ED assay simulator.
\item \textbf{S2 Appendix: Data Tables} --- Experimental data tables for SC RNA, MC RNA, SC ED, and MC ED measurements.
\item \textbf{S3 Appendix: Theoretical Justification of Discrepancy Measures} --- Full proofs connecting our discrepancy to sufficient statistics for RNA and ED data.
\item \textbf{S4 Appendix: Additional Figures} --- Supplementary diagnostics of the BOLFI inference procedure, including GP surrogate model diagnostics and normality assessments of RNA residuals.
\item \textbf{S5 Appendix: Detailed Calibration Results for All Data Configurations} --- Complete calibration results for all five data configurations at the 50\%, 90\%, and 95\% credible interval levels.
\end{itemize}

\newpage
\section*{Supporting Information}

\setcounter{figure}{0}
\setcounter{table}{0}
\setcounter{equation}{0}
\renewcommand{\thefigure}{S\arabic{figure}}
\renewcommand{\thetable}{S\arabic{table}}
\renewcommand{\theequation}{S\arabic{equation}}

\subsection*{S1 Appendix. Simulator Implementation}
\label{sec:sim_RNA_ED}

\subsubsection*{Virus Dynamics Simulator (RNA Data)}
The virus dynamics simulator follows the ordinary differential equation model developed by Quirouette et al. \cite{quirouette2024does, Quirouette2023}. The full system of equations is provided in the main text (Section 2.1). Here we summarize key implementation details for reproducibility.

\paragraph{Initial Conditions}

In the ODE model, the SC infection is initialized at \( t = -1 \, \text{h} \) with \( T = N_{\text{cells}} = 1.9 \times 10^6 \) cells, \( E_i = I_j = 0 \), and \( V_{\text{RNA}} \) set to the qRT-PCR measurement at that time multiplied by the supernatant volume \( s = 10 \, \text{ml} \). The initial \( V \) is set to achieve MOI = 3 within the 1 h incubation.
For the MC infection, the model is initialized at \( t = 0 \, \text{h} \) with \( T = N_{\text{cells}} \), and both \( V \) and \( V_{\text{RNA}} \) set to the geometric mean of experimental measurements up to 7 hpi, multiplied by \( s = 10 \, \text{ml} \).

\paragraph{Implementation Details}
At each sampling time, 0.5 ml of the 10 ml supernatant was removed and replaced with fresh media. In the ODE model, this is implemented by multiplying \( V \) and \( V_{\text{RNA}} \) by \( f_{\text{sampling}} = 0.95 \).

Following Simon et al. \cite{Simon2016}, \( n_E = n_I = 60 \), making the eclipse and infectious phase durations follow an Erlang distribution. Virus clearance rates were fixed to \( c = 0.0573 \, \text{h}^{-1} \) and \( c_{\text{RNA}} = 0.001 \, \text{h}^{-1} \). Table IV in Quirouette et al. \cite{Quirouette2023} summarizes all fixed parameters.

\subsubsection*{ED Assay Simulation}

The ED assay simulator follows the stochastic process described in \cite{quirouette2024does, Quirouette2023}. Each assay uses 8 dilution columns with 4 replicate wells per column. The simulator deposits virions into each well based on the sample's infectious concentration and the dilution factor; the number of virions deposited follows a Binomial distribution determined by the well volume, dilution factor, and virion volume \cite{Cresta2021}.

A well is classified as infected if at least one deposited virion successfully establishes infection. The probability that a deposited virion fails to establish a productive infection (the extinction probability) depends on the model parameters through the burst size and the probability of successful cell entry; the full expression for ED assay is provided in \cite{quirouette2024does, Quirouette2023}. For each well, a uniform random number \( r \in [0,1) \) is drawn; the well is uninfected if \( r < (P_{V \rightarrow \text{Extinction}})^{V_{0}^{i,j}} \), and infected otherwise.

We use raw infection counts directly as data, avoiding additional modeling layers such as TCID$_{50}$ conversion \cite{Cresta2021}. For complete details, see \cite{quirouette2024does, Quirouette2023}.

\newpage
\subsection*{S2 Appendix. Data tables}
\label{sec:data_tables}

This appendix contains the experimental data used in this study, originally published in Simon et al. \cite{Simon2016}. The data consist of four tables: single-cycle (SC) and multiple-cycle (MC) RNA measurements (Tables~\ref{tab:SC RNA data} and \ref{tab:MC RNA data}), and single-cycle and multiple-cycle endpoint dilution (ED) assay outcomes (Tables~\ref{tab:SC_ED_data} and \ref{tab:MC_ED_data}). The RNA data are presented as vRNA concentrations in units of vRNA/ml, while the ED data show the number of infected wells (out of 4) for each dilution column at each time point.

\begin{table}[H]
    \centering
    \caption{Data table of sH1N1 SC RNA data \cite{Simon2016}}
    \begin{tabular}{cc}
        \toprule
        Time post infection [h] & Virus [vRNA/ml] \\
        \midrule
        1 & $4.05 \times 10^6$ \\
        1 & $6.73 \times 10^6$ \\
        2 & $3.14 \times 10^6$ \\
        2 & $3.61 \times 10^6$ \\
        2 & $6.22 \times 10^6$ \\
        3.5 & $3.13 \times 10^7$ \\
        3.5 & $5.12 \times 10^6$ \\
        3.5 & $4.39 \times 10^6$ \\
        4.63 & $6.03 \times 10^6$ \\
        4.63 & $1.41 \times 10^7$ \\
        4.63 & $5.14 \times 10^7$ \\
        5.65 & $4.96 \times 10^7$ \\
        5.65 & $2.37 \times 10^7$ \\
        5.65 & $4.02 \times 10^7$ \\
        7.17 & $1.38 \times 10^8$ \\
        7.17 & $6.54 \times 10^7$ \\
        7.17 & $1.24 \times 10^8$ \\
        8.62 & $2.98 \times 10^8$ \\
        8.62 & $5.87 \times 10^7$ \\
        8.62 & $3.33 \times 10^7$ \\
        9.9 & $6.81 \times 10^7$ \\
        9.9 & $1.20 \times 10^8$ \\
        9.9 & $1.25 \times 10^8$ \\
        11 & $1.08 \times 10^9$ \\
        11 & $1.49 \times 10^9$ \\
        11 & $1.74 \times 10^9$ \\
        12 & $8.53 \times 10^9$ \\
        12 & $4.22 \times 10^9$ \\
        12 & $2.28 \times 10^9$ \\
        15.5 & $5.14 \times 10^9$ \\
        15.5 & $4.78 \times 10^9$ \\
        15.5 & $4.83 \times 10^9$ \\
        18 & $5.92 \times 10^9$ \\
        18 & $1.07 \times 10^{10}$ \\
        18 & $6.06 \times 10^9$ \\
        \bottomrule
    \end{tabular}
    \label{tab:SC RNA data}
\end{table}

\begin{table}[H]
    \centering
    \caption{Data table of sH1N1 MC RNA data \cite{Simon2016}}
    \begin{tabular}{cc}
        \toprule
        Time post infection [h] & Virus [vRNA/ml] \\
        \midrule
        0.0 & $3.86 \times 10^4$ \\
        0.0 & $1.19 \times 10^4$ \\
        0.0 & $8.22 \times 10^3$ \\
        3.166667 & $7.25 \times 10^3$ \\
        3.166667 & $2.72 \times 10^3$ \\
        3.166667 & $2.99 \times 10^4$ \\
        6.0 & $6.99 \times 10^3$ \\
        6.0 & $1.54 \times 10^4$ \\
        6.0 & $9.04 \times 10^3$ \\
        18.58333 & $1.79 \times 10^4$ \\
        18.58333 & $1.21 \times 10^4$ \\
        18.58333 & $1.22 \times 10^4$ \\
        28.5 & $4.42 \times 10^5$ \\
        28.5 & $6.43 \times 10^5$ \\
        28.5 & $1.63 \times 10^6$ \\
        42.86666 & $2.30 \times 10^8$ \\
        42.86666 & $7.97 \times 10^7$ \\
        42.86666 & $9.04 \times 10^7$ \\
        53.3 & $4.47 \times 10^8$ \\
        53.3 & $6.77 \times 10^8$ \\
        53.3 & $2.30 \times 10^8$ \\
        66.58334 & $6.82 \times 10^9$ \\
        66.58334 & $1.88 \times 10^{10}$ \\
        66.58334 & $3.57 \times 10^9$ \\
        77.83334 & $5.88 \times 10^9$ \\
        77.83334 & $1.39 \times 10^{10}$ \\
        77.83334 & $1.02 \times 10^{10}$ \\
        77.83334 & $1.98 \times 10^9$ \\
        77.83334 & $5.09 \times 10^9$ \\
        77.83334 & $3.92 \times 10^9$ \\
        91.66666 & $7.56 \times 10^9$ \\
        91.66666 & $1.25 \times 10^{10}$ \\
        91.66666 & $4.02 \times 10^9$ \\
        99.08334 & $1.01 \times 10^{10}$ \\
        99.08334 & $1.46 \times 10^{10}$ \\
        99.08334 & $4.72 \times 10^9$ \\
        99.08334 & $2.10 \times 10^{10}$ \\
        99.08334 & $5.27 \times 10^{10}$ \\
        99.08334 & $1.63 \times 10^{10}$ \\
        147.8333 & $1.02 \times 10^{10}$ \\
        147.8333 & $2.09 \times 10^{10}$ \\
        147.8333 & $1.00 \times 10^{10}$ \\
        \bottomrule
    \end{tabular}
    \label{tab:MC RNA data}
\end{table}

\begin{table}[H]
    \centering
    \caption{Multiple Cycle ED Data Table \cite{Simon2016}}
    \begin{tabular}{c|c|c}
        \toprule
        Time post infection [h] & The dilution factor [$10^{n}$] & ED assay infection counts  \\ 
        \midrule
        0 & -1 -2 -3 -4 -5 -6 -7 -8 & 0 0 0 0 0 0 0 0 \\
        0 & -1 -2 -3 -4 -5 -6 -7 -8 & 1 0 0 0 0 0 0 0 \\
        0 & -1 -2 -3 -4 -5 -6 -7 -8 & 0 0 0 0 0 0 0 0 \\
        3.16667 & -1 -2 -3 -4 -5 -6 -7 -8 & 2 0 0 0 0 0 0 0 \\
        3.16667 & -1 -2 -3 -4 -5 -6 -7 -8 & 1 0 0 0 0 0 0 0 \\
        3.16667 & -1 -2 -3 -4 -5 -6 -7 -8 & 0 0 0 0 0 0 0 0 \\
        6 & -1 -2 -3 -4 -5 -6 -7 -8 & 0 0 0 0 0 0 0 0 \\
        6 & -1 -2 -3 -4 -5 -6 -7 -8 & 0 0 0 0 0 0 0 0 \\
        6 & -1 -2 -3 -4 -5 -6 -7 -8 & 1 0 0 0 0 0 0 0 \\
        18.5833 & -1 -2 -3 -4 -5 -6 -7 -8 & 4 1 0 0 0 0 0 0 \\
        18.5833 & -1 -2 -3 -4 -5 -6 -7 -8 & 4 1 0 0 0 0 0 0 \\
        18.5833 & -1 -2 -3 -4 -5 -6 -7 -8 & 4 2 0 0 0 0 0 0 \\
        28.5 & -3 -4 -5 -6 -7 -8 -9 -10 & 3 1 0 0 0 0 0 0 \\
        28.5 & -3 -4 -5 -6 -7 -8 -9 -10 & 3 1 0 0 0 0 0 0 \\
        28.5 & -3 -4 -5 -6 -7 -8 -9 -10 & 2 0 0 0 0 0 0 0 \\
        42.8667 & -1 -2 -3 -4 -5 -6 -7 -8 & 4 4 4 4 1 0 0 0 \\
        42.8667 & -1 -2 -3 -4 -5 -6 -7 -8 & 4 4 4 4 4 1 0 0 \\
        42.8667 & -1 -2 -3 -4 -5 -6 -7 -8 & 4 4 4 4 0 0 0 0 \\
        53.3 & -4 -5 -6 -7 -8 -9 -10 -11 & 4 4 2 0 0 0 0 0 \\
        53.3 & -4 -5 -6 -7 -8 -9 -10 -11 & 4 4 1 0 0 0 0 0 \\
        53.3 & -4 -5 -6 -7 -8 -9 -10 -11 & 4 4 1 0 0 0 0 0 \\
        66.5833 & -4 -5 -6 -7 -8 -9 -10 -11 & 4 4 2 0 0 0 0 0 \\
        66.5833 & -4 -5 -6 -7 -8 -9 -10 -11 & 4 4 3 0 0 0 0 0 \\
        66.5833 & -4 -5 -6 -7 -8 -9 -10 -11 & 4 4 2 0 0 0 0 0 \\
        77.8333 & -1 -2 -3 -4 -5 -6 -7 -8 & 4 4 4 4 4 3 0 0 \\
        77.8333 & -1 -2 -3 -4 -5 -6 -7 -8 & 4 4 4 4 4 1 1 0 \\
        77.8333 & -1 -2 -3 -4 -5 -6 -7 -8 & 4 4 4 4 4 0 0 0 \\
        91.6667 & -1 -2 -3 -4 -5 -6 -7 -8 & 4 4 4 4 4 0 0 0 \\
        91.6667 & -1 -2 -3 -4 -5 -6 -7 -8 & 4 4 4 4 4 0 0 0 \\
        91.6667 & -1 -2 -3 -4 -5 -6 -7 -8 & 4 4 4 4 3 1 0 0 \\
        99.0833 & -1 -2 -3 -4 -5 -6 -7 -8 & 4 4 4 4 3 1 0 0 \\
        99.0833 & -1 -2 -3 -4 -5 -6 -7 -8 & 4 4 4 4 4 2 0 0 \\
        99.0833 & -1 -2 -3 -4 -5 -6 -7 -8 & 4 4 4 4 4 1 0 0 \\
        147.833 & -1 -2 -3 -4 -5 -6 -7 -8 & 4 4 4 4 2 1 0 0 \\
        147.833 & -1 -2 -3 -4 -5 -6 -7 -8 & 4 4 4 4 1 0 0 0 \\
        147.833 & -1 -2 -3 -4 -5 -6 -7 -8 & 4 4 4 4 1 0 0 0 \\
        \bottomrule
    \end{tabular}
    \label{tab:MC_ED_data}
\end{table}        

\begin{table}[H]
    \centering
    \caption{Single Cycle ED Data Table \cite{Simon2016}}
    \begin{tabular}{c|c|c}
        \toprule
        Time post infection [h] & The dilution factor [$10^{n}$] & ED assay infection counts  \\ 
        \midrule
        1 & -1 -2 -3 -4 -5 -6 -7 -8 & 4 4 4 4 0 0 0 0 \\
        1 & -1 -2 -3 -4 -5 -6 -7 -8 & 4 4 4 4 3 1 0 0 \\
        1 & -1 -2 -3 -4 -5 -6 -7 -8 & 4 4 4 4 2 0 0 0 \\
        2 & -1 -2 -3 -4 -5 -6 -7 -8 & 4 4 4 4 1 0 0 0 \\
        2 & -1 -2 -3 -4 -5 -6 -7 -8 & 4 4 4 4 2 0 0 0 \\
        2 & -1 -2 -3 -4 -5 -6 -7 -8 & 4 4 4 4 0 0 0 0 \\
        3.5 & -1 -2 -3 -4 -5 -6 -7 -8 & 4 4 4 4 0 0 0 0 \\
        3.5 & -1 -2 -3 -4 -5 -6 -7 -8 & 4 4 4 4 4 0 0 0 \\
        3.5 & -1 -2 -3 -4 -5 -6 -7 -8 & 4 4 4 4 1 0 0 0 \\
        4.63 & -1 -2 -3 -4 -5 -6 -7 -8 & 4 4 4 4 2 0 0 0 \\
        4.63 & -1 -2 -3 -4 -5 -6 -7 -8 & 4 4 4 4 0 0 0 0 \\
        4.63 & -1 -2 -3 -4 -5 -6 -7 -8 & 4 4 4 4 2 0 0 0 \\
        5.65 & -1 -2 -3 -4 -5 -6 -7 -8 & 4 4 4 4 4 2 0 0 \\
        5.65 & -1 -2 -3 -4 -5 -6 -7 -8 & 4 4 4 4 4 0 0 0 \\
        5.65 & -1 -2 -3 -4 -5 -6 -7 -8 & 4 4 4 4 4 0 0 0 \\
        7.17 & -1 -2 -3 -4 -5 -6 -7 -8 & 4 4 4 4 4 0 0 0 \\
        7.17 & -1 -2 -3 -4 -5 -6 -7 -8 & 4 4 4 4 3 1 0 0 \\
        7.17 & -1 -2 -3 -4 -5 -6 -7 -8 & 4 4 4 4 4 1 0 0 \\
        8.62 & -1 -2 -3 -4 -5 -6 -7 -8 & 4 4 4 4 4 1 0 0 \\
        8.62 & -1 -2 -3 -4 -5 -6 -7 -8 & 4 4 4 4 3 1 0 0 \\
        8.62 & -1 -2 -3 -4 -5 -6 -7 -8 & 4 4 4 4 4 1 0 0 \\
        9.9 & -2 -3 -4 -5 -6 -7 -8 -9 & 4 4 4 4 0 0 0 0 \\
        9.9 & -2 -3 -4 -5 -6 -7 -8 -9 & 4 4 4 4 1 0 0 0 \\
        9.9 & -2 -3 -4 -5 -6 -7 -8 -9 & 4 4 4 4 3 0 0 0 \\
        11 & -2 -3 -4 -5 -6 -7 -8 -9 & 4 4 4 4 2 0 0 0 \\
        11 & -2 -3 -4 -5 -6 -7 -8 -9 & 4 4 4 4 1 0 0 0 \\
        11 & -2 -3 -4 -5 -6 -7 -8 -9 & 4 4 4 3 1 0 0 0 \\
        12 & -2 -3 -4 -5 -6 -7 -8 -9 & 4 4 4 4 2 0 0 0 \\
        12 & -2 -3 -4 -5 -6 -7 -8 -9 & 4 4 4 4 1 0 0 0 \\
        12 & -2 -3 -4 -5 -6 -7 -8 -9 & 4 4 4 4 2 0 0 0 \\
        15.5 & -2 -3 -4 -5 -6 -7 -8 -9 & 4 4 4 4 3 0 0 0 \\
        15.5 & -2 -3 -4 -5 -6 -7 -8 -9 & 4 4 4 4 3 0 0 0 \\
        15.5 & -2 -3 -4 -5 -6 -7 -8 -9 & 4 4 4 4 1 0 0 0 \\
        18 & -2 -3 -4 -5 -6 -7 -8 -9 & 4 4 4 4 1 0 0 0 \\
        18 & -2 -3 -4 -5 -6 -7 -8 -9 & 4 4 4 4 3 0 0 0 \\
        18 & -2 -3 -4 -5 -6 -7 -8 -9 & 4 4 4 4 3 1 0 0 \\
        \bottomrule
    \end{tabular}
    \label{tab:SC_ED_data}
\end{table}

\FloatBarrier

\newpage
\subsection*{S3 Appendix. Theoretical Justification of Discrepancy Measures}

\subsubsection*{A1. Theoretical Justification of the Discrepancy for RNA Data}

We provide here the theoretical justification for our RNA discrepancy measure: the Euclidean distance on $\log_{10}$-transformed, pooled data across replicates.

\paragraph{Problem Formulation and Notation}

Let $\{t_1, \dots, t_m\}$ denote distinct experimental time points. At each time $t_i$, we have $n_i \geq 1$ replicate measurements (in our data, $n_i = 2$ to $6$). Let $y_{ij}$ denote the observed vRNA concentration for replicate $j$ at time $t_i$, and let $V(t_i, \theta)$ denote the ODE model prediction at time $t_i$ given parameters $\boldsymbol{\theta} \in \Theta \subset \mathbb{R}^d$.

Define the $\log_{10}$-transformed observations:
\begin{equation}
z_{ij}^{\text{obs}} = \log_{10}(y_{ij}), \qquad \mu_i(\boldsymbol{\theta}) = \log_{10}(V(t_i, \boldsymbol{\theta}))
\end{equation}

Our discrepancy measure is:
\begin{equation}
d(D, D_{\text{obs}}) = \left[ \sum_{i=1}^m \sum_{j=1}^{n_i} (z_{ij} - z_{ij}^{\text{obs}})^2 \right]^{1/2}
\label{eq:s1_discrepancy}
\end{equation}

where $D = \{z_{ij}\}$ denotes a simulated dataset, $z_{ij} = \mu_i(\boldsymbol{\theta}) +\epsilon_{ij}$,  and $D_{\text{obs}} = \{z_{ij}^{\text{obs}}\}$ denotes the observed data.
Here, $\epsilon_{ij}$ is the noise we add as described in the main text in Section~\ref{sec:simulator}.

\paragraph{Sufficient Statistics in Likelihood-Free Inference}

Before presenting the theoretical results, we briefly review the concept of sufficient statistics and explain their relevance to likelihood-free inference (LFI).

\begin{definition}[Sufficient Statistic]
A statistic $S(D)$ is said to be \emph{sufficient} for a parameter $\theta$ if the conditional distribution of the data $D$ given $S(D)$ does not depend on $\theta$. Equivalently, $S(D)$ captures all information about $\theta$ contained in the data.
\end{definition}

The Neyman-Fisher factorization theorem provides a practical criterion for identifying sufficient statistics: a statistic $S(D)$ is sufficient for $\theta$ if and only if the likelihood function can be factorized as:
\begin{equation}
L(\theta; D) = h(D) \cdot g(S(D), \theta)
\label{eq:s1_factorization}
\end{equation}
where $h(D)$ is a non-negative function that does not depend on $\theta$, and $g(S(D), \theta)$ depends on the data only through $S(D)$.

In traditional Bayesian inference, sufficiency implies that the posterior distribution satisfies $\pi(\theta \mid D) = \pi(\theta \mid S(D))$; the observed data $D$ can be replaced by $S(D)$ without any loss of information about $\theta$. This dimension reduction is particularly valuable when $S(D)$ has much lower dimension than $D$ itself.

In likelihood-free inference, the role of sufficient statistics is more subtle. LFI methods (including Approximate Bayesian Computation, ABC) approximate the posterior by comparing summary statistics or discrepancies computed from simulated and observed data. A key insight is that if one uses a discrepancy function $d(D, D_{\text{obs}})$ that depends on the data \emph{only through a sufficient statistic} $S(D)$ (and similarly for $D_{\text{obs}}$), then:
\begin{equation}
d(D, D_{\text{obs}}) = \tilde{d}(S(D), S(D_{\text{obs}}))
\end{equation}
for some function $\tilde{d}$. In this case, the ABC posterior converges to the true posterior as the acceptance threshold $\epsilon \to 0$ and the number of simulations $\to \infty$, without requiring additional assumptions about the noise distribution beyond those encoded in the sufficient statistic.

However, when the sufficient statistic is not directly used as the discrepancy (as in our approach), we must verify that the chosen discrepancy preserves the essential information about $\theta$. The following theorems establish that our Euclidean discrepancy on pooled log-transformed data, while not a function solely of the sufficient statistic, is asymptotically equivalent to comparing the sufficient statistic and does not introduce bias in parameter estimation.

\paragraph{Sufficient Statistics for the Log-Normal Model}

We first characterize the sufficient statistics under the assumed data-generating process. Our empirical validation (Supplementary Figures~\ref{fig:MC_normality_check} and~\ref{fig:SC_normality_check}) demonstrated that the residuals on the $\log_{10}$ scale are approximately normally distributed. See also discussion in the main text (Section~\ref{sec:simulator}). Thus, we model:

\begin{equation}
z_{ij} = \mu_i(\boldsymbol{\theta}) + \varepsilon_{ij}, \qquad \varepsilon_{ij} \overset{\text{i.i.d.}}{\sim} \mathcal{N}(0, \sigma^2)
\label{eq:s1_data_model}
\end{equation}
where $\sigma$ is known (estimated from pooled residuals across all time points; see Section 2.2). Equivalently, on the original scale, $y_{ij}$ follows a log-normal distribution.

\begin{theorem}[Sufficient Statistic for Log-Normal Model]
For the model in (\ref{eq:s1_data_model}) with known variance $\sigma^2$, the vector of time-point-specific sample means
\begin{equation}
S(D) = (\bar{z}_1, \bar{z}_2, \dots, \bar{z}_m), \qquad \bar{z}_i = \frac{1}{n_i}\sum_{j=1}^{n_i} z_{ij}
\end{equation}
is a sufficient statistic for the parameter vector $\boldsymbol{\theta}$.
\end{theorem}

\begin{proof}
The joint likelihood function for the data $D$ given $\boldsymbol{\theta}$ is:
\begin{align}
L(\boldsymbol{\theta}) &= \prod_{i=1}^m \prod_{j=1}^{n_i} \frac{1}{\sqrt{2\pi\sigma^2}} \exp\left(-\frac{(z_{ij} - \mu_i(\boldsymbol{\theta}))^2}{2\sigma^2}\right) \nonumber \\
&= (2\pi\sigma^2)^{-N/2} \exp\left(-\frac{1}{2\sigma^2} \sum_{i=1}^m \sum_{j=1}^{n_i} (z_{ij} - \mu_i(\boldsymbol{\theta}))^2\right)
\end{align}
where $N = \sum_{i=1}^m n_i$ is the total number of observations.

For each time point $i$, decompose the sum of squares:
\begin{align}
\sum_{j=1}^{n_i} (z_{ij} - \mu_i(\boldsymbol{\theta}))^2 &= \sum_{j=1}^{n_i} \left[(z_{ij} - \bar{z}_i) + (\bar{z}_i - \mu_i(\boldsymbol{\theta}))\right]^2 \nonumber \\
&= \underbrace{\sum_{j=1}^{n_i} (z_{ij} - \bar{z}_i)^2}_{s_i^2} + n_i(\bar{z}_i - \mu_i(\boldsymbol{\theta}))^2 \label{eq:s1_decomposition}
\end{align}
where the cross-term vanishes because $\sum_j (z_{ij} - \bar{z}_i) = 0$. $s_i^2$ is the within-replicate sums of squares.

Substituting (\ref{eq:s1_decomposition}) into the likelihood:
\begin{align}
L(\boldsymbol{\theta}) &= (2\pi\sigma^2)^{-N/2} \exp\left(-\frac{1}{2\sigma^2} \sum_{i=1}^m s_i^2\right) \times \exp\left(-\frac{1}{2\sigma^2} \sum_{i=1}^m n_i(\bar{z}_i - \mu_i(\boldsymbol{\theta}))^2\right) \nonumber \\
&= h(D) \cdot g(S(D), \boldsymbol{\theta})
\end{align}
where $h(D) = (2\pi\sigma^2)^{-N/2} \exp\left(-\frac{1}{2\sigma^2} \sum_{i=1}^m s_i^2\right)$ does not depend on $\boldsymbol{\theta}$, and $g(S(D), \boldsymbol{\theta}) = \exp\left(-\frac{1}{2\sigma^2} \sum_{i=1}^m n_i(\bar{z}_i - \mu_i(\boldsymbol{\theta}))^2\right)$ depends on the data only through $S(D) = (\bar{z}_1, \dots, \bar{z}_m)$. By the Neyman-Fisher factorization theorem, $S(D)$ is sufficient for $\boldsymbol{\theta}$.
\end{proof}

\begin{corollary}[Ancillary Statistics]
The within-replicate sums of squares $\{s_i^2\}_{i=1}^m$ are \emph{ancillary} for $\boldsymbol{\theta}$: their distribution does not depend on $\boldsymbol{\theta}$. Consequently, they contain no information about the ODE parameters and can be safely ignored or marginalized over in the inference procedure.
\end{corollary}

This result is fundamental: in the context of likelihood-free inference, any discrepancy function that depends primarily on the sample means $\bar{z}_i$ (or equivalently, that is insensitive to the within-replicate variance in expectation) will preserve the information about $\boldsymbol{\theta}$. Our Euclidean discrepancy on pooled log-transformed data satisfies this property.

\paragraph{Relationship Between the Discrepancy and Sufficient Statistics}

We now establish the connection between our discrepancy $d$ and the sufficient statistic $S(D)$.

\begin{theorem}[Expectation of the Squared Discrepancy]
For data generated from model (\ref{eq:s1_data_model}) with known $\sigma^2$, the expected squared discrepancy between a simulated dataset $D \sim P_\theta$ and the fixed observed dataset $D_{\text{obs}}$ is:
\begin{equation}
\mathbb{E}_{D \sim P_\theta}\left[d^2(D, D_{\text{obs}})\right] = C + \sum_{i=1}^m n_i(\bar{z}_i^{\text{obs}} - \mu_i(\theta))^2
\label{eq:s1_expectation}
\end{equation}
where 
\begin{equation}
C = N\sigma^2 + \sum_{i=1}^m s_i^{\text{obs},2}
\end{equation}
is a constant independent of $\boldsymbol{\theta}$, and $s_i^{\text{obs},2} = \sum_j (z_{ij}^{\text{obs}} - \bar{z}_i^{\text{obs}})^2$.
\end{theorem}

\begin{proof}
Expand the squared discrepancy:
\begin{align}
d^2 &= \sum_{i=1}^m \sum_{j=1}^{n_i} (z_{ij} - z_{ij}^{\text{obs}})^2 \nonumber \\
&= \underbrace{\sum_{i,j} (z_{ij} - \mu_i(\theta))^2}_{A} + \underbrace{\sum_{i,j} (z_{ij}^{\text{obs}} - \mu_i(\theta))^2}_{B} + 2\underbrace{\sum_{i,j} (z_{ij} - \mu_i(\theta))(z_{ij}^{\text{obs}} - \mu_i(\theta))}_{C}
\end{align}

Take expectations with respect to $D \sim P_\theta$ (i.e., over the simulated $z_{ij}$, treating $z_{ij}^{\text{obs}}$ as fixed constants):

\begin{itemize}
\item For term $A$: $\mathbb{E}[(z_{ij} - \mu_i(\theta))^2] = \sigma^2$; therefore, $\mathbb{E}[A] = N\sigma^2$.
\item For term $B$: $B$ is a constant (depends only on observed data); thus $\mathbb{E}[B] = B$.
\item For term $C$: $\mathbb{E}[z_{ij} - \mu_i(\theta)] = \mathbb{E}[z_{ij}] - \mu_i(\theta) = \mu_i(\theta) - \mu_i(\theta) = 0$; hence $\mathbb{E}[C] = 0$.
\end{itemize}

Thus:
\begin{equation}
\mathbb{E}[d^2] = N\sigma^2 + \sum_{i=1}^m \sum_{j=1}^{n_i} (z_{ij}^{\text{obs}} - \mu_i(\theta))^2
\end{equation}

Now decompose the observed data term using the same identity as in (\ref{eq:s1_decomposition}):
\begin{equation}
\sum_{j=1}^{n_i} (z_{ij}^{\text{obs}} - \mu_i(\theta))^2 = s_i^{\text{obs},2} + n_i(\bar{z}_i^{\text{obs}} - \mu_i(\theta))^2
\end{equation}

Summing over $i$ yields (\ref{eq:s1_expectation}).
\end{proof}

\begin{corollary}[Parameter Ranking Equivalence]
For any two parameter values $\theta_1, \theta_2 \in \Theta$,
\begin{equation}
\mathbb{E}[d^2(D, D_{\text{obs}}) \mid \theta_1] \leq \mathbb{E}[d^2(D, D_{\text{obs}}) \mid \theta_2]
\end{equation}
if and only if
\begin{equation}
\sum_{i=1}^m n_i(\bar{z}_i^{\text{obs}} - \mu_i(\theta_1))^2 \leq \sum_{i=1}^m n_i(\bar{z}_i^{\text{obs}} - \mu_i(\theta_2))^2
\end{equation}

Thus, ranking parameters by expected discrepancy is equivalent to ranking them by distance between the sufficient statistic and model predictions.
\end{corollary}

This corollary establishes that our discrepancy, despite using the full pooled data, preserves the relative ordering of parameters induced by the sufficient statistic. The constant offset $C$ does not affect parameter comparisons.

\paragraph{Robustness to Non-Gaussian Noise}

The most significant advantage of our discrepancy-based approach over traditional parametric methods is its robustness to deviations from the Gaussian assumption.

\begin{proposition}[Non-Parametric Validity]
The discrepancy $d(D, D_{\text{obs}})$ defined in (\ref{eq:s1_discrepancy}) is well-defined and remains a valid measure of dissimilarity between datasets for \emph{any} noise distribution with finite second moments, without requiring the Gaussian assumption in (\ref{eq:s1_data_model}).
\end{proposition}

\begin{proof}
The Euclidean distance is a proper metric on $\mathbb{R}^N$. For any two datasets $D$ and $D_{\text{obs}}$, $d(D, D_{\text{obs}}) \geq 0$, with $d(D, D_{\text{obs}}) = 0$ if and only if $z_{ij} = z_{ij}^{\text{obs}}$ for all $i,j$. The expectation calculation in Theorem 2 generalizes:
\begin{align}
\mathbb{E}[d^2] &= \sum_{i,j} \mathbb{E}[z_{ij}^2] + \sum_{i,j} (z_{ij}^{\text{obs}})^2 - 2\sum_{i,j} \mathbb{E}[z_{ij}] z_{ij}^{\text{obs}} \\
&= \sum_i n_i(\mathbb{E}[z_{i1}^2] + \mathbb{E}[z_{i1}^{\text{obs},2}]) - 2\sum_i n_i \mathbb{E}[z_{i1}] \bar{z}_i^{\text{obs}}
\end{align}
No Gaussian assumption is required for these expectations to exist; only finite second moments are needed, which is satisfied for any data with bounded measurement ranges.
\end{proof}

This robustness property is critical because:
\begin{enumerate}
\item Our Shapiro-Wilk tests ($p = 0.78$ for MC RNA, $p = 0.54$ for SC RNA) indicate no significant deviation from normality, but with only $n=35$ or $42$ residuals, the test has limited power. The normality assumption remains an approximation, not a proven property.
\item Traditional MCMC with Gaussian likelihood would be \emph{misspecified} if the true noise deviates from normality, potentially leading to biased parameter estimates and invalid credible intervals.
\item Our LFI approach remains valid regardless of the exact noise distribution, as long as the discrepancy captures dissimilarity between datasets.
\end{enumerate}

The practical advantages of this robustness property are summarized in Table~\ref{tab:drna_robustness}, which compares our LFI approach with parametric MCMC under various assumptions relevant to the small-replicate regime of viral kinetic experiments.

\begin{table}[H]
\centering
\caption{Comparison of robustness properties between parametric MCMC with Gaussian likelihood and our LFI approach with Euclidean discrepancy}
\label{tab:drna_robustness}
\begin{tabular}{lcc}
\toprule
Property & MCMC (Gaussian) & Our LFI \\
\midrule
Assumes exact normality & Required & Not required \\
Valid if normality holds & Optimal efficiency & Good (slightly lower) \\
Valid if normality is approximate & Unknown bias & Robust \\
Valid for any finite-variance noise & No & Yes \\
Requires specifying likelihood & Yes & No \\
\midrule
\multicolumn{3}{c}{\textit{With $n_i = 2\text{--}6$ replicates}} \\
Sample mean variance & $\sigma^2/n_i$ & $\sigma^2/n_i$ (via all replicates) \\
Robustness to outliers & Low & High \\
\bottomrule
\end{tabular}
\end{table}

\subsubsection*{A2. Theoretical Justification of the Discrepancy for ED data}
Endpoint dilution (ED) data measure virus infectivity through counts of infected wells across serial dilutions. Unlike RNA data, which we model with log-normal noise, ED data follow a known statistical distribution derived from the Poisson process underlying viral infection. We design a discrepancy measure that respects this structure while remaining compatible with our likelihood-free inference framework.

\paragraph{Structure of ED Data}
At each time point $t_i$, the ED assay tests $R = 3$ independent replicates across $K = 8$ dilution factors $10^{n_k}$ (where $n_k = -1, -2, \dots, -8$ for multi-cycle data, with adjusted ranges for single-cycle data as shown in Tables \ref{tab:MC_ED_data} and \ref{tab:SC_ED_data}). For replicate $r$ at dilution $10^{n_k}$, we observe $c_{ik}^{(r)} \in \{0,1,2,3,4\}$, the number of infected wells out of 4 wells. Under the standard TCID$_{50}$ model, the probability of infection in a well is:
\begin{equation}
p_{ik}(\theta) = 1 - \exp\left(-V(t_i, \theta) \cdot 10^{n_k} \cdot \ln(2)\right)
\label{eq:ed_prob}
\end{equation}
where $V(t_i, \theta)$ is the ODE-predicted virus concentration and $\ln(2)$ converts between TCID$_{50}$ units and infectious virion concentration. The observed counts follow a binomial distribution:
\begin{equation}
c_{ik}^{(r)} \mid \theta \sim \text{Binomial}(4, p_{ik}(\theta))
\label{eq:ed_distribution}
\end{equation}
independently across dilutions, time points, and replicates.

\paragraph{Sufficient Statistics for ED Data}
For a fixed dilution $k$ and time point $i$, the data consist of $R = 3$ independent Binomial$(4, p_{ik})$ observations $\{c_{ik}^{(r)}\}_{r=1}^3$. The joint likelihood is:
\begin{align}
L(p_{ik}) &= \prod_{r=1}^3 \binom{4}{c_{ik}^{(r)}} p_{ik}^{c_{ik}^{(r)}} (1-p_{ik})^{4 - c_{ik}^{(r)}} \nonumber \\
&= \underbrace{\left(\prod_{r=1}^3 \binom{4}{c_{ik}^{(r)}}\right)}_{h(D)} \cdot \underbrace{p_{ik}^{s_{ik}} (1-p_{ik})^{12 - s_{ik}}}_{g(s_{ik}, p_{ik})}
\end{align}
where $s_{ik} = \sum_{r=1}^3 c_{ik}^{(r)}$ is the total number of infected wells across replicates. By the Neyman-Fisher factorization theorem, $s_{ik}$ is a \emph{sufficient statistic} for $p_{ik}(\theta)$. The specific pattern of counts across replicates contains no additional information about $\theta$.

For ED data, we compared four candidate discrepancy measures as presented in Table~\ref{tab:d_ed_comparison}.

\begin{table}[H]
\centering
\scriptsize
\setlength{\tabcolsep}{2pt}
\caption{Comparison of the discrepancy measures for ED count data. Formulas: $L_1$: $\sum |c^{\text{sim}} - c^{\text{obs}}|$; 
$L_2$: $\sqrt{\sum (c^{\text{sim}} - c^{\text{obs}})^2}$; 
Poisson deviance: $2\sum [c^{\text{obs}}\log(c^{\text{obs}}/c^{\text{sim}}) - (c^{\text{obs}}-c^{\text{sim}})]$; 
Binary Hamming: $\sum \mathbf{1}_{c^{\text{sim}} \neq c^{\text{obs}}}$. The $L_1$ distance is selected in this work.}
\label{tab:d_ed_comparison}
\begin{tabular}{lcc}
\toprule
\textbf{Measure} & \textbf{Advantages} & \textbf{Disadvantages} \\
\midrule
$\boldsymbol{L_1}$ & Robust; preserves sufficient statistic; fast & Less sensitive to large deviations than $L_2$
 \\
$L_2$ & Higher efficiency for large counts & Dominated by saturated wells \\
Poisson deviance & Matches true likelihood & Assumes Poisson; unstable for zeros \\
Hamming & Simple; robust & Loses magnitude information \\
\bottomrule
\end{tabular}
\end{table}
The $L_1$ distance was selected because it satisfies three key properties that make it particularly well-suited for ED count data, as we now demonstrate.
\paragraph{Why $L_1$ Distance Is Optimal for ED Data.}

For binomial count data with small sample sizes ($n=4$ wells per dilution, $R=3$ replicates), the $L_1$ distance satisfies all required criteria:

For a single replicate with observed count $c^{\text{obs}}$, define:
\begin{equation}
f(p; c^{\text{obs}}) = \mathbb{E}_{c \sim \text{Binomial}(4,p)}[|c - c^{\text{obs}}|] = \sum_{x=0}^4 |x - c^{\text{obs}}| \binom{4}{x} p^x (1-p)^{4-x}
\end{equation}

This function has three crucial properties:

\begin{enumerate}
\item \textbf{Global minimization at the truth:} For fixed $c^{\text{obs}}$, $f(p; c^{\text{obs}})$ is strictly convex and attains its unique minimum at $p = c^{\text{obs}}/4$, the maximum likelihood estimate.
\item \textbf{Connection to total variation:} To first order in $|p - p'|$, the expected $L_1$ distance between two binomial distributions approximates the total variation distance: $\mathbb{E}[|c - c'|] = 4|p - p'| + O(|p-p'|^2)$.
\item \textbf{Preservation of sufficiency:} For $R=3$ replicates, the total $L_1$ discrepancy $F(p) = \sum_{r=1}^3 f(p; c_r^{\text{obs}})$ is minimized at $p = \frac{1}{3}\sum_r c_r^{\text{obs}}/4 = s^{\text{obs}}/12$, where $s^{\text{obs}} = \sum_{r=1}^3 c_r^{\text{obs}}$ is the sufficient statistic (total infected wells across replicates).
\end{enumerate}

Thus, ranking candidate parameters by their expected $L_1$ discrepancy is equivalent, up to the approximation error inherent in LFI, to ranking them by the distance between the sufficient statistic and its model-predicted expectation.

The $L_2$ distance is problematic because wells with $c=4$ (saturated) produce large squared differences ($16$ when mismatched with $0$) that overwhelm contributions from informative intermediate dilutions where $c \in \{0,1,2,3\}$. The Poisson deviance assumes a Poisson likelihood, which is not strictly correct for binomial counts with $n=4$, and becomes numerically unstable when $c^{\text{obs}} = 0$ (common at low concentrations). The Hamming distance discards the difference between 1 and 4 infected wells, losing sensitivity precisely where the assay provides information.

\subsubsection*{A3. Summary of Theoretical Justification}

We have established that our discrepancy measures for RNA and ED data:

\begin{itemize}
\item Are connected to sufficient statistics via expectation, preserving parameter rankings.
\item Do not require parametric noise assumptions to be exact; they remain valid for any finite-variance noise distribution.
\item Are particularly well-suited for the small-replicate regime ($n_i = 2\text{--}6$) typical of viral kinetic experiments.
\item For ED data, the $L_1$ distance preserves the sufficient statistic, is robust to saturation, and requires no distributional assumptions beyond finite moments.
\end{itemize}

Thus, for both the multi-cycle (MC) and single-cycle (SC) RNA and ED data analyzed in this work, our discrepancy design is theoretically justified for likelihood-free inference.

\newpage

\subsection*{S4 Appendix. Additional figures}
\label{sec:additional_figures}

The following figures provide supplementary diagnostics of the BOLFI inference procedure, including the Gaussian process surrogate model, posterior distributions, and normality assessments of the RNA residuals. These figures support the results presented in Table~\ref{tab:BOLFI inferred parameters} and Fig~\ref{BOLFI_conerReorder} in the main text.

\begin{figure}[!ht]
\centering
\includegraphics[width=\textwidth]{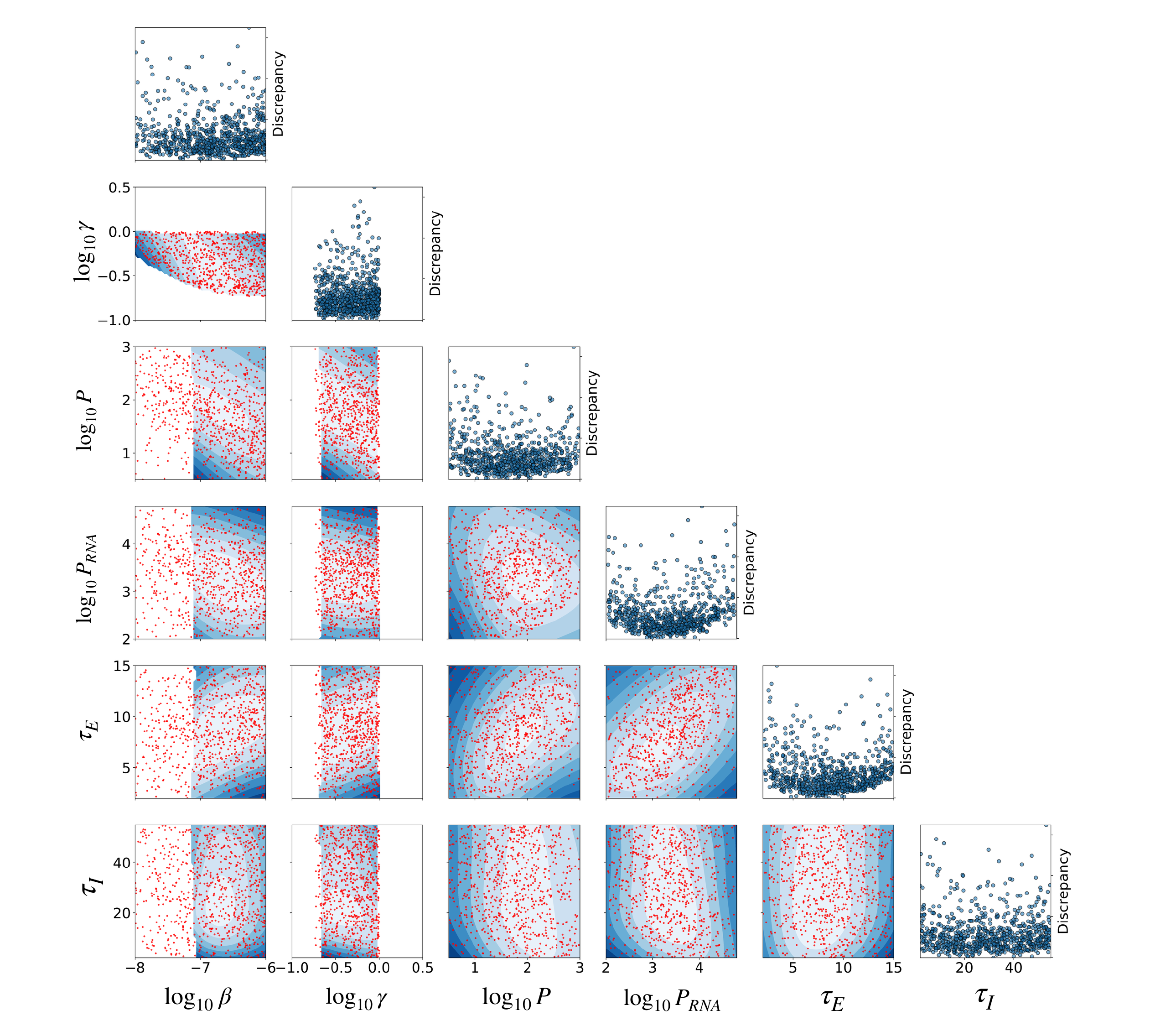}
\caption{\textbf{GP surrogate model pairwise diagnostics.} Matrix of pairwise contour plots of the GP-predicted discrepancy surface. Lighter regions indicate lower discrepancy (better fit), darker regions higher discrepancy (worse fit). Steep gradients indicate well-identified parameters; flat regions indicate weak sensitivity. Red dots: sampled parameter sets; white regions: classifier-excluded invalid areas. Values for $\beta$, $\gamma$, $P$, and $P_{\text{RNA}}$ are shown on the $\log_{10}$ scale; their original units are ml/(cell$\cdot$h), cell/[V], [V]/(cell$\cdot$h), and vRNA/(cell$\cdot$h), respectively. $\tau_E$ and $\tau_I$ are on a linear scale with unit [h].}
\label{fig:gp_pairwise}
\end{figure}

\begin{figure}[!ht]
\centering
\includegraphics[width=\textwidth]{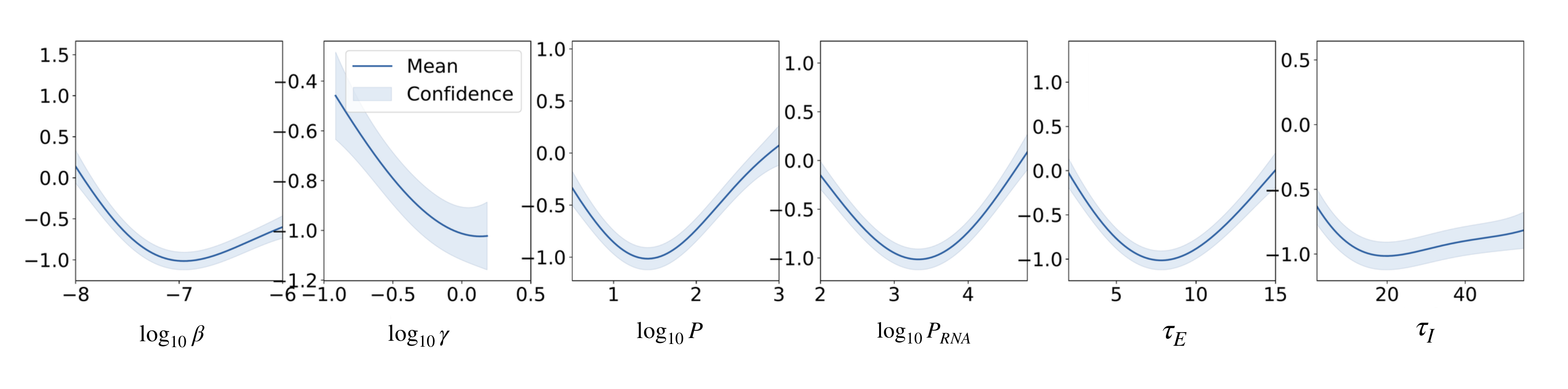}
\caption{\textbf{One-dimensional slices of the GP surrogate.} Solid lines show the predicted mean discrepancy for each parameter, with all other parameters fixed at their optimal values. Shaded regions indicate 95\% confidence intervals. The minima correspond to the MAP estimates. Units are the same as in Fig~\ref{fig:gp_pairwise}.}
\label{fig:gp_1d_slices}
\end{figure}

\begin{figure}[!ht]
\centering
\includegraphics[width=\textwidth]{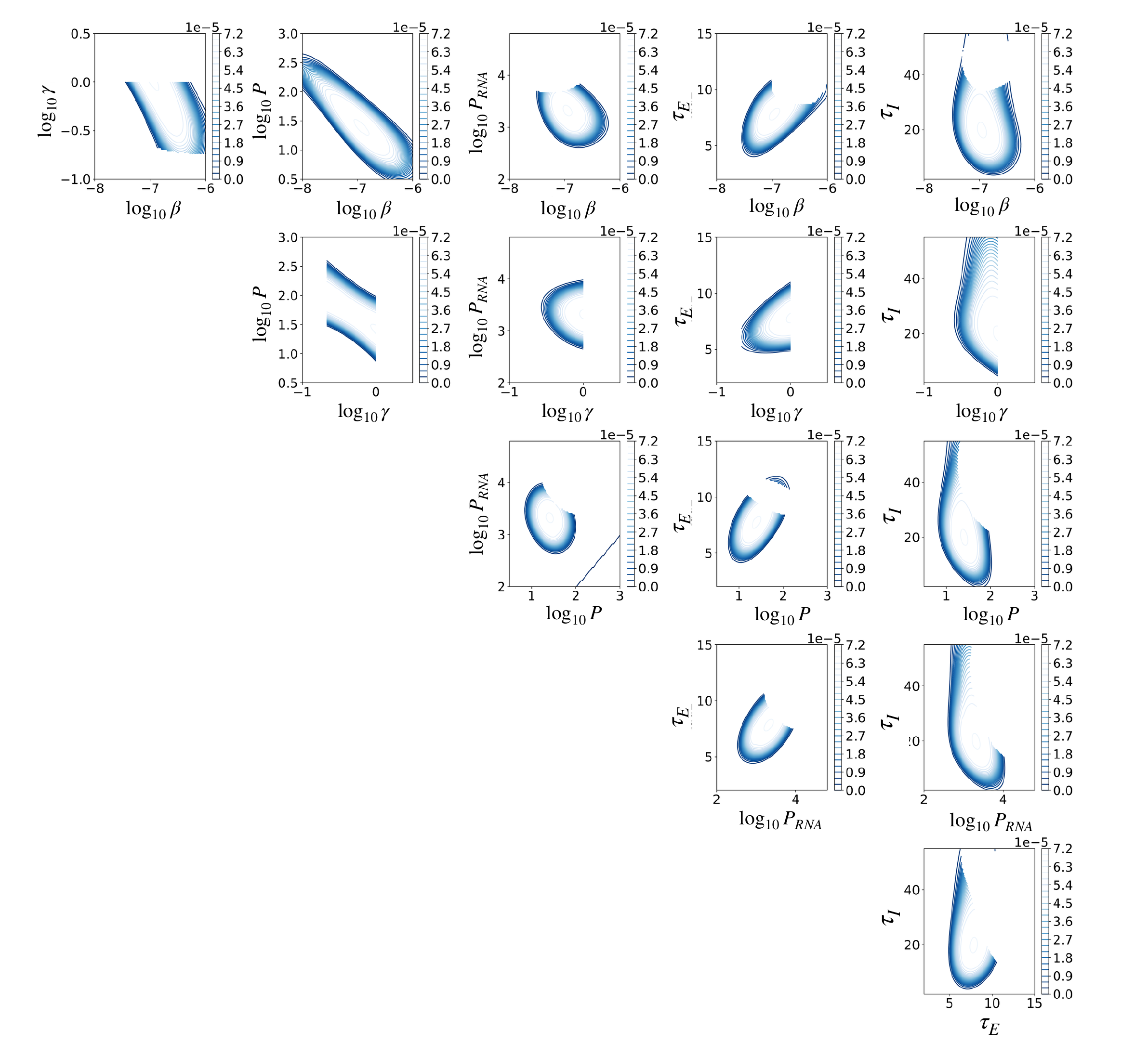}
\caption{\textbf{Two-dimensional posterior density slices.} Slices through the posterior probability density for each parameter pair, with other parameters marginalized. White regions indicate classifier-excluded invalid areas. Units are the same as in Fig~\ref{fig:gp_pairwise}.}
\label{fig:posterior_2d_slices}
\end{figure}

\paragraph{GP surrogate model diagnostics.}
Figs~\ref{fig:gp_pairwise} -- \ref{fig:posterior_2d_slices} provide diagnostic visualizations of the Gaussian process surrogate model learned by BOLFI. Fig~\ref{fig:gp_pairwise} shows pairwise contour plots of the GP-predicted discrepancy surface across all parameter combinations. Lighter regions indicate lower discrepancy (better fit), while darker regions indicate higher discrepancy (poorer fit). Steep gradients in the contour lines indicate well-identified parameters; flat regions indicate weak sensitivity. Red dots show parameter sets sampled during BOLFI's active learning, and white regions indicate areas excluded by the classifier where the simulator fails. The 1D histograms at the top of each column show the discrepancy variation along each individual parameter, while the 2D contour plots below show the corresponding joint discrepancy surfaces. Together, these plots reveal how the discrepancy varies across the parameter space and which regions have been explored.

Figure~\ref{fig:gp_1d_slices} shows one-dimensional slices of the GP surrogate for each parameter, with all other parameters fixed at their optimal values. The minima of these curves correspond to the maximum a posteriori (MAP) estimates, and the width of the shaded 95\% confidence intervals reflects the local uncertainty in the surrogate model. Narrow intervals indicate that the GP has high confidence in the discrepancy surface in that region, while wider intervals suggest areas requiring further exploration.

Figure~\ref{fig:posterior_2d_slices} presents two-dimensional slices through the posterior probability density for each parameter pair, with other parameters marginalized. The white regions again indicate classifier-excluded areas (see Fig.~\ref{fig:gp_pairwise}). The split observed in the $P$ vs $\tau_E$ panel reflects the structure of the parameter space: if the eclipse phase is short ($\tau_E$ small), a lower production rate ($P$) is sufficient to generate the observed viral load; if the eclipse phase is long ($\tau_E$ large), a higher production rate is needed to compensate. Thus, two distinct combinations — short $\tau_E$ with low $P$, and long $\tau_E$ with high $P$ — can produce similar experimental measurements. The adaptive constraint $\hat{P} < \hat{P}_{\text{RNA}}$ further restricts the valid region, creating a boundary that appears as a split in the GP-predicted posterior density. We interpret this split as a real feature of the parameter space, reflecting the trade-off between the timing and magnitude of viral production.

\begin{figure}[!ht]
    \centering
    \includegraphics[width=.8\textwidth]{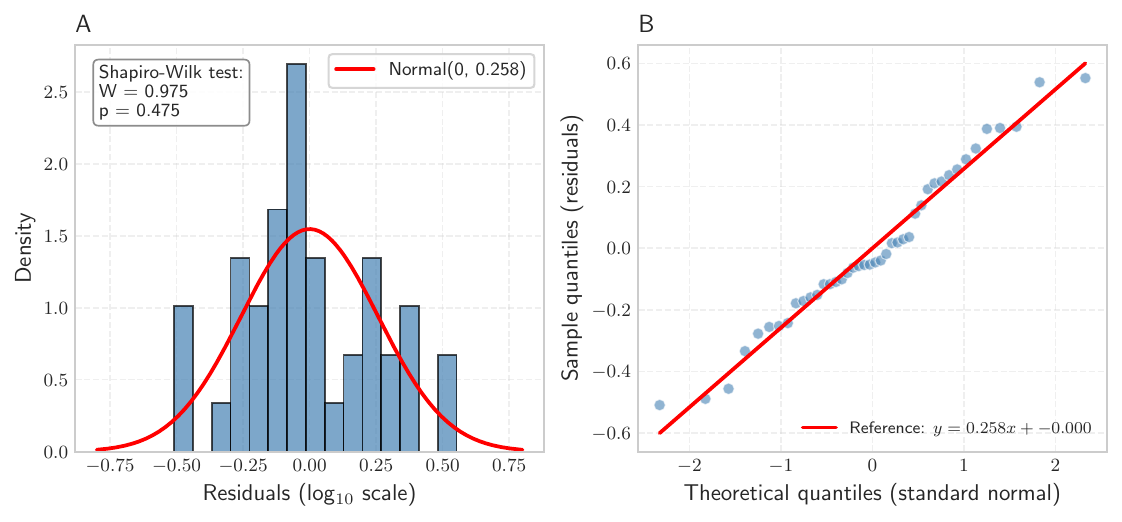}
    \caption{\textbf{Normality assessment of $\log_{10}$-transformed RNA residuals for the multiple-cycle infection data.} 
    (A) Histogram of pooled residuals with overlaid normal distribution ($\sigma = 0.258$, dashed red line) and Shapiro-Wilk test results ($W = 0.982$, $p = 0.78$). 
    (B) Q-Q plot with theoretical reference line ($y = 0.258x$). }
    \label{fig:MC_normality_check}
\end{figure}

\begin{figure}[!ht]
    \centering
    \includegraphics[width=.8\textwidth]{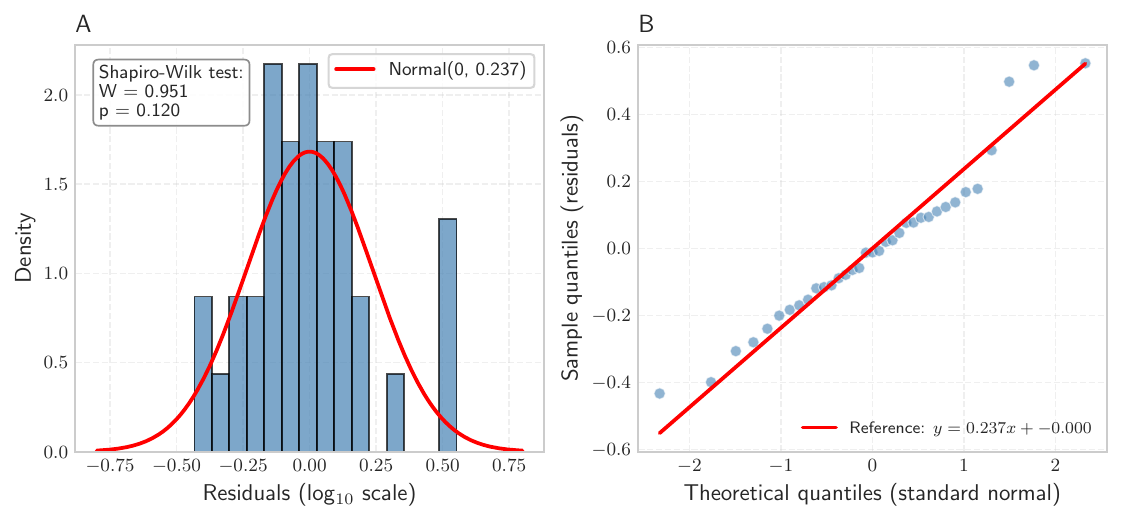}
    \caption{\textbf{Normality assessment of $\log_{10}$-transformed RNA residuals for the single-cycle infection data.} 
    (A) Histogram of pooled residuals with overlaid normal distribution ($\sigma = 0.237$, dashed red line) and Shapiro-Wilk test results ($W = 0.976$, $p = 0.54$). 
    (B) Q-Q plot with theoretical reference line ($y = 0.237x$). }
    \label{fig:SC_normality_check}
\end{figure}

\paragraph{Normality assessment of RNA residuals.}
Figs~\ref{fig:MC_normality_check} and \ref{fig:SC_normality_check} assess the normality assumption for the $\log_{10}$-transformed RNA residuals used in the simulator's noise model. Figure~\ref{fig:MC_normality_check} shows the results for the multiple-cycle (MC) RNA data, with (A) a histogram of the pooled residuals overlaid with a normal distribution ($\sigma = 0.258$, dashed red line) and (B) the corresponding Q-Q plot. The Shapiro-Wilk test result ($W = 0.982$, $p = 0.78$) indicates no significant deviation from normality. Similarly, Fig~\ref{fig:SC_normality_check} shows the results for the single-cycle (SC) RNA data, with a Shapiro-Wilk test result ($W = 0.976$, $p = 0.54$) also indicating no significant deviation from normality. In both cases, the residuals are approximately centered at zero and follow the theoretical normal quantiles. As noted in the main text (Section 2.2), the normality assumption should be viewed as a practical approximation that facilitates parameter estimation, rather than a statistically validated property of the data, given the limited power of normality tests with small sample sizes ($n=35$ for SC, $n=42$ for MC).


\newpage

\subsection*{S5 Appendix. Detailed Calibration Results for All Data Configurations}
\label{sec:Calibration Results all}

\begin{figure}[ht]
\centering
\includegraphics[width=0.9\textwidth]{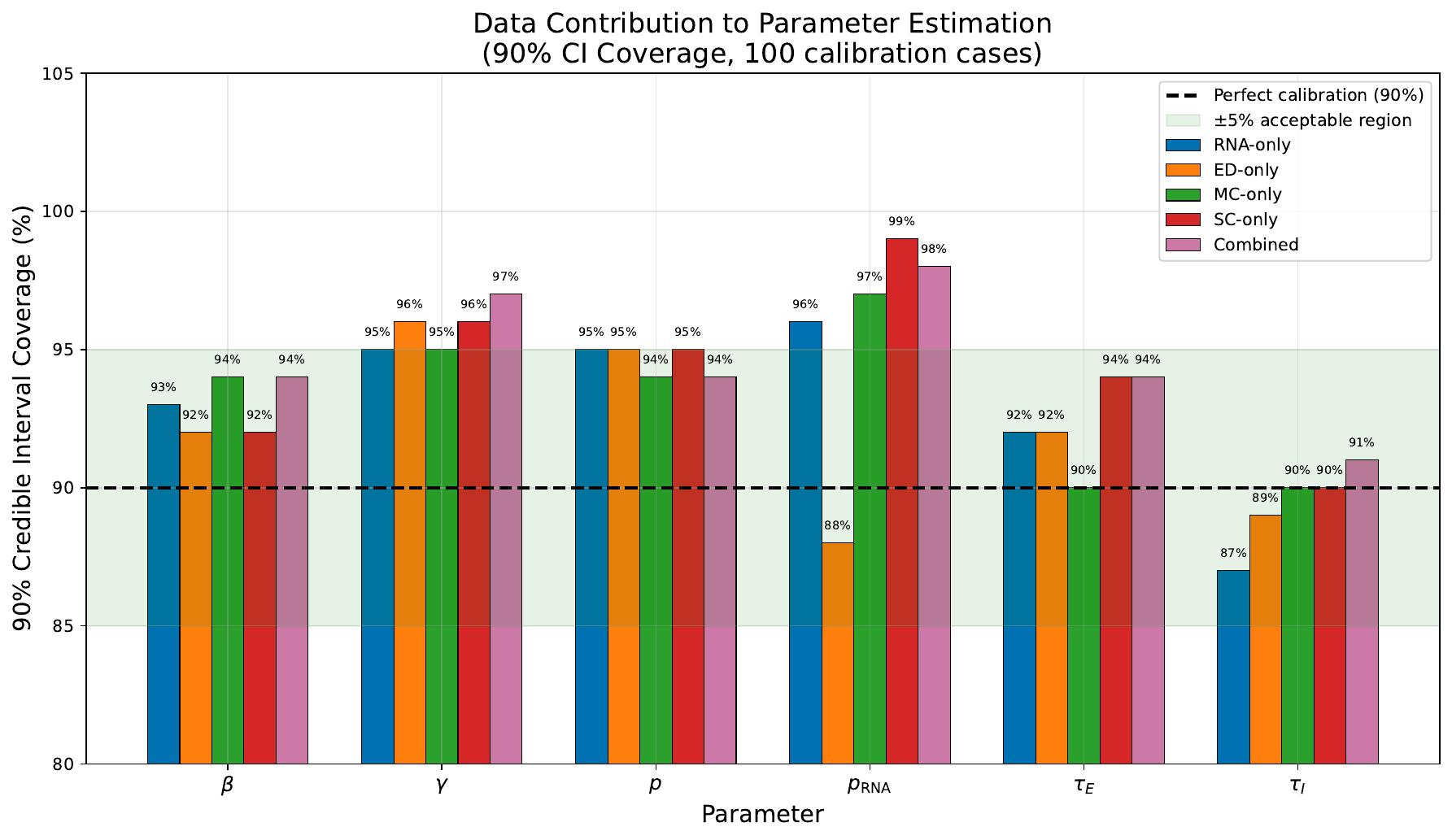}
\caption{{\bf{Comparison of 90\% credible interval coverage}} across five data configurations. The dashed line at 90\% indicates perfect calibration; coverage values within the green shaded region ($\pm 5\%$) are considered statistically consistent with well-calibrated inference.}
\label{fig:data_contribution_90}
\end{figure}

\begin{figure}[ht]
\centering
\includegraphics[width=0.9\textwidth]{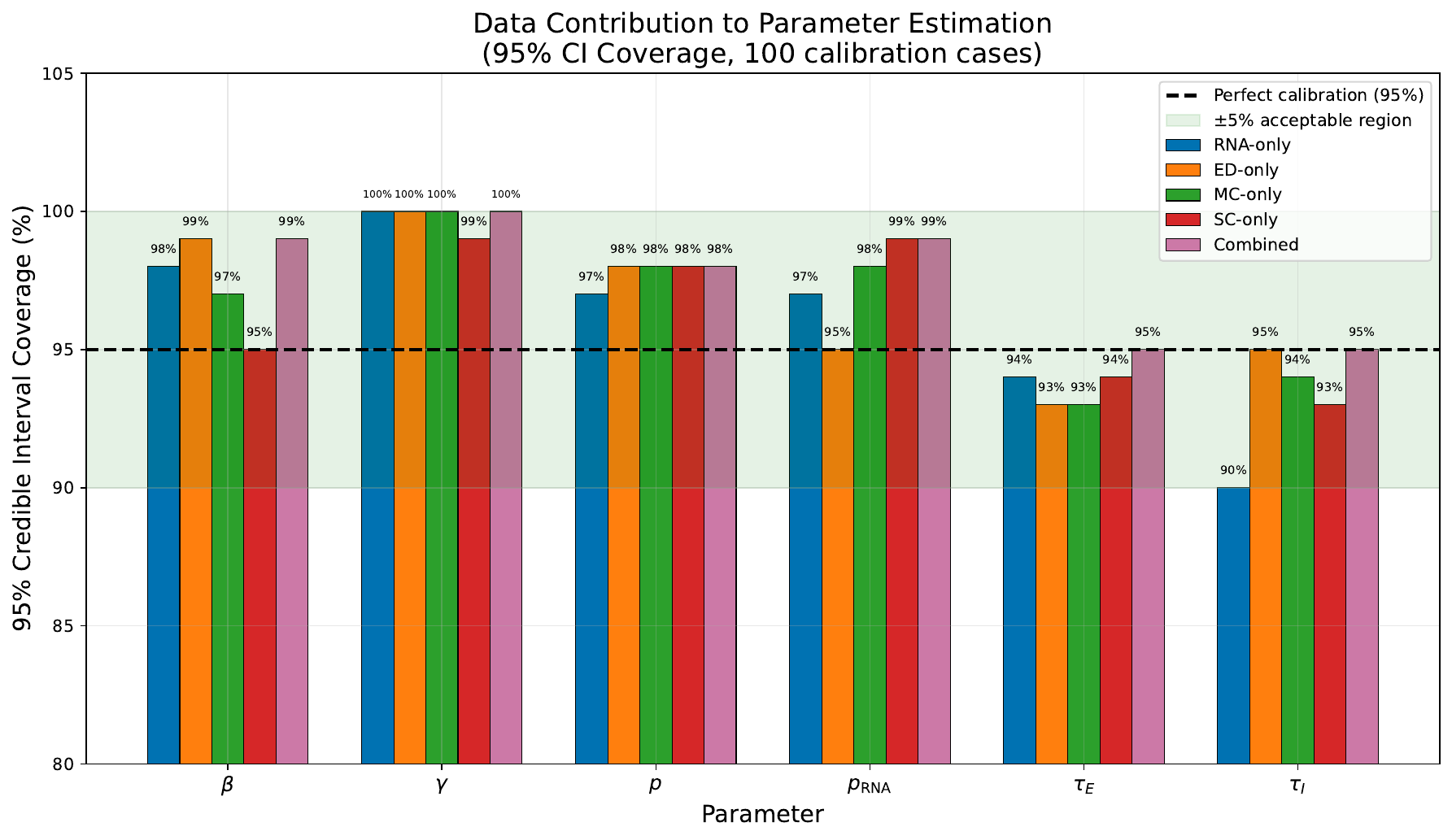}
\caption{{\bf{Comparison of 95\% credible interval coverage}} across five data configurations. The dashed line at 95\% indicates perfect calibration; coverage values within the green shaded region ($\pm 5\%$) are considered statistically consistent with well-calibrated inference.}
\label{fig:data_contribution_95}
\end{figure}

Table~\ref{tab:calibration_all_levels} provides the full numerical results underlying the main text analysis. The 50\% coverage (primary metric) is discussed in the main text; the 90\% and 95\% coverage figures are shown in Figs~\ref{fig:data_contribution_90} and \ref{fig:data_contribution_95}.

\begin{table}[!ht]
\centering
\caption{{\bf{Complete calibration results for each data configuration.}} Values show the percentage of 100 synthetic datasets where the true parameter fell within the 50\%, 90\%, and 95\% credible intervals.}
\label{tab:calibration_all_levels}
\resizebox{\textwidth}{!}{%
\begin{tabular}{l|ccc|ccc|ccc|ccc|ccc}
\hline
 & \multicolumn{3}{c|}{RNA-only} & \multicolumn{3}{c|}{MC-only} & 
\multicolumn{3}{c|}{SC-only} & \multicolumn{3}{c|}{ED-only} & \multicolumn{3}{c}{Combined} \\
\cline{2-16}
 & 50 & 90 & 95 & 50 & 90 & 95 & 50 & 90 & 95 & 50 & 90 & 95 & 50 & 90 & 95 \\
\hline
$\beta$ & 58 & 93 & 98 & 53 & 94 & 97 & 51 & 92 & 95 & 58 & 92 & 99 & 53 & 94 & 99 \\
$\gamma$ & 52 & 95 & 100 & 53 & 95 & 100 & 52 & 96 & 99 & 49 & 96 & 100 & 55 & 97 & 100 \\
$P$ & 60 & 95 & 97 & 54 & 94 & 98 & 58 & 95 & 98 & 45 & 95 & 98 & 49 & 94 & 98 \\
$P_{\text{RNA}}$ & 74 & 96 & 97 & 76 & 97 & 98 & 82 & 99 & 99 & 55 & 88 & 95 & 81 & 98 & 99 \\
$\tau_E$ & 79 & 92 & 94 & 49 & 90 & 93 & 79 & 94 & 94 & 64 & 92 & 93 & 77 & 94 & 95 \\
$\tau_I$ & 46 & 87 & 90 & 54 & 90 & 94 & 47 & 90 & 93 & 57 & 89 & 95 & 49 & 91 & 95 \\
\hline
\end{tabular}%
}
\end{table}


\end{document}